\documentclass[11pt]{article}

\pdfoutput=1

\usepackage[T1]{fontenc}
\usepackage[latin9]{inputenc}
\usepackage[a4paper]{geometry}
\usepackage[active]{srcltx}
\usepackage{amsmath}
\usepackage{amssymb}
\usepackage{esint}
\usepackage{ulem}

\makeatletter

\usepackage{textcomp}

\pdfoutput=1 

\usepackage{jheppub}

\usepackage{etoolbox}
    
    \patchcmd{\maketitle}{\@fpheader}{}{}{}

\newcommand*\xbar[1]{%
  \hbox{%
    \vbox{%
      \hrule height 0.5pt 
      \kern0.3ex
      \hbox{%
        \kern-0.0em
        \ensuremath{#1}%
        \kern-0.0em
      }%
    }%
  }%
}

\usepackage{amsfonts}

\usepackage{bbm}

\newcommand{\be}{\begin{equation}}
\newcommand{\ee}{\end{equation}}
\newcommand{\bea}{\begin{eqnarray}}
\newcommand{\eea}{\end{eqnarray}}

\renewcommand{\d}{\partial}

\def\cL{\mathcal{L}}

\title{\boldmath Asymptotic structure of the gravitational field in five spacetime dimensions: Hamiltonian analysis}

\author[a]{Oscar Fuentealba,}
\author[a,b]{Marc Henneaux,}
\author[c]{Javier Matulich,}
\author[d]{and C\'{e}dric Troessaert}

\affiliation[a]{Universit\'e Libre de Bruxelles and International Solvay Institutes, ULB-Campus Plaine CP231, B-1050 Brussels, Belgium}
\affiliation[b]{Coll\`ege de France, 11 place Marcelin Berthelot, 75005 Paris, France}
\affiliation[c]{Institute for Theoretical Physics, TU Wien, Wiedner Hauptstrasse 8-10/136, A-1040
Vienna, Austria}
\affiliation[d]{Haute-Ecole Robert Schuman, Rue Fontaine aux M\^{u}res, 13b, B-6800, Belgium}

\emailAdd{oscar.fuentealba@ulb.be}
\emailAdd{marc.henneaux@ulb.be}
\emailAdd{javier.matulich@tuwien.ac.at}
\emailAdd{cedric.troessaert@hers.be}

\preprint{}

\abstract{We develop the analysis of the asymptotic properties of gravity in higher spacetime dimensions $D$, with a particular emphasis on the case $D=5$. Our approach deals with spatial infinity and is Hamiltonian throughout. It is shown that the asymptotic symmetry algebra BMS$_5$, which is realized non linearly, contains a four-fold family of angle-dependent supertranslations.  The structure of this non-linear algebra is investigated and a presentation in which the Poincar\'e subalgebra is linearly realized is constructed. Invariance of the energy is studied.  Concluding comments on higher dimensions $D \geq 6$ are also given.}

\makeatother

\begin{document}
  \maketitle \flushbottom

 \newpage{}

\section{Introduction}

The  Bondi-Metzner-Sachs  algebra BMS$_4$ is the algebra of asymptotic symmetries of four-dimensional Einstein gravity in the asymptotically flat context \cite{Bondi:1962px,Sachs:1962wk,Sachs:1962zza,Penrose:1962ij,Madler:2016xju,Alessio:2017lps,Ashtekar:2018lor}.  This infinite-dimensional symmetry has been connected to soft graviton theorems through the corresponding Ward identities \cite{Strominger:2013jfa,He:2014laa,Strominger:2017zoo}, a remarkable result that sparked a lot of activity leading to a deeper physical understanding of classical and quantum properties of gravity.

There has been some debate on how the BMS algebra gets generalized in higher spacetime dimension $D$, especially in odd spacetime dimensions where an expansion of the gravitational radiation field as one goes to infinity along null directions  (``null infinity'') involves inverse fractional powers of the radial distance \cite{Hollands:2003ie,Hollands:2003xp,Hollands:2004ac,Hollands:2016oma}.  The status of the supertranslations, which are the characteristic feature of the BMS$_4$ enlargement of the Poincar\'e algebra, was in particular questioned in some work, leading to a tension with the validity of soft theorems in any $D$ \cite{Schwab:2014xua,Afkhami-Jeddi:2014fia,Kalousios:2014uva,Zlotnikov:2014sva}.   This question was clarified  in the articles \cite{Kapec:2015vwa,Mao:2017wvx,Campiglia:2017xkp,Pate:2017fgt,Aggarwal:2018ilg,Campoleoni:2020ejn,Capone:2021ouo}, which gave boundary conditions at null infinity that accomodated supertranslations (mostly in even spacetime dimensions, except for reference \cite{Capone:2021ouo}, which also considers odd spacetime dimensions). 

In a recent paper \cite{Fuentealba:2021yvo}, we indicated how the asymptotic symmetry structure could be completely understood at spatial infinity in higher spacetime dimension independently of the parity of the dimension, which can be even or odd.  We presented the results in five spacetime dimensions to stress the point that odd spacetime dimensions raised no particular difficulty.  The BMS$_5$ algebra that emerged with our boundary conditions exhibited new interesting features, which were also presented.  

These new features are:
\begin{itemize}
\item The BMS$_5$ algebra contains more supertranslations than anticipated, i.e., supertranslations depend on four (and not one) independent functions of the angles.  The supertranslations are furthermore of two different types: leading ($\mathcal{O}(1)$) and subleading ($\mathcal{O}(r^{-1})$), both yielding non trivial conserved quantities.
\item The ADM expression of the energy (more generally, of the supertranslation charges) is modified by non-linear contributions, which are important for the correct transformation properties of the energy and makes it free from supertranslation ambiguities.
\item The Poisson  bracket algebra of the asymptotic symmetry generators is also modified by non-linear terms.  The form of these non-linear terms depends on which ``basic'' charges are chosen to describe the polynomial algebra of conserved charges.  In \cite{Fuentealba:2021yvo},  the Poisson brackets of the boost charges between themselves were found in particular to get contributions that are cubic in the supertranslation charges with the choices of generators made there.  One might wonder these non-linearities can be removed by redefinitions.
\item There is also a non-vanishing central charge in the Poisson brackets of leading with subleading supertranslations.
\end{itemize}

The purpose of this paper is to expand the analysis of \cite{Fuentealba:2021yvo} by providing detailed information on the derivations.  We analyze further the key new properties of the BMS$_5$ algebra and in particular show that the homogeneous Lorentz subalgebra can be realized linearly, although (apparently unremovable) non-linearities remain in  Poisson brackets involving the boosts and leading supertranslations if one insists on imposing some natural conditions that we outline.  We also comment on how the analysis generalizes to spacetime dimensions bigger than five.  

Our approach  to the BMS symmetry is based on the Hamiltonian formulation of Einstein theory on spacelike hypersurfaces asymptoting spacelike infinity \cite{Dirac:1958sc,Dirac:1958jc,Arnowitt:1962hi,Regge:1974zd}. We follow the methods of  \cite{Henneaux:2018cst,Henneaux:2018hdj,Henneaux:2019yax} where finiteness and invariance of the off-shell action are central ingredients for determining the asymptotic conditions and the asymptotic symmetries. 

More precisely, we impose the requirement that the boundary conditions defining the gravitational phase space in the asymptotic flat case fulfill the following consistency features:
\begin{itemize}
\item The boundary conditions make the action, and in particular the kinetic ``$p \dot{q}$'' term finite so that a well-defined symplectic structure exists.
\item The asymptotic symmetries contain the BMS algebra\footnote{A symmetry is a phase space transformation that preserves the boundary conditions {\em and} leaves the action invariant up to surface terms at the time boundaries.  An asymptotic symmetry is a symmetry that takes the form of a diffeomorphism (rewritten in Hamiltonian form). It is generated by a combination of the constraints plus a surface term at infinity, so that its non trivial physical content is purely determined by its asymptotic behaviour.}.
\end{itemize} 
Using the well-defined symplectic structure, one can then associate to any symmetry a charge-generator in the standard way: symmetries are canonical transformations.   The charge-generators are defined up to the trivial addition of a constant but are otherwise unambiguous.  The arbitrary constants can be fixed by requesting for instance that the charges all vanish for the background Minkowski metric. 

\subsubsection*{A brief review of the $D=4$ results}

Precise boundary conditions that fulfill these requirements were given in  four spacetime dimensions in the papers \cite{Henneaux:2018cst,Henneaux:2018hdj,Henneaux:2019yax}.  The boundary conditions of \cite{Henneaux:2018cst} and those of \cite{Henneaux:2018hdj,Henneaux:2019yax} are actually distinct, but both differ from those of \cite{Regge:1974zd} by a twist (that takes a different form in each case) in the parity conditions on the leading term in the expansion of the gravitational variables near spatial infinity.   The effect of this twist is to elevate the BMS transformations to the status of non trivial gauge symmetries, i.e., gauge symmetries with non-vanishing charges -- while the original boundary conditions of  \cite{Regge:1974zd} effectively factored out the BMS supertranslations, which had zero charge.  A gauge transformation with non-vanishing generator is called ``improper'' \cite{Benguria:1976in} (the terminology ``large gauge transformations''  is sometimes also used). 

We shall focus on the boundary conditions of \cite{Henneaux:2018hdj,Henneaux:2019yax}, which yield a smooth match with null infinity (to leading order) \cite{Henneaux:2018hdj}.  In asymptotically cartesian coordinates, these are,
\begin{align}
g_{ij}&=\delta_{ij}+\frac{\overline{h}_{ij}}{r}+\mathcal{O}\left(r^{-2}\right)\,,\label{eq:001}\\
\pi^{ij}&=\frac{\overline{\pi}^{ij}}{r^2}+\mathcal{O}\left(r^{-3}\right)\,.\label{eq:002}
\end{align}
where $g_{ij}$ and $\pi^{ij}$ are the components of the spatial metric and its conjugate momentum.  The leading orders $\overline{h}_{ij}$ and $\overline{\pi}^{ij}$, which are functions of the angles only, are required to have the following parity properties under the antipodal map:
\begin{itemize}
\item The even part of $\overline{h}_{ij}$ is unrestricted, but its odd part must be given by (the leading order of) the change of the metric under an even spatial diffeomorphism $\zeta^i$ of order $\mathcal{O}(1)$, i.e. $\overline{h}_{ij}^{odd}/r = \zeta_{i,j} + \zeta_{j,i}$.
\item The odd part of $\overline{\pi}^{ij}$ is unrestricted, but its even part must be given by (the leading order) of the change of the momentum under an even normal diffeomorphism $V^\perp$ of order $\mathcal{O}(1)$, i.e. $\overline{\pi}^{ij}_{even}/r^2 = \partial^i \partial^j V^\perp - \delta^{ij} \triangle V^\perp$.
\end{itemize}
We note that  a spatial diffeomorphism of order $\mathcal{O}(1)$ has no action on the leading order $\mathcal{O}(r^{-2})$ of $\pi^{ij}$, while a normal diffeomorphism of order $\mathcal{O}(1)$ has no action on the leading order $\mathcal{O}(r^{-1})$ of $h_{ij}$, which explains why these terms do not appear in the boundary conditions.

If we were to set $\overline{h}_{ij}^{odd} = 0$ and $\overline{\pi}^{ij}_{even}=0$, one would recover the stricter boundary conditions of \cite{Regge:1974zd}.  But $\overline{h}_{ij}^{odd} = 0 =\overline{\pi}^{ij}_{even}$ are gauge conditions that fix improper gauge symmetries and so cannot be imposed.  

It was shown moreover in \cite{Henneaux:2018hdj} that finiteness of the kinetic term (which is not automatic when the leading orders of the fields do not obey strict parity conditions) required that the constraints be of order $\mathcal{O}(r^{-4})$.  Furthermore, the extra condition that the radial-angular components of the coefficient $\overline{h}_{ij}$ of the leading term vanish had also to be imposed in order to guarantee that the asymptotic Lorentz boosts be asymptotic symmetries, which forced $\zeta_i$ to take the form $\zeta_i = \partial_i(r U)$, where $U$ is an odd function of the angles.  These two additional requirements completed the set of  boundary conditions.  

The full set of boundary conditions yields an asymptotic symmetry algebra which is the BMS$_4$ algebra found at null infinity.  The condition on the leading radial-angular components of the metric guarantees in particular that the sizes of the supertranslation subalgebras match.

We shall now generalize the boundary conditions at spatial infinity to higher dimensions, by enforcing the same requirement that these boundary conditions should include diffeomorphisms of order $\mathcal{O}(1)$ as in $D=4$, which cannot not be eliminated by proper gauge fixing. We consider explicitly  the case $D = 5$, which, while simpler than the higher dimensional case, presents already some technical intricacies.
 
\subsubsection*{Organization of paper}

Our paper is organized as follows. In  Section \ref{sec:Theory}, we give the boundary conditions at spatial infinity on the canonical variables and recall the phase space description of the diffeomorphisms.  We also verify  that the boundary conditions make the action, and hence the symplectic form, finite.  In Section \ref{sec:AsympSymm}, we display the asymptotic symmetries and construct their canonical generators.  We show that the supertranslations are of two types (``leading'' and ``subleading'') and involve four functions of the angles on the $3$-sphere. We also stress the presence of non-linear contributions to the generators, including the energy and linear momentum. Section \ref{sec:BMS5} displays the non-linear algebra of these generators.  In Section \ref{sec:Simpli}, we simplify this algebra by non-linear redefinitions of the generators. We succeed in making the Poincar\'e subalgebra (including the boosts) linearly realised, but non-linear terms remain in the brackets of the boosts with the leading supertranslations, which we have not been able to eliminate.  Section \ref{sec:energy} is devoted to the transformation properties of the energy under supertranslations and emphasizes the importance of the non-linear contributions to the charges. In Section \ref{sec:Matching}, we comment on the matching with null infinity, which can proceed along the lines of \cite{Henneaux:2019yqq}.  Finally, we give some remarks on the extension to higher dimensions $D \geq 6$ in the conclusions (Section \ref{sec:Conclusions}).  Technical formulas and derivations, which are sometimes cumbersome, are relegated to appendices.

\section{Hamiltonian action and  asymptotic conditions in five spacetime dimensions}
\label{sec:Theory}

\subsection{Action in Hamiltonian form}

The Hamiltonian action of General Relativity in five spacetime dimensions reads
\be
S[g_{ij},\pi^{ij},N,N^i]=\int dt \left[\int d^{4}x \left(\pi^{ij}\dot{g}_{ij}-N \mathcal{H}- N^{i} \mathcal{H}_{i}\right)-B_{\infty}\right]\,.
\ee
Here, $\pi^{ij}$ is again the conjugate momentum of the four-dimensional spatial metric $g_{ij}$, while $N$ and $N^i$ stand for the lapse and the shift, respectively. The surface integral on the $3$-sphere at spatial infinity $B_\infty$, which depends on the values of the lapse and the shift at infinity, coincides with the energy when $N\rightarrow 1,\, N^i \rightarrow 0$. It will be given below.

Variation with respect to the Lagrange multipliers $N$ and $N^i$ implies the "Hamiltonian" and "momentum" constraints
\begin{align}
\mathcal{H} & =\frac{1}{\sqrt{g}}\left(\pi^{ij}\pi_{ij}-\frac{\pi^{2}}{3}\right)-\sqrt{g}R\approx 0\,,\label{eq:HamG}\\
\mathcal{H}_{i} & =-2\nabla^{j}\pi_{ij}\approx 0 \,.\label{eq:MomG}
\end{align}

Variation with respect to the canonical variables $g_{ij}$ and $\pi^{ij}$ yields the dynamical Einstein equations in Hamiltonian form.

\subsection{Hamiltonian description of diffeomorphisms}

Diffeomorphisms act infinitesimally on the spacetime fields through the Lie derivative,
\be
\delta_{\xi^\mu} \Phi^A = \mathcal{L }_{\xi^\mu} \Phi^A.
\ee
These transformations form a Lie algebra under the Lie bracket of vector fields, which is the vector field associated with the commutator of two infinitesimal diffeomorphisms.

While this set of transformations (with $\xi^\mu$ depending only on $x$) forms a complete set of gauge symmetries of a diffeomorphism-invariant theory, it is by no means the only one.  Indeed, the presentation of the gauge symmetries of  any gauge theory allows changes of a given ``complete set'' by redefinitions of the gauge parameters involving the fields as well as the addition of ``on-shell trivial gauge symmetries'' \cite{Henneaux:1992ig}.  Such changes in the description of the gauge symmetries generically modify the algebra of the transformations  and necessarily appear in the transition to the Hamiltonian description when the symmetry transformations involve the time derivatives of the fields \cite{Henneaux:1992ig}. 

This is in particular the case for diffeomorphisms.  The detailed derivation of the Hamiltonian reformulation of the diffeomorphisms was given in \cite{Fradkin:1977hw}.  The corresponding constraint algebra was geometrically related in \cite{Teitelboim:1972vw} to the ``algebra'' of deformations of spacelike hypersufaces in a riemannian geometry.  It is important to stress that the Lagrangian and Hamiltonian descriptions of the diffeomorphisms are different but equivalent, in the sense that the change of description can be implemented by a canonical transformation in the extended phase space involving also the ghost variables \cite{Henneaux:1992ig}.  How this equivalence is formally achieved in the path integral was established in \cite{Fradkin:1977hw}.  

In the Hamiltonian description, the transformation law of the canonical variables $(g_{ij},\pi^{ij})$ under diffeomorphisms is generated by $\int d^4x (\xi^\perp \mathcal{H} + \xi^i \mathcal{H}_i )+ B$, i.e., $$\delta_{\xi,\xi^i} F = [F, \int d^4x (\xi^\perp \mathcal{H} + \xi^i \mathcal{H}_i )+ B]$$ where $\xi^\perp \equiv \xi$ and $\xi^i$ describe the deformation of the spatial hypersurface and are such that the boundary term $B$ that makes the generator well-defined exists (see below).  The transformations are local in space and thus  the variations $\delta_{\xi,\xi^i} g_{ij}$ and $\delta_{\xi,\xi^i}\pi^{ij}$ at a point $\mathbf{x}$ at finite distance do not depend on the asymptotic behaviour of $(\xi, \xi^i)$, and can be computed assuming that they vanish asymptotically so that $B=0$.  One finds explicitly
\begin{align}
\delta_{\xi,\xi^i} g_{ij} & =\frac{2\xi}{\sqrt{g}}\left(\pi_{ij}-\frac{1}{3}g_{ij}\pi\right)+\mathcal{L}_{\xi}g_{ij}\,,\label{eq:dh-diff}\\
\delta_{\xi,\xi^i}\pi^{ij} & =-\xi\sqrt{g}\left(R^{ij}-\frac{1}{2}g^{ij}R\right)+\frac{\xi}{2\sqrt{g}}g^{ij}\left(\pi^{mn}\pi_{mn}-\frac{\pi^{2}}{3}\right)\nonumber \\
 & \quad-\frac{2\xi}{\sqrt{g}}\left(\pi^{im}\pi_{m}^{j}-\frac{1}{3}\pi^{ij}\pi\right)+\sqrt{g}\left(\xi^{|ij}-g^{ij}\xi_{\quad|m}^{|m}\right)+\mathcal{L}_{\xi}\pi^{ij}\,,\label{eq:dp-diff}
\end{align}
where $\pi\equiv g_{ij}\pi^{ij}$. The spatial Lie derivatives of $g_{ij}$ and $\pi^{ij}$ read
\begin{align}
\mathcal{L}_{\xi}g_{ij} & =\xi^{k}\partial_{k}g_{ij}+g_{ki}\partial_{j}\xi^{k}+g_{kj}\partial_{i}\xi^{k}\,,\\
\mathcal{L}_{\xi}\pi^{ij} & =\partial_{k}\left(\xi^{k}\pi^{ij}\right)-\partial_{k}\xi^{i}\pi^{jk}-\partial_{k}\xi^{j}\pi^{ik}\,.
\end{align}
We shall interchangeably refer to the transformations (\ref{eq:dh-diff})-(\ref{eq:dp-diff}) as the changes of the canonical variables under the ``diffeomorphisms'' or the ``surface deformations''  parametrized by $(\xi, \xi^i)$.

There are additional unwritten terms proportional to the constraints in (\ref{eq:dh-diff}) and (\ref{eq:dp-diff})  if $\xi$ and $\xi^i$ depends also on the fields.  The deformation parameters $(\xi, \xi^i)$ may indeed depend on the canonical variables, and should in fact allow to do so in the most general formulation.  The only restriction is that $(\xi, \xi^i)$ preserve the boundary conditions and have a well-defined canonical generator, i.e., that the corresponding $B$ in $\int d^4x (\xi^\perp \mathcal{H} + \xi^i \mathcal{H}_i )+ B$ actually exists.  Field-dependent redefinitions of the deformation parameters $(\xi, \xi^i)$ may be viewed as field-dependent redefinitions of the constraints (which evidently change their Poisson bracket algebra) accompanied by a redefinition of the improper gauge symmetries when the deformation parameters do not vanish at infinity (in a way compatible with the above restriction).  If the redefinition does involve the fields asymptotically, the resulting redefinition of the canonical symmetry  generators will  in general be non-linear and modify the asymptotic algebra.

\subsection{Boundary conditions}
\label{sec:NewBC}
In five spacetime dimensions, the Schwarzchild solution and its boosted form behave asymptotically as 
\begin{align}
g_{ij}&=\delta_{ij}+\frac{\overline{h}_{ij}}{r^2}+\frac{h^{(2)}_{ij}}{r^3}+\mathcal{O}\left(r^{-4}\right)\,,\label{eq:core1}\\
\pi^{ij}&=\frac{\overline{\pi}^{ij}}{r^3}+\frac{\pi^{(2)ij}}{r^4}+\mathcal{O}\left(r^{-5}\right)\,.\label{eq:core2}
\end{align}
i.e., decay one inverse power of $r$ faster than in four dimensions.   This is also true for the rotating solutions.  

We shall adopt boundary conditions that allow, on top of the ``core behavior'' (\ref{eq:core1}) and (\ref{eq:core2}), improper diffeomorphism terms parametrized by vector fields that contain a piece of order $\mathcal{O}(1)$.   We shall also require, as in four dimensions, that the spatial part of the vector fields generating these improper diffeomorphisms  takes the special form $\zeta_i = \partial_i (r U)$ so that $h_{rA} = 0$ to leading $\mathcal{O}(1)$ order ($x^A$ are angular coordinates on the $3$-sphere).   As in the four-dimensional case, the vanishing of the leading order of $h_{rA}$ is imposed in order to tame the Lorentz boosts. 

The resulting fall-off is given in spherical coordinates by
\begin{align}
    g_{rr} &= 1 + \frac {2 \xbar \lambda} {r^2} +\frac {h^{(2)}_{rr}} {r^3} + \mathcal{O}\left(r^{-4}\right)\,, \label{eq:Agrr}\\
    g_{rA} &= \frac{\xbar \lambda_A}{r} + \frac{h^{(2)}_{rA}}{r^2} + \mathcal{O}\left(r^{-3}\right)\,,\label{eq:AgrA}\\
    g_{AB} &= r^2 \xbar g_{AB} + r \theta_{AB} + \xbar h_{AB} +\frac {h^{(2)}_{AB}} {r} + \mathcal{O}\left(r^{-2}\right)\,, \label{eq:AgAB}\\
    \pi^{rr} &= r \kappa^{rr} + \xbar \pi^{rr} +\frac{\pi^{(2)}_{rr}}{r}+ \mathcal{O}\left(r^{-2}\right)\,, \label{eq:prr}\\
    \pi^{rA} &= \kappa^{rA} + \frac{\xbar \pi^{rA}}{r} + \frac{\pi^{(2)rA}}{r^2} + \mathcal{O}\left(r^{-3}\right)\,,\label{eq:prA}\\
    \pi^{AB} &= \frac{\kappa^{AB}}{r} + \frac{\xbar \pi^{AB}}{r^2}+  \frac{\pi^{(2)AB}}{r^3}+ \mathcal{O}\left(r^{-4}\right)\,.\label{eq:pAB}
\end{align} 
 where the index $A$ refers to the angular coordinates.
 
The terms parametrized by $\xbar \lambda$, $\xbar \lambda_A$, $\xbar h_{AB}$, $\xbar \pi^{rr}$, $\xbar \pi^{rA}$ and $\xbar \pi^{AB}$ constitute the ``core'' (\ref{eq:core1}) and (\ref{eq:core2}).  
They are subleading with respect to the improper diffeomorphism terms, which  involve $\theta_{AB}$, $\kappa^{rr}$, $\kappa^{rA}$ and $\kappa^{AB}$ with explicit expressions
\begin{align}
    \theta_{AB} &= \xbar D_A \xbar D_B U + \xbar g_{AB} U\,, \label{thetaAB} \\
    \kappa^{rr} &= \sqrt {\xbar g} \, \xbar D_A \xbar D^A V\,,\\
    \kappa^{rA} &= \sqrt {\xbar g} \, \xbar D^A V\,,\label{kapparA}\\
    \kappa^{AB} &= \sqrt {\xbar g} \, (\xbar g^{AB} \xbar \triangle V - \xbar D^A \xbar D^B V)\,,\label{kappaAB}
\end{align}
where $\xbar D_A$ denotes the covariant derivative associated to the unit metric $\xbar g_{AB}$ on the $3$-sphere at spatial infinity and $\xbar \triangle\equiv \xbar D_A \xbar D^A$ is its Laplacian.  Here, $U$ and $V$ stand for arbitrary functions of the coordinates of the $3$-sphere which parametrize the vector field of the improper diffeomorphism terms. Specifically, $V$ is its normal component while $U$ determines its tangential component through $\zeta_i = \partial_i(rU)$.  Note that to leading order in the $r^{-1}$-expansion, the diffeomorphisms linearize.  Therefore, the above terms with $U$ and $V$ finite (and not infinitesimal) are  indeed the leading expressions of the diffeomorphism variations of the fields.  The nonlinearities arising because of the non-abelian nature of the diffeomorphism group, only appear in the subleading terms and can be absorbed in the ``core'' part.

A property that will be important later on is that the operator relating $U$ to $\theta_{AB}$ has a non-trivial kernel.  The solutions of $\xbar D_A \xbar D_B f + \xbar g_{AB} f = 0$ can be easily identified by expanding $f$ in spherical harmonics and turn out to be linear combinations of the first spherical harmonics $Y^{1 \ell m}$ ($\ell= 0,1$, $m =-\ell, \cdots, \ell$) (vector representation of the rotation group),
\be
\xbar D_A \xbar D_B f + \xbar g_{AB} f = 0 \qquad \Leftrightarrow \qquad f = \sum_{\ell=0,1}\, \sum_{m=-\ell }^{\ell} a_{\ell m} Y^{1 \ell m} \label{eq:kernel1}
\ee
Functions $f$ in the kernel obey also
\be
(\xbar \triangle + 3) f = 0 \label{eq:kernel2}
\ee
and conversely, if $f$ fulfills (\ref{eq:kernel2}), then it is a linear combination of the first spherical harmonics and is therefore in the kernel of the operator $\xbar D_A \xbar D_B  + \xbar g_{AB} $.

The operator that determines $\kappa^{ij}$ in terms of $V$ has also a non trivial kernel given by the zero mode of $V$, i.e., $V=$ constant.

We emphasize that contrary to the situation in four spacetime dimensions, we impose no parity condition of any kind on the leading orders of the field.  This is because these conditions are not needed for making the symplectic structure finite, as shown below.    In four dimensions, the ``core'' part of the fields and their improper gauge part were of the same order in the expansion in inverse powers of $r$ but differed by parity.  Here, they are at different orders. The same property was found for electromagnetism in dimensions $>4$ \cite{Henneaux:2019yqq}, with which our treatment shares many conceptual features.

The boundary conditions (\ref{eq:Agrr})-(\ref{eq:pAB}) must be completed by further restrictions  on the asymptotic behavior of the constraints, a question to which we now turn.

\subsection{Asymptotic conditions on the constraints}

 To analyse how $\mathcal H$ and $\mathcal H_k$  behave at infinity, it is useful to note that the functions generated by the improper diffeomorphisms parametrized by $U$ and $V$ obey the following identities,
\begin{align}
\overline{D}_{A}\theta_{BC}-\overline{D}_{B}\theta_{AC} & =0\,,\label{Dthetha}\\
\overline{D}_{A}\kappa^{rA}+\kappa_{rr}-\kappa_{A}^{A} & =0\,, \label{kappa1}\\
\overline{D}_{B}\kappa_{A}^{B} +2\kappa_{rA}& =0\,, \label{kappa2}\\
\kappa_{A}^{A}- 2\kappa^{rr}& =0\,.
\end{align}
 Because of these identities, we shall show that the constraint functions behave at infinity  as $\mathcal H = \mathcal O\left(r^{-1}\right)$, $\mathcal H_r = \mathcal O\left(r^{-1}\right)$,  $\mathcal H_A = \mathcal O\left(1\right)$ and not as  $\mathcal H = \mathcal O\left(1\right)$, $\mathcal H_r = \mathcal O\left(1\right)$,  $\mathcal H_A = \mathcal O\left(r^1\right)$ as one might have naively worked out from a direct - and superficial -  counting of the powers of $r$ of the different terms appearing in the constraints. The automatic vanishing of the leading order of the constraints ultimately results from the fact that the leading terms in the fields take the form of a diffeomorphism transformation under which the constraints are invariant, be the diffeomorphisms proper or improper.
 
 To establish explicitly this faster-than-naively-expected decay, we expand the constraints using the asymptotic form of the fields. 
 One then finds that the asymptotic behaviour of the Hamiltonian density is given by
\begin{align}
    \mathcal H &= \sqrt{\xbar g}\left(\xbar \triangle \theta^A_A-\xbar D_A \xbar D_B \theta^{AB}\right) \nonumber \\
    &\quad -\frac 1 {r} \sqrt {\xbar g} \Big(\xbar D^A(\xbar D^B \xbar h_{AB} - \xbar D_A \xbar h)- 2 \xbar \triangle\, \xbar \lambda + 2\xbar D_A \xbar \lambda^A \nonumber \\
    &\qquad \qquad\quad+ \frac 3 4 \theta^{AB} \theta_{AB}
       -  \frac 1 4\theta^2 
       + \frac 1 4 \xbar D^B \theta^{AC} \xbar D_A \theta_{BC} - \frac 1 4 \xbar D_A \theta^{AC} \xbar D^B \theta_{BC}\Big)\nonumber \\
        &\quad+ \frac{1}{r\sqrt{\xbar g}} \Big( 2 \xbar g_{AB}\kappa^{rA}\kappa^{rB}  + 
        \kappa^{AB} \kappa_{AB} - \frac 1 2 (\kappa^A_A)^2\Big)+ \mathcal O\left(r^{-2}\right)\,.
\end{align}
The leading $\mathcal{O}(1)$-term of $\mathcal H$ vanishes by virtue of the identity  \eqref{Dthetha}. Thus,  $\mathcal H \sim r^{-1}$.

For the components of the momentum density, one gets
\begin{align}
      \mathcal H_{r} & = -2 \Big( \xbar D_A \kappa^{rA} +\kappa^{rr} - \kappa^A_A \Big)
          - \frac 2 r \Big(\xbar D_A \xbar \pi^{rA} - \xbar \pi^A_A - \frac 1 2 \theta_{AB} \kappa^{AB}\Big)
          + \mathcal O\left(r^{-2}\right),\\
              \mathcal H_A &  = -2(r \xbar g_{AB} + \theta_{AB} ) \Big( \xbar D_C \kappa^{CB}+2 \kappa^{rB}\Big) 
        \nonumber \\
        & \quad
        -2 \Big(  \xbar D_B \xbar \pi^B_A +\xbar g_{AB} \xbar \pi^{rB}
            - \theta_{AB} \kappa^{rB}+ \frac 1 2 (2\xbar D_B \theta_{AC}
            - \xbar D_A \theta_{BC}) \kappa^{BC}\Big) + \mathcal O\left(r^{-1}\right).
\end{align}
The leading terms of these constraints also vanish, this time by virtue of the identities \eqref{kappa1} and \eqref{kappa2}, which implies  that $\mathcal H_r\sim r^{-1}$ and $\mathcal H_A\sim 1$ as announced.

As we explicitly show below, finiteness of the symplectic structure and of the (off-shell) canonical generators imposes an even stronger decay of the constraints, which is
\begin{equation} \label{eq:faster-fall-off}
    \mathcal H = \mathcal O\left(r^{-3}\right)\,, \quad 
    \mathcal H_r = \mathcal O\left(r^{-3}\right)\,, \quad 
    \mathcal H_A = \mathcal O\left(r^{-2}\right)\,. 
\end{equation}
We thus strengthen the boundary conditions by imposing also these conditions.

The set (\ref{eq:Agrr})-(\ref{eq:pAB}) with the diffeomorphism terms given by (\ref{thetaAB})-(\ref{kappaAB}), supplemented by (\ref{eq:faster-fall-off}), define completely our boundary conditions in five spacetime dimensions.

\subsection{Finiteness of the symplectic structure}

Because of the slower decay of the fields compared with the standard ``core'' part, one must check that the action is finite.  Finiteness of the bulk part of the Hamiltonian is immediate due to the fast decay of the constraints.  The boundary term will be shown to be finite when we discuss the charges. Finiteness of the symplectic structure is more subtle.

With our boundary conditions, the kinetic term in the action might  a priori possess two types of divergences, to wit, linear and logarithmic ones. Indeed, if one replaces the asymptotic expression of the fields into the kinetic term,  one gets (in spherical coordinates),
\begin{equation}
    \int d^4x \, \pi^{ij} \dot{g}_{ij} =
        \int dtdrd^3 x \, \left\{\kappa^{AB}\dot{\theta}_{AB} + \frac{1}{r}\left(2 \kappa^{rr}  \dot{\xbar \lambda}
        + 2 \kappa^{rA} \dot{\xbar \lambda}_A + \kappa^{AB} \dot{\xbar h}_{AB}
        +\xbar \pi^{AB} \dot{\theta}_{AB}\right)+ O(r^{-2})\right\}\,. \label{Kinetic}
\end{equation}

The linear divergence in \eqref{Kinetic} can be shown to be zero by using the definition of the functions $\theta_{AB}$ and $\kappa^{AB}$ in equations \eqref{thetaAB} and \eqref{kappaAB}, respectively.  When this substitution is performed, one gets indeed after integration by parts on $3$-sphere,  that
\begin{equation}
 \int dtdrd^3 x \kappa^{AB}\dot{\theta}_{AB}=\int dtdr\oint d^3x \sqrt {\xbar g}\, 
            \xbar D^C V\left(\left[\xbar D_A ,\xbar D_C\right] \xbar D^A \dot U - 2\xbar D_C \dot U \right)\,, 
 \end{equation}
 which identically vanishes due to the commutator of two covariant derivatives acting on an arbitrary vector field $T_A$ defined on the $3$-sphere $\left[\xbar D_A,\xbar D_C\right]T^B= \delta_{A}^{B}T_C-\delta_{C}^B T_A$. 
 
The treatment of the logarithmically divergent term in \eqref{Kinetic} also makes use of the definition of the fields $\theta_{AB}$, $\kappa^{rr}$, $\kappa^{rA}$ and $\kappa^{AB}$. After some algebra and integration by parts on the $3$-sphere, one gets that the coefficient of the logarithmically divergent term turns out to be a total time derivative term, explicitly
 \begin{equation}
        \oint d^3 x \left(2 \kappa^{rr}  \dot{\xbar \lambda}
        + 2 \kappa^{rA} \dot{\xbar \lambda}_A + \kappa^{AB} \dot{\xbar h}_{AB}
        +\xbar \pi^{AB} \dot{\theta}_{AB}\right)=\partial_t \left(\oint d^3 x\sqrt{\xbar g}\, \xbar D_A V \xbar D^A V \xbar \triangle V \right)\,.
 \end{equation}
Such a term can be removed by adding appropriate surface terms at the time boundaries ({\it and at the time boundaries only}), namely
\be
-\left[\int \frac{dr}{r}\left(\oint d^3 x\sqrt{\xbar g}\, \xbar D_A V \xbar D^A V \xbar \triangle V \right)\right]_{t_0}^{t_1}
\ee
Such a modification is irrelevant for the symplectic structure since it amounts to adding a total derivative $d_V \theta$ to the prepotential $p \, d_Vq$ of the symplectic $2$-form $\Omega = d_V(p \, d_V q) = d_V p \,  d_V q$, leading to the same  $2$-form $\Omega$ ($d_V^2 \theta = 0$).  Here $q \leftrightarrow g_{ij}$, $p \leftrightarrow \pi^{ij}$ and 
$$\Omega = d_V  \int d^4x  \pi^{ij} d_V g_{ij} = \int d^4x d_V \pi^{ij} d_V g_{ij}.$$ 

We note in that context that the equations of motion (with $N\rightarrow 1$ and $N^i\rightarrow 0$) imply $\dot{V}=0$ so that it is natural to take as boundary condition $V(t_0)=V(t_1)$. In that case the divergent contributions from both space-like boundaries $t = t_0$ and $t=t_1$, cancel each other making the logarithmic term actually zero, and removing the need for adding a total time derivative.

The conclusion is that the symplectic structure is completely devoid of divergent terms, and it is thus finite.

\section{Asymptotic symmetries and canonical generators}
\label{sec:AsympSymm}

\subsection{Asymptotic symmetries}

The set of asymptotic conditions at spatial infinity  is preserved under asymptotic diffeomorphisms generated by the following vector fields
\begin{align}
\xi & =br+T  
+\frac{1}{r} T^{(1)} + \hbox{``more''} + \mathcal{O}(r^{-2})\,,  \label{eq:Trans1} \\
\xi^{r} & =W+\frac{1}{r}W^{(1)}+\mathcal{O}(r^{-2})\,,  \label{eq:Trans2}\\
\xi^{A} & =Y^{A}+\frac{1}{r} I^A 
+\frac{1}{r^{2}}I_{(1)}^{A} + \hbox{``more''} +\mathcal{O}(r^{-3})\,, \quad I^A = \overline{D}^{A}W \, .\label{eq:Trans3}
\end{align}
These vector fields contain three relevant orders in the expansion of inverse powers of $r$, which are (in Cartesian coordinates): $\mathcal{O}(r)$, $\mathcal{O}(1)$ and $\mathcal{O}(r^{-1})$.

At order $\mathcal{O}(r)$,  $br= b_i  x^i$ describes the Lorentz boosts while the vectors $Y^A \frac{\partial}{\partial x^A} $ are the Killing vectors of spatial rotations. The function $b=b_i n^i$ where $n^i$ is the outward pointing unit normal to the sphere and the vector components $Y^A$, which depend only on the angles,  obey equations that express  the flat metric Killing equations,
\begin{equation}
\overline{D}_{A}\overline{D}_{B}b+\overline{g}_{AB}b=0\,,\qquad
\mathcal{L}_{Y}\overline{g}_{AB}=Y^{C}\partial_{C}\overline{g}_{AB}+\partial_{A}Y^{C}\overline{g}_{BC}+\partial_{B}Y^{C}\overline{g}_{AC}=0\,.\label{eq:Killings}
\end{equation}

At order $\mathcal{O}(1)$, the functions $T$ and $W$ are arbitrary functions on the $3$-sphere and describe the natural generalization to five dimensions of  four-dimensional supertranslations.  Ordinary time translations correspond to the zero mode $T_0$ while spatial translations correspond to the first spherical harmonics of $W$. 
Since there is no parity condition, $W$ has also a zero mode.  According to (\ref{eq:Trans1})-(\ref{eq:Trans3}), this zero mode defines an asymptotic global shift of the radial coordinate by an angle-independent $\mathcal{O}(1)$-term, a transformation which is permitted in our approach. We shall call ``leading supertranslations'' these $\mathcal{O}(1)$ transformations parametrized by the functions $T$ and $W$ subject to no parity conditions.  As discussed below in the section on the algebra, they commute up to subleading supertranslations, to which we now turn\footnote{By including appropriate field-dependent subleading supertranslations, the leading supertranslations can be redefined to commute exactly, see also below.}.
 
 At order $\mathcal{O}(r^{-1})$, the functions $T^{(1)}$, $W^{(1)}$ and $I^A_{(1)}$  are also arbitrary functions on the $3$-sphere.  They are explicitly written because they define independent non trivial symmetries with non-vanishing charges.  
 However, as we shall show below, only $T^{(1)}$ and the combination $I^{(1)}=\xbar D_A I^{(1)A}-\xbar \triangle W^{(1)}$  actually appear in the expression of the charges, so that transformations for which $I^{(1)} = 0 $ are proper gauge transformations.  These subleading transformations depend therefore effectively on two independent functions of the angles.  Because the transformations parametrized by $T^{(1)}$ and  $I^{(1)} $ commute between themselves and with the supertranslations  (up to central terms, see below), they are also called (subleading) ``supertranslations''. 
 
 Finally,  ``more'' denotes correcting terms of same order as the supertranslations, which are completely determined by $b$ and $W$ and which must be included in order to preserve the boundary conditions ($I^A_{(b)}$) and make the charges integrable ($T_{(b)}$, $T^{(1)}_{(b,W)}$ and $I^{(1)A}_{(b,W)}$).  These terms vanish when $b$ and $W$ are zero. They read
 \be
 \xi(\textrm{``more''}) = T_{(b)} + \frac{1}{r}T^{(1)}_{(b,W)}, \qquad \xi^r(\textrm{``more''}) = 0, \qquad \xi^A(\textrm{``more''}) = \frac{1}{r} I^A_{(b)} + \frac{1}{r^2} I^{(1)A}_{(b,W)},
 \ee
where the correcting term necessary to preserve the asymptotic conditions reads (see end of Appendix \ref{App0})
\begin{equation}
    I^A _{(b)}= \frac {2b}{\sqrt {\xbar g}} \kappa^{rA} = 2 b \xbar D^A V\,. \label{Eq:CorrIA0}
\end{equation}
and  where the explicit expressions of $T_{(b)}$, $T^{(1)}_{(b,W)}$ and $ I^{(1)A}_{(b,W)}$  are respectively given in  (\ref{Tb}), (\ref{T1bW}) and (\ref{I1bW}) below. Similar terms appear already in four spacetime dimensions \cite{Henneaux:2018cst,Henneaux:2018hdj,Henneaux:2019yax}. As (\ref{Eq:CorrIA0}) and (\ref{Tb}), (\ref{T1bW}) and (\ref{I1bW}) indicate, the correcting terms depend on the fields even though the parameters $b$,$Y^A$, $T$, $W$, $T^{(1)}$, $W^{(1)}$ and $I^A_{(1)}$ are taken to be field-independent.

\subsection{Canonical generators}

Given a phase space vector field $X$ (i.e.,  infinitesimal transformations of the canonical variables), the corresponding generator $F_X$ - which exists if the vector field is Hamiltonian, namely, leaves the symplectic form $\Omega$ invariant - is determined through the equation $\iota_X \Omega = - d_V F_X$.
In our case, we know the vector field $X$ (up to asymptotic corrections) and the bulk part of $F_X$, which is $\int d^4x (\xi \mathcal{H} + \xi^i \mathcal{H}_i )$ and actually determines the form of $X$.  What we do not know yet are the correcting terms to be included in the surface deformation parameters $\xi$ and $\xi^i$ (upon which the infinitesimal transformation $X$ clearly depends), and the corresponding surface terms to be added to $\int d^4x (\xi \mathcal{H} + \xi^i \mathcal{H}_i )$.

Because the symplectic $2$-form of five-dimensional gravity takes the standard canonical form $\int d^4 x d_V \pi^{ij} \wedge d_V g_{ij}$, the  Hamiltonian generators of the asymptotic symmetries are obtained through the  method of \cite{Regge:1974zd}, i.e., the surface terms to be added to the bulk part of the generators must be such that these have a well-defined functional derivative (see \cite{Henneaux:2018gfi} for considerations valid in the more general case when the symplectic form itself contains a surface contribution). 

 In order to determine the surface terms to be included in the canonical generators of the surface deformations (\ref{eq:Trans1})-(\ref{eq:Trans3}), one needs to know how the asymptotic fields behave under these transformations.  This can be derived from (\ref{eq:dh-diff}) and (\ref{eq:dp-diff}) using the boundary conditions and the asymptotic form (\ref{eq:Trans1})-(\ref{eq:Trans3}) of the deformation vector fields $(\xi, \xi^i)$.  Since the expressions are rather cumbersome, these are given in appendices, first without the correcting terms (Appendix \ref{App0}) and second with the correcting terms included (Appendix \ref{subsec:improved}).  When the deformation vector fields $(\xi, \xi^i)$ depend on the canonical variables, there are additional terms proportional to the constraints in (\ref{eq:dh-diff}) and (\ref{eq:dp-diff}), but these are irrelevant asymptotically because of the fast decay of $\mathcal H$ and $\mathcal H_k$. 
 
The asymptotic expansion of the extrinsic curvature $K_{ij}$ and other useful quantities are needed for the subsequent derivations.  These can be found in appendix \ref{App2}.

 We start carrying the computations with the transformations of the fields improved to include only the boundary-conditions-preserving correction terms coming from (\ref{Eq:CorrIA0})), in order to stress the need for the other improvement terms ``more'' in the asymptotic form of the transformations.  

The surface integral $Q_{\xi}[g_{ij},\pi^{ij}]$ that must be added to the weakly vanishing bulk part $\int d^4 x (\xi \mathcal H + \xi^i \mathcal H_i)$ to make its functional derivatives well-defined must be such that $\delta Q_{\xi}[g_{ij},\pi^{ij}]$ is given by
\begin{equation}
\delta Q_{\xi}[g_{ij},\pi^{ij}]=\oint d^3x\bigg\{2\sqrt{g}\xi\delta K+\sqrt{g} g^{BC}\delta g_{CA}\Big(\xi K^A_B+\frac{1}{\lambda}(\partial_r\xi-\lambda^D\partial_D\xi)\delta^A_B  \Big)+2\xi^i\delta\pi^r_i-\xi^r\delta g_{jk}\pi^{jk} \bigg\}\,,\label{dQ1}
\end{equation}
since this is indeed minus the term that arises from the integrations by parts necessary to bring the variation of the bulk part $\int d^4 x (\xi \mathcal H + \xi^i \mathcal H_i)$ to the required form.

Replacing the asymptotic form of the fields, of their variations and of $\xi$, $\xi^i$ in \eqref{dQ1}, we obtain 
\begin{align}
\delta Q_{\xi}[g_{ij},\pi^{ij}]&=r^2 \oint d^3x \Big(\sqrt{\xbar g} \, b \delta \theta+2Y^A\delta \kappa^r_B\Big) \label{eq:quadratic} \\
&\quad+ r \oint d^3x\Big[\sqrt{\xbar g}b\Big(2\delta \xbar k+\frac{1}{2}\theta\delta \theta+\frac{1}{2}\theta_{AB}\delta \theta^{AB}\Big)\nonumber\\
&\qquad \qquad\qquad+2Y^A\delta \left(\xbar \pi^r_A+\theta_{AB}\kappa^{rB}\right)+2\Big(\xbar D^A W+I^A_{(b)}\Big)\delta \kappa^r_A+2W\delta \kappa^{rr}\Big] \label{eq:linear}\\
&\quad+ \oint d^3x \sqrt{\xbar g}\Big[2 b \delta k^{(2)} + 2 \Big(T + \frac {b\theta} 2 \Big)\delta \xbar k + \Big(T^{(1)} + b e^{(2)} + \frac {T\theta} 2 \Big) \delta \theta \nonumber\\
&\qquad \qquad\qquad -2 T \delta e^{(2)}- \Big( 2 T^{(1)} + \xbar \lambda^B\partial_B b\Big) \delta \theta + \frac 1 2 \Big( T + \frac {b\theta} 2 \Big)\theta_{AB} \delta \theta^{AB} \nonumber\\
&\qquad \qquad\qquad+ \frac b 2 \theta_{AB} (\delta \xbar h^{AB} - \theta^A_C \delta \theta^{BC}) + b \xbar k_{AB} \delta \theta^{AB}\Big] \nonumber\\
&\quad+\oint d^3x \Big[ 2Y^A\delta \xbar \pi^{(2)r}_A+2\Big(\xbar D^A W+I^A_{(b)}\Big) \delta\left(\xbar \pi^r_A+\theta_{AB}\kappa^{rB}\right)\nonumber\\
&\qquad \qquad\qquad+ 2W\delta \xbar \pi^{rr} -W\kappa^{AB}\delta \theta_{AB}+2I^{(1)A}\delta \kappa^r_{A}+2W^{(1)}\delta \kappa^{rr}\Big]\,,
\end{align}
with $\theta\equiv \theta^A_A$, and
\begin{align}
\bar{k}_{A}^{A} & =\overline{h}_{A}^{A}+\overline{D}^{A}\overline{\lambda}_{A}+3\overline{\lambda}-\frac{1}{2}\theta_{A}^{B}\theta_{B}^{A}\,,\\
k^{(2)} & =\frac{3}{2} h^{(2)}_{rr}+\frac{3}{2}h^{(2)A}_A-\frac{3}{2}\xbar h_{AB} \theta^{AB}-\frac{1}{2} \xbar \lambda  \theta+\frac{1}{2} \theta_{A}^{B}\theta^{C}_{B}\theta^{A}_{C} \nonumber \\
&\quad+\xbar D_{A}h^{(2)A}_r+\frac{1}{2}\xbar \lambda^A \xbar D_{A}\theta- \xbar \lambda^A \xbar D_{B}\theta^{B}_{A}-\theta^{AB}\xbar D_{A}\xbar \lambda_{B} \,,\\
e^{(2)} & =\frac{1}{2}\xbar h^A_{A}-\frac{1}{4}\theta_{AB} \theta^{AB}+\frac{1}{8} \theta^2\,,\\
\xbar \pi^{(2)r}_A&= \pi^{(2)r}_A+\theta_{AB}\xbar \pi^{rB}+h_{AB}\kappa^{rB}+\xbar \lambda_{A}\kappa^{rr}\,.
\end{align}
As stated above,  we have included the contribution $I^A_{(b)}$ that maintains the asymptotic form of the fields, but not the contributions coming from the other correcting terms, the necessity of which comes from integrability and is addressed below Eq (\ref{Eq:VarQFinite}).

 Apparently, the variation of the canonical generator possesses quadratic and linear divergences. However, the quadratic divergent term in \eqref{eq:quadratic} is actually zero by virtue of the definitions of $\theta_{AB}$ and $\kappa^{rA}$, and the asymptotic Killing equations for $b$ and $Y^A$ in \eqref{eq:Killings}. The proof of the vanishing of the linear divergence \eqref{eq:linear} can also be established and can be found in appendix \ref{App3}.   This demands the faster fall-off of the constraints \eqref{eq:faster-fall-off}. 
 
 Accordingly, the charge is free of divergent terms, and its (finite) variation is given by
 \begin{align}
   \delta Q_{\xi}[g_{ij},\pi^{ij}]&  =  \oint d^3x \,\sqrt{\xbar g} \, \Big[
        2 b \delta k^{(2)} + \frac 1 2 \Big( T + \frac {b\theta} 2 \Big)\theta^A_B \delta \theta^B_A 
        + 2 \Big( T + \frac {b\theta} 2 \Big)\delta \xbar k \nonumber \\
       &\quad -2 \Big(T + \frac {b\theta} 2 \Big) \delta e^{(2)}+ b \delta (\theta e^{(2)}) -\Big( T^{(1)} + \xbar \lambda^B\partial_B b\Big)\delta \theta \nonumber \\ 
      &\quad    +  \frac 1 2 \Big(T + \frac {b\theta} 2\Big)\theta\delta \theta - \frac b 4 \theta^2 \delta \theta - b \theta^A_C \theta^C_B \delta \theta^B_A + \frac b 2 \delta \Big(\theta^A_B \xbar h^B_A\Big) \Big]
      \nonumber \\ 
   &\quad     +\oint d^3x \, \frac{4b}{\sqrt{\xbar g}}\Big[  \delta \left(\kappa^r_A \xbar \pi^{rA}+\theta_{AB}\kappa^{rA}\kappa^{rB}\right)
       - \left(\xbar \pi^{rA} +  \theta^A_B\kappa^{rB}\right) \delta \kappa^r_A\Big]
     \nonumber    \\ 
 &  \quad    + \oint d^3x \,\sqrt{\xbar g} \, \Big( \frac b 2 \xbar h^A_B\delta \theta^B_A + b\xbar D_B \xbar \lambda^A \delta \theta^B_A \Big) \nonumber \\
 &\quad+\oint d^3x \Big[ 2Y^A\delta \xbar \pi^{(2)r}_A+2\xbar D^A W \delta\left(\xbar \pi^r_A+\theta_{AB}\kappa^{rB}\right)
+ 2W\delta \xbar \pi^{rr} \nonumber \\
&\qquad \qquad \qquad -W\kappa^{AB}\delta \theta_{AB}+2I^{(1)A}\delta \kappa^r_{A}+2W^{(1)}\delta \kappa^{rr}\Big]\,. \label{Eq:VarQFinite}
\end{align}

In this expression, the terms proportional to the boost and spatial supertranslation parameters $b$ and $W$ are non-integrable, which would imply that neither the boosts nor the spatial supertranslations define Hamiltonian vector fields in phase space. However, we can render these transformations canonical by adding to them correcting diffeomorphisms with suitably chosen parameters $T_{(b)}$, $T^{(1)}_{(b,W)}$ and $I^{(1)A}_{(b,W)}$.  These terms  remove the non-integrable contributions and are subleading with respect to the boosts (for $b$) and the spatial supertranslations (for $W$).  They are the other terms under the label  ``more'' in (\ref{eq:Trans1}) and (\ref{eq:Trans3}), and read explicitly
\begin{align}
T_{(b)}&=-\frac{b\theta}{2} \label{Tb}\,,\\ 
T^{(1)}_{(b,W)}&=- \xbar \lambda^B\partial_B b+\left(\xbar \triangle+3\right)^{-1}\left[\Big(\xbar D^A\xbar D^B+\xbar g^{AB}\Big)\Big(\frac{b}{2}\xbar h_{AB}+b \xbar D_A\xbar \lambda_B\Big)\right]+\frac{1}{\sqrt{\xbar g}}\xbar D_A W \kappa^{rA} \label{T1bW}\,,\\ 
I^{(1)A}_{(b,W)}&=\frac{2b}{\sqrt{\xbar g}}\left(\xbar \pi^{rA} +  \theta^A_B\kappa^{rB}\right)-\frac{1}{2}\theta^A_{B}\xbar D^B W \label{I1bW}\,. 
\end{align}

To be pointed out is the fact that $T^{(1)}_{(b,W)}$ involves the inverse operator $(\xbar \triangle+3)^{-1}$.  Because $\xbar \triangle+3$ has a non trivial kernel, the corresponding expression might not exist.  It does exist in our case, however, because the quantity in square bracket in (\ref{T1bW}) has no component along the $k=1$ spherical harmonics $Y^{1 \ell m}$ since it involves the operator  $\xbar D^A\xbar D^B+\xbar g^{AB}$ which projects out such components.  Although $T^{(1)}_{(b,W)}$ exists, it is not unique and determined up to the addition of an arbirary element $\sum_{\ell, m} \beta_{\ell m} Y^{1 \ell m}$ in the kernel of $\xbar \triangle+3$.  As we shall see, this defines a proper gauge symmetry, i.e., the corresponding generator vanishes when the constraints hold.

Once the correcting diffeomorphisms have been included,  the variation $ \delta Q_{\xi}[g_{ij},\pi^{ij}]$  is  a total derivative in field space.  The corresponding integrated-in-field space surface integral at infinity necessary to complete the bulk term to get a well-defined canonical generator is then given by
 \begin{align}
  Q_{\xi}[g_{ij},\pi^{ij}]&  =  \oint d^3x \,\sqrt{\xbar g} \, b\Big[
        2 k^{(2)} + \theta e^{(2)} - \frac{1}{12} \theta^3 - \frac{1}{3} \theta^A_C \theta^C_B\theta^B_A \nonumber \\
        &\qquad \qquad \qquad \quad+ \frac{1}{2} \theta^A_B \xbar h^B_A+\frac{4}{\xbar g}\Big(\kappa^r_A  \xbar \pi^{rA}+\theta_{AB}\kappa^{rA} 
        \kappa^{rB}\Big)\Big]  \nonumber \\
       &\quad+\oint d^3x\sqrt{\xbar g}\Big[
         2T\Big(\xbar k -  e^{(2)}+  \frac 1 8 \theta^2 + \frac 1 8 \theta^A_B \theta^B_A\Big) -T^{(1)} \theta \Big] \nonumber \\ 
 &\quad+\oint d^3x \Big[ 2Y^A \xbar \pi^{(2)r}_A
+2I^{(1)A}\kappa^r_{A}+2W^{(1)} \kappa^{rr} \nonumber \\
 &\qquad \qquad \qquad \quad+ 2W\Big( \xbar \pi^{rr}-\xbar D_A \xbar \pi^{rA}-\frac{1}{2}\xbar D^A(\theta_{AB}\kappa^{rB}) \Big)\Big]\,.
\end{align}

One can rewrite this surface integral in a manner closer to standard notations for the homogeneous Lorentz group.  To that end, we introduce the standard basis $\{Y^{ijA} \frac{\partial}{\partial x^A}\}$ of rotation Killing vectors 
\be
Y^{ijA} \frac{\partial}{\partial x^A} \equiv x^{[i}\delta^{j]k} \frac{\partial}{\partial x^k} = x^{[i} e^{j]A} \frac{\partial}{\partial x^A} = - Y^{jiA} \frac{\partial}{\partial x^A}
\ee
where $e^{jA}(x^B)$ are the following tangent vectors to the $3$-sphere,
\be
e^{jA}(x^B) = \delta^{jk} \frac{\partial x^A}{\partial x^k}
\ee
so that
\be
\frac{\partial}{\partial x^k} = n_k \frac{\partial}{\partial r} + e^A_k \frac{\partial}{\partial x^A}, \qquad n_k(x^A) = \frac{\partial r}{\partial x^k} = \frac{x_k}{r} \, .
\ee
(Indices referring to the asymptotically cartesian coordinates $x^k$ are raised and lowered  in these formulas with $\delta^{ij}$ and $\delta_{ij}$.)  We thus have
\be
Y^A \frac{\partial}{\partial x^A} = \frac{1}{2}b_{ij} Y^{ijA} \frac{\partial}{\partial x^A}, \qquad b_{ij} = - b_{ji}, \qquad Y^A=\frac{1}{2}b_{ij}x^i e^{jA}.
\ee

In these notations, the surface integral $Q_{\xi}[g_{ij},\pi^{ij}]$ takes the familiar form of ``one charge for each independent asymptotic algebra parameter'',
\begin{equation}
Q_{\xi}[g_{ij},\pi^{ij}]=b_iB^i+\frac{1}{2}b_{ij}M^{ij}+Q_{T}+Q_{W} + Q_{T^{(1)}}+Q_{I^{(1)}}\,,\label{eq:Qxigpi}
\end{equation}
where
\be
 b=b_i n^i (x^A), \qquad  I^{(1)}=\xbar D_A I^{(1)A}-\xbar \triangle W^{(1)}
 \ee
 as we already indicated and 
\begin{align}
& B^i = \oint d^3x \,\sqrt{\xbar g} \, n^i \Big[
        2 k^{(2)} + \theta e^{(2)} - \frac{1}{12} \theta^3 - \frac{1}{3} \theta^A_C \theta^C_B\theta^B_A \nonumber \\
        &\qquad \qquad \qquad \quad+ \frac{1}{2} \theta^A_B \xbar h^B_A+\frac{4}{\xbar g}\Big(\kappa^r_A  \xbar \pi^{rA}+\theta_{AB}\kappa^{rA} 
        \kappa^{rB}\Big)\Big]\,,  \\
& M^{ij} =  \oint d^3x  \,  x^{[i} e^{j]A} \, \xbar \pi^{(2)r}_A \, , \\
&Q_{T}=\oint d^3x \sqrt{\xbar g }\,T\, \mathcal T \,,\quad Q_{W}=\oint d^3x\,W\, \mathcal W\,, \label{eq:QTandQW}\\
&Q_{T^{(1)}}=-\oint d^3x \sqrt{\xbar g }\, T^{(1)}\theta = -\oint d^3x \sqrt{\xbar g }\, T^{(1)}\Big((\xbar \triangle+3)U\Big)\,, \\
& Q_{I^{(1)}}= 2 \oint d^3x  \Big(I^{(1)A}\kappa^r_{A}+W^{(1)} \kappa^{rr} \Big)= -2\oint d^3x \sqrt{\xbar g} I^{(1)}V \, .
\end{align}
Here we have set
\be
 \mathcal T= 2  \Big(\xbar k -  e^{(2)}+  \frac 1 8 \theta^2 + \frac 1 8 \theta^A_B \theta^B_A\Big)\, , \qquad
 \mathcal W =  2\Big( \xbar \pi^{rr}-\xbar D_A \xbar \pi^{rA}-\frac{1}{2}\xbar D^A(\theta_{AB}\kappa^{rB}) \Big)
\ee

Since the charges can be expressed in terms of $\theta_{AB}$ and $\kappa^{ij}$, they are free from the ambiguities that plague the functions $U$ and $V$.  However, the  expressions of $Q_{T^{(1)}}$ and $Q_{I^{(1)}}$ in terms of $U$ and $V$ exhibit two features which will play an important role below.

\begin{itemize}
\item The writing of the charge $Q_{T^{(1)}}$ in terms of $U$ makes it clear that it vanishes if $U$ is in the kernel of the operator $\xbar \triangle+3$.  But the symmetry of the operator $\xbar \triangle+3$  implies at the same time that $Q_{T^{(1)}}$ also vanishes when $T^{(1)}$ is in the same kernel, i.e., is a linear combination of the first spherical harmonics, as announced. This means that these transformations are pure gauge. We thus see that the ambiguity in the function $U$, which describes part of the improper gauge components of the fields at leading order (associated with spatial supertranslations), is paired with the absence of a true improper gauge freedom at the subleading order (in the time supertranslations $T^{(1)}$), described by the same function space (first spherical harmonics).  

\item A somewhat similar phenomenon (pairing of leading and subleading orders) characterizes $Q_{I^{(1)}}$.  To see this, we observe that the expression of $Q_{I^{(1)}}$ in terms of $V$ explicitly shows that only the combination $I^{(1)}=\xbar D_A I^{(1)A}-\xbar \triangle W^{(1)}$ defines an improper gauge symmetry.  Now, $I^{(1)}$ is requested to have no zero mode but is otherwise an arbitrary function on the sphere.  Indeed, it follows from its definition that $I^{(1)}$ has no zero mode.  And conversely, given a function $I^{(1)}$ with no zero mode, one can always find a vector field $I^{(1)A}$ and a function $W^{(1)}$ such that $I^{(1)}=\xbar D_A I^{(1)A}-\xbar \triangle W^{(1)}$.  Take for example $I^{(1)A} = 0$, $W^{(1)} = - \xbar \triangle^{-1} I^{(1)}$. It follows that the ambiguity in $V$, described by the same functional space (constants), is indeed irrelevant in the charge $Q_{I^{(1)}}$.  The ambiguity in the function $V$, which describes the other part of the improper gauge components of the fields at leading order (associated with time supertranslations), is paired with the absence of a true  improper gauge freedom at the subleading order (in the space supertranslations $I^{(1)}$).

The ambiguity in the leading terms $U$ and $V$ has thus an interesting ``dual''  impact on the subleading supertranslations through the form of their charges, which are such that the first spherical harmonics of the subleading {\it time }supertranslations parametrized by $T^{(1)}$ and the zero mode of the subleading {\it space} supertranslations parametrized by $ I^{(1)}$ are pure gauge and do not define non trivial improper diffeomorphisms with a non trivial physical action.
\end{itemize}

Although the first spherical harmonics of the leading spatial supertranslations $W$ (ordinary spatial translations) and the zero mode of the leading time supertranslations (ordinary time translations) have no physical action on the improper diffeomorphism components of the fields parametrized by $U$ and $V$, they do act non trivially on the ``core''.  The corresponding charges are the linear momentum and the energy.  There is no action on the improper diffeomorphism components because the associated vector fields are exact Killing vectors of the flat metric. This distinction between ordinary spacetime translations and pure supertranslations has a clear origin in the linear theory, where the Killing vectors of the background flat metric define global charges of the Pauli-Fierz theory and become improper gauge diffeomorphisms upon switching on the interactions, while the pure supertranslations are improper diffeomorphisms already in the linear theory \cite{Fuentealba:2020ghw}.

The asymptotic symmetry algebra is spanned by  the homogeneous Lorentz generators $B^i$ and $M^{ij}$and the integrands $\mathcal T (x^A)$, $\mathcal W (x^A)$, $\theta(x^A) = (\xbar \triangle+3)U(x^A)$ and $V(x^A)$ which appear in the expressions of the integrated supertranslation charges.  These integrands are arbitrary functions on the sphere, modulo the fact stressed already a couple of times that $(\xbar \triangle+3)U(x^A)$ has no vector mode ($k=1$ in the spherical harmonics expansion) and the zero mode ($k=0$ in the spherical harmonics expansion) of $V(x^A)$ is pure gauge.
In fact, it will be more convenient to use $U(x^A)$ itself and
\be
K_A = \xbar D_A V = \frac{\kappa^r_{\; A}}{\sqrt{\xbar{g}}}
\ee
which is permissible provided $U$ always appears with the projector $(\xbar \triangle+3)$ or with the operator $(\xbar D_A \xbar D_B+\xbar g_{AB})$ acting on it,  which both remove the $k=1$ mode.  The use of $K_A$ instead of $V$ is also allowed, since $K_A$ contains the same physical information as $V$: the knowledge of $K_A$ is equivalent to the knowledge of $V$ up to its irrelevant zero mode $k=0$.

\section{Asymptotic symmetry algebra BMS$_5$}
\label{sec:BMS5}

\subsection{Poisson versus Dirac brackets}
We have derived that the canonical generators of the asymptotic symmetries read
\begin{equation}
P_{\xi}[g_{ij},\pi^{ij}]=\int d^4x \left(\xi\mathcal{H}+\xi^i \mathcal{H}_i \right)+Q_\xi[g_{ij},\pi^{ij}]\,,  \label{eq:B+SGen}
\end{equation}
with $Q_\xi[g_{ij},\pi^{ij}]$ given by (\ref{eq:Qxigpi}).
We now turn to the question of computing the algebra $\{P_{\xi}[g_{ij},\pi^{ij}], P_{\eta}[g_{ij},\pi^{ij}]\}$ of these generators, where we use in this paper the notation  $\{,\}$ for the Poisson bracket.  We will call this algebra ``the BMS$_5$ algebra''.

Two generators that differ solely in the bulk -- and hence, which are equal when the constraints hold --, define the same observable and must be identified.  Only the asymptotic value is relevant. Furthermore, it is known on general grounds that the bracket $\{P_{\xi}[g_{ij},\pi^{ij}], P_{\eta}[g_{ij},\pi^{ij}]\}$ of two well-defined generator is a well defined generator \cite{Brown:1986ed}. Therefore, if one knows its asymptotic form, one knows that it is accompanied by an appropriate bulk term $ \int d^4x \left(\zeta\mathcal{H}+\zeta^i \mathcal{H}_i \right)$ such that the sum is a well-defined generator - in particular, the bulk vector field $(\zeta, \zeta^i)$ asymptotically matches the asymptotic symmetry parameters of the bracket.

 For that reason, it is customary to focus exclusively on the physically relevant surface integrals at infinity and not write explicitly (any one of) the accompanying bulk terms $\int d^4x (\xi\mathcal{H}+\xi^i \mathcal{H}_i )$, where $(\xi, \xi^i)$ are bulk vector fields that asymptotically matches the given parameters of the asymptotic symmetries. To evaluate the asymptotic part of the bracket $\{P_{\xi}[g_{ij},\pi^{ij}], P_{\eta}[g_{ij},\pi^{ij}]\}$, one observes that  $\{P_{\xi}[g_{ij},\pi^{ij}], P_{\eta}[g_{ij},\pi^{ij}]\}$ is equal to $\delta_\eta P_{\xi}[g_{ij},\pi^{ij}] = - \delta_\xi P_{\eta}[g_{ij},\pi^{ij}]$ and that $\delta_\eta \int d^4x \left(\xi\mathcal{H}+\xi^i \mathcal{H}_i \right)$ is equal to zero when the constraints hold, so that the boundary term in $\{P_{\xi}[g_{ij},\pi^{ij}], P_{\eta}[g_{ij},\pi^{ij}]\}$ is equal to $\delta_\eta Q_\xi[g_{ij},\pi^{ij}]$ (equivalently, $-\delta_\xi Q_\eta[g_{ij},\pi^{ij}]$).  This method provides an efficient way to automatically compute the complete brackets of the generators, which include two types of contributions : (i) the contributions coming from the surface deformation algebra of the deformation vector fields; and (ii) the additional  contributions due to the fact that the deformation vector fields $(\xi, \xi^i)$ are field-dependent.  It is the method that we shall adopt below.
 
 It should be stressed that we are allowed to use these well-known properties of the symplectic formalism in the derivation of the charge algebra because our boundary conditions guarantee a well-defined (finite) off-shell symplectic structure.  The standard canonical formalism applies.

One could alternatively fix the gauge in the bulk and work with the Dirac bracket.  This procedure is equivalent to the one just outlined because  Poisson brackets and Dirac brackets of observables coincide when the constraints and gauge conditions hold.

\subsection{Explicit form of the algebra}
A direct but somewhat tedious computation of $\delta_{\xi_2} Q_{\xi_1}[g_{ij},\pi^{ij}]$ shows that the bracket 
$\{ P_{\xi_1}[g_{ij},\pi^{ij}],P_{\xi_2}[g_{ij},\pi^{ij}]\}$ is the sum of a familiar linear term in the asymptotic charges, augmented by central terms and non-linear contributions that are polynomial in the asymptotic charges,
\begin{equation}
\Big\{ P_{\xi_1}[g_{ij},\pi^{ij}],P_{\xi_2}[g_{ij},\pi^{ij}]\Big\} =P_{\hat \xi}[g_{ij},\pi^{ij}]+C_{\lbrace \xi_1,\xi_2\rbrace}+\Lambda_{\lbrace \xi_1,\xi_2\rbrace}[g_{ij},\pi^{ij}]\,.
\end{equation}
We will describe in turn each of these contributions by giving only their asymptotic form, aware that there are of course also bulk contributions proportional to the constraints that accompany them so that the generators are well-defined, but not writing them explicitly because they carry no particular physical information relevant to our analysis.  The resulting ``ignorance'' corresponds to a pure gauge generator.

Before turning to the explicit display of the various terms in the algebra, we mention that we verified the consistency check that the Jacobi identities held for all double Poisson brackets.

\subsubsection{Linear terms}
\label{subsec:LinearTerms}
The linear term $P_{\hat \xi}[g_{ij},\pi^{ij}]$ takes the form (\ref{eq:B+SGen}) with
\begin{eqnarray}
\hat{Y}^A&=&Y_1^{B}\partial_{B}Y_2^A+\xbar g^{AB}b_1\partial_B b_2-(1\leftrightarrow2)\,,\\
\hat b&= &Y_1^B\partial_B b_2-(1\leftrightarrow2)\,,\\
\hat T&=& Y_1^B\partial_B T_2-4b_1 W_2-\partial_A b_1 \xbar D^AW_2-b_1\xbar D_A\xbar D^AW_2-(1\leftrightarrow2)\,,\\
\hat W&=& Y_1^B\partial_B W_2-b_1 T_2-(1\leftrightarrow2)\,,\\
\hat{T}^{(1)} & =&Y_{1}^{B}\partial_{B}T_{2}^{(1)}+I_{2}^{(1)A}\partial_{A}b_{1}\nonumber \\
  &&\quad-(\xbar\triangle+3)^{-1}\Big[\Big(\xbar D_{A}\xbar D_{B}+\xbar g_{AB}\Big)\Big(b_{1}\xbar D^{A}I_{2}^{(1)B}\Big)\Big]+2b_{1}W_{2}^{(1)}-\partial_{A}b_{1}\xbar D^{A}W_{2}^{(1)}+b_{1}\xbar\triangle W_{2}^{(1)} \nonumber \\
  && \quad -(1\leftrightarrow2)\,, \label{eq:hatT1}\\
\qquad & \Leftrightarrow &  \quad \hat{\tilde{T}}^{(1)}= Y_1^B\partial_B \tilde{T}^{(1)}_2-4b_1I^{(1)}_2-\partial_A b_1 \xbar D^A \tilde{I}^{(1)}_2-b_1\xbar D_A\xbar D^A \tilde{I}^{(1)}_2-(1\leftrightarrow2)\,,\\
\hat I^{(1)A}&=&Y_1^B\partial_B I^{(1)A}_2-I^{(1)B}_2\partial_BY_1^A+\xbar g^{AB}\Big(T^{(1)}_2\partial_B b_1-b_1\partial_B T^{(1)}_2\Big) +T_{1}\xbar D^{A}T_{2} -(1\leftrightarrow2) \,. \\
\hat{W}^{(1)} & = &Y_{1}^{B}\partial_{B}W_{2}^{(1)}-(1\leftrightarrow2)\,, \\
\qquad & \Rightarrow & \hat{I}^{(1)}= Y_1^B\partial_B I^{(1)}_2-b_1 \tilde{T}^{(1)}_2 + T_{1}\xbar \triangle T_{2}-(1\leftrightarrow2) \,.
\end{eqnarray}
where we have set
 \be
 \tilde T^{(1)} = (\xbar \triangle+3) T^{(1)}
 \ee
 
 The function $\hat T^{(1)}$ in (\ref{eq:hatT1})  is well defined because the operator $(\xbar \triangle+3)^{-1}$ acts on a function with no component involving the first spherical harmonics.  Of course, it is defined only up to a linear combination of the first spherical harmonics.  The function $ \hat{\tilde{T}}^{(1)}$ carries no such ambiguity since it involves the projector $(\xbar \triangle+3)$.  
  
One can also easily verify directly that  the function $\hat{I}^{(1)}$ has no zero mode. Both $\tilde{I}^{(1)}_1$ and $\tilde{I}^{(1)}_2$ transform in representations of the rotation group that do not involve the trivial representation and hence there is no zero mode in $Y_1^B\partial_B I^{(1)}_2$ or $Y_2^B\partial_B I^{(1)}_1$. Furthermore the tensor product of the vector representation ($b_1$ or $b_2$) with the representation in which $\tilde{T}^{(1)}_2$ or $\tilde{T}^{(1)}_1$ transform, which does not contain the vector representation, does not contain itself the trivial representation.

\subsubsection{Central charges}

Central charges appear as in three-dimensional AdS gravity \cite{Brown:1986nw}. 
These central charges are non-zero only for the brackets between the leading supertranslation charges $Q_{T}$, $Q_{W}$ and the subleading supertranslation charges $Q_{T^{(1)}}$, $Q_{I^{(1)}}$.  One finds explicitly that the non-vanishing central charges are
\begin{align}
\mathcal{C}_{\{W,T^{(1)}\}} & =-\mathcal{C}_{\{T^{(1)},W\}}=2\oint d^{3}x\sqrt{\xbar g}\,W\left(\xbar\triangle+3\right)T^{(1)} = 2\oint d^{3}x\sqrt{\xbar g}\,W\, \tilde{T}^{(1)}\,,\\
\mathcal{C}_{\{I^{(1)},T\}} & =-\mathcal{C}_{\{T,I^{(1)}\}}=2\oint d^{3}x\sqrt{\xbar g}\,I^{(1)}T\,.
\end{align}

These formulas exhibit one important feature: that the central charge is zero when  $Q_{T}$ and $Q_{W}$ are the generators of ordinary spacetime translations. Indeed, if $T$ reduces to its zero mode  $T_0$ (ordinary time translation), the central charge $\mathcal{C}_{\{I^{(1)},T_{0}\}}$ vanishes since $I^{(1)}$ has no zero mode. Similarly, if $W$ reduces to its vector component $W_{k=1}$, the central charge $\mathcal{C}_{\{W_{k=1},T^{(1)}\}}$ vanishes since it involves the projection operator $(\xbar\triangle+3)$.

\subsubsection{Non-linear terms}

A direct but tedious computation shows that the non-linear terms are given by
\begin{eqnarray}
\Lambda_{\lbrace b_1,b_2\rbrace}&=&\oint d^3x \xbar \,(b_1\partial_A b_2-b_2\partial_A b_1)\Big[8\xbar g^{\,-1}\kappa^{rA} \kappa^{rB}\kappa_{rB}+\frac{1}{2}\theta\big(3\theta \kappa^{rA}+4\theta^A_B \kappa^{rB}\big)\Big]\,,\label{eq:BruteNonLinear0}\\
\Lambda_{\lbrace b,T\rbrace}& = & -\Lambda_{\lbrace T,b\rbrace} = \oint d^{3}x\kappa^{rA}\left[T\left(\theta_{A}^{B}\partial_{B}b+\partial_{A}(b\theta)\right)-b\left(\theta_{A}^{B}\partial_{B}T+3\theta\partial_{A}T\right)\right]\,,\\
\Lambda_{\lbrace b,W\rbrace}&=&-\Lambda_{\lbrace W,b\rbrace}\\
& = & \oint d^{3}x\frac{1}{\sqrt{\xbar g}}\Big[bW(10\kappa^{rA}\kappa_{A}^{r}-2\xbar D_{A}\kappa^{rA}\xbar D_{B}\kappa^{rB}+\kappa^{AB}\kappa_{AB}) \nonumber \\
 &  & +4b\xbar\triangle W\kappa^{rA}\kappa_{A}^{r}-4b\xbar D^{A}W\big(\kappa_{AB}\kappa^{rB}-\kappa_{B}^{r}\xbar D_{A}\kappa^{rB}+\kappa^{rB}\xbar D_{B}\kappa_{A}^{r}-\frac{1}{2}\kappa_{A}^{r}\xbar D_{B}\kappa^{rB}\big)\Big] \nonumber \\
 &  & +\oint d^{3}x\sqrt{\xbar g}\,\Big[\frac{1}{4}bW\big(2\theta^{2}-7\theta_{AB}\theta^{AB}+\xbar D_{A}\theta\xbar D^{A}\theta-\xbar D_{A}\theta_{BC}\xbar D^{A}\theta^{BC}\big) \nonumber \\
 &  & +\frac{1}{2}b\xbar D^{A}W\big(\theta_{AB}\xbar D^{B}\theta-2\theta^{BC}\xbar D_{A}\theta_{BC}\big)-\frac{1}{4}\partial_{A}b\xbar D^{A}W\theta_{BC}\theta^{BC}\nonumber \\
 &  & +\frac{1}{4}b\xbar\triangle W\big(\theta^{2}-3\theta_{BC}\theta^{BC}\big)+\frac{1}{2}b\xbar D^{A}\xbar D^{B}W\big(\theta_{A}^{C}\theta_{CB}-\theta_{AB}\theta\big)\Big]\,.
 \label{eq:BruteNonLinear}
\end{eqnarray}

These non-linear terms are quadratic functions of the charges $U$ and $V$ of the subleading supertranslations.   Our task now is to translate the above formulas in terms of the brackets of the generators.

\subsubsection*{Brackets involving the homogeneous  Lorentz generators}

We  then find from the linear terms in Subsection \ref{subsec:LinearTerms} and from (\ref{eq:BruteNonLinear0}) that the brackets of the homogeneous Lorentz generators read
\begin{align}
\{B^i,B^j\}&=2M^{ij}+\Lambda^{ij}\,,\\
\{B^i,M^{jk}\}&= \frac{1}{2}\Big(\delta^{ik}B^j-\delta^{ij}B^k\Big)\,,\\
\{M^{ij},M^{kl}\}&=\frac{1}{2}\Big(\delta^{ik}M^{jl}-\delta^{il}M^{jk}-\delta^{jk}M^{il}+\delta^{jl}M^{ik}\Big)\,.
\end{align}
where the non-linear term $\Lambda^{ij}$ is cubic in the supertranslation generators and given by
\begin{equation}
\Lambda^{ij}=2\oint d^3x \sqrt{\xbar g}\,x^{[i}e^{j]A}\Big[8 K_A K_B+2(\xbar D_A \xbar D_BU+\xbar g_{AB}U)(\xbar \triangle U+3U)+\frac{3}{2}\xbar g_{AB}(\xbar \triangle U+3U)^2\Big]K^B\,.
\end{equation}
There is no non-linear contribution in the brackets involving the angular momentum $M^{ij}$.  

The brackets of the subleading supertranslation generators $(\xbar \triangle+3)U$ and $K_A$ with the homogeneous Lorentz generators are linear.  They read
\begin{align}
\{B^{i},(\xbar\triangle+3)U\} & =-2\xbar D^{A}(n^{i}K_{A})\,,\\
\{M^{ij},(\xbar\triangle+3)U\} & =-\xbar D_{A}\big[x^{[i}e^{j]A}(\xbar\triangle+3)U\big]\,,\\
\{B^{i},K_{A}\} & =\frac{1}{2}\xbar D_{A}\big[n^{i}(\xbar\triangle+3)U\big]+\frac{1}{2}\xbar D^{B}n^{i}(\xbar D_{A}\xbar D_{B}U+\xbar g_{AB}U)\,,\\
\{M^{ij},K_{A}\} & =\xbar D_{B}(x^{[i}e^{j]B}K_{A})-x^{[i}\xbar D_{A}e^{j]B}K_{B}\,.
\end{align}
They are equivalent, up to relevant projections and unphysical ambiguities, to the following Lorentz brackets of $U$ and $V$,
\begin{align}
\{B^i,U\}&=-2n^i V\,,\\
\{M^{ij},U\}&=-\xbar D_B(x^{[i}e^{j]B} U)\,, \\
\{B^i,V\}&=\frac{1}{2}\Big[4n^i U+\partial_A n^i \xbar D^A U+n^i \xbar \triangle U \Big]\,,\\
\{M^{ij},V\}&=-\xbar D_B(x^{[i}e^{j]B} V)\,.
\end{align}

The brackets of the leading supertranslations with the homogeneous Lorentz generators read
\begin{align}
\{B^i,\mathcal T\}&=-n^i \mathcal W+\Lambda^i_{\mathcal T}\,,\\
\{M^{ij},\mathcal T\}&=-\xbar D_B(x^{[i}e^{j]B} \mathcal T)\,, \\
\{B^i,\mathcal W\}&=-4n^i \mathcal T-\partial_A n^i \xbar D^A \mathcal T-n^i \xbar \triangle \mathcal T+\Lambda^i_{\mathcal W}\,,\\
\{M^{ij},\mathcal W\}&=-\xbar D_B(x^{[i}e^{j]B} \mathcal W)\,.
\end{align}

The non-linear contributions $\Lambda^i_{\mathcal T}$ and $\Lambda^i_{\mathcal W}$ are given by

\begin{align}
\Lambda_{\mathcal{T}}^{i}& =K^{A}\xbar D^{B}n^{i}(\xbar D_{A}\xbar D_{B}U+\xbar g_{AB}U)+K^{A}\xbar D_{A}[n^{i}(\xbar\triangle U+3U)]\nonumber \\
 & \quad+\xbar D_{A}[K_{B}n^{i}(\xbar D^{A}\xbar D^{B}U+\xbar g^{AB}U)]+3\xbar D_{A}[K^{A}n^{i}(\xbar\triangle U+3U)]\,,
\end{align}

\begin{align}
\Lambda_{\mathcal{W}}^{i} & =10n^{i}K^{A}K_{A}-n^{i}(\xbar D_{C}K^{C})^{2}\nonumber \\
 & \quad+n^{i}\xbar D^{A}K^{B}\xbar D_{A}K_{B}+4\xbar D_{A}[n^{i}(\xbar D_{C}K^{C}K^{A}-\xbar D^{A}K^{B}K_{B}-\frac{1}{2}K_{A}\xbar D_{B}K^{B})]\nonumber \\
 & \quad+4\xbar\triangle(n^{i}K^{A}K_{A})+\frac{1}{4}n^{i}\Big[2(\xbar\triangle U+3U)]^{2}-7(\xbar D_{A}\xbar D_{B}U+\xbar g_{AB}U)(\xbar D^{A}\xbar D^{B}U+\xbar g^{AB}U)\nonumber \\
 & \quad+\xbar D_{A}(\xbar\triangle U+3U)\xbar D^{A}(\xbar\triangle U+3U)-\xbar D_{A}(\xbar D_{B}\xbar D_{C}U+\xbar g_{BC}U)\xbar D^{A}(\xbar D^{B}\xbar D^{C}U+\xbar g^{BC}U)\Big]\nonumber \\
 & \quad-\frac{1}{2}\xbar D^{A}\Big[n^{i}(\xbar D_{A}\xbar D_{B}U+\xbar g_{AB}U)\xbar D^{B}(\xbar\triangle U+3U)\Big]+\frac{1}{4}\xbar\triangle\Big[n^{i}(\xbar\triangle U+3U)^{2}\Big]\nonumber \\
 & \quad+\frac{1}{4}\xbar D^{A}\Big[\xbar D_{A}n^{i}(\xbar D_{B}\xbar D_{C}U+\xbar g_{BC}U)(\xbar D^{B}\xbar D^{C}U+\xbar g^{BC}U)\Big]\nonumber \\
 & \quad+\xbar D^{A}\Big[n^{i}(\xbar D^{B}\xbar D^{C}U+\xbar g^{BC}U)\xbar D_{A}(\xbar D_{B}\xbar D_{C}U+\xbar g_{BC}U)\Big]\nonumber \\
 & \quad+\frac{1}{2}\xbar D^{A}\xbar D^{B}\Big[n^{i}(\xbar D^{C}\xbar D_{A}U+\delta_{A}^{C}U)(\xbar D^{B}\xbar D^{C}U+\xbar g^{BC}U)\Big]\nonumber \\
 & \quad-\frac{3}{4}\xbar\triangle\Big[n^{i}(\xbar D_{B}\xbar D_{C}U+\xbar g_{BC}U)(\xbar D^{B}\xbar D^{C}U+\xbar g^{BC}U)\Big]\nonumber \\
 & \quad-\frac{1}{2}\xbar D^{A}\xbar D^{B}\Big[n^{i}(\xbar D_{A}\xbar D_{B}U+\xbar g_{AB}U)(\xbar\triangle U+3U)\Big]\,.
\end{align}

\subsubsection*{Brackets of the supertranslation generators}

The supertranslation generators form a subalgebra, which we can read off by collecting the information from the above formulas.  We find that the non-vanishing brackets  are
\begin{align}
 \{Q_{T_1},Q_{T_2}\}& = Q_{{\hat{I}^{(1)}}_{\{T_1,T_2\}}} = -2\oint d^{3}x\sqrt{\xbar g}\,\xbar D^A(T_1 \xbar D_A T_2 - T_2 D_A T_1) V\, , \label{eq:T1T2}\\
\{Q_{W},Q_{T^{(1)}}\} & =2\oint d^{3}x\sqrt{\xbar g}\,W\left(\xbar\triangle+3\right)T^{(1)} = 2\oint d^{3}x\sqrt{\xbar g}\,W\, \tilde{T}^{(1)}\,,\\
 \{Q_{T},Q_{I^{(1)}}\}& =-2\oint d^{3}x\sqrt{\xbar g}\,I^{(1)}T\,.
\end{align}
with 
\be
\hat{I}^{(1)}_{\{T_1,T_2\}} = T_{1}\xbar \triangle T_{2}-T_{2}\xbar \triangle T_{1}
\ee

The algebra of the leading and subleading supertranslations is thus non abelian and centrally extended.  The non-abelian commutator is $\{Q_{T_1},Q_{T_2}\} \sim Q_{{\hat{I}^{(1)}}}$, showing that the Poisson bracket of two leading supertranslations in time yields a subleading supertranslation in space.  

This is not too surprising, in view of the fact that the Lie bracket of two vector fields of order $\mathcal{O}(1)$ will generically be a vector field of order $\mathcal{O}(r^{-1})$.  We should stress, however, that the supertranslations are accompanied by field-dependent correcting terms. These terms play an important role in the specific form of the algebra encountered above. They are in particular key in the emergence of the central charges.

The presence of $Q_{{\hat{I}^{(1)}}_{\{T_1,T_2\}}} $ in $\{Q_{T_1},Q_{T_2}\}$ was overlooked in \cite{Fuentealba:2021yvo}, where the abelian relation $\{Q_{T_1},Q_{T_2}\}=0$ was given instead. It turns out that one can add field-dependent diffeomorphisms to $T$ such that the leading supertranslations commute.  This is done in the next section. With the choice made above leading to the definite expression (\ref{eq:QTandQW}) for $Q_{T}$, however, one gets $\{Q_{T_1},Q_{T_2}\}\not=0$.

How to simplify the asymptotic symmetry algebra by non-linear redefinitions of the generators is a question to which we now turn.

\section{Simplification of the presentation of the BMS$_5$ algebra}
\label{sec:Simpli}

\subsection{Non-linear algebras and Poisson manifolds}

A striking feature of the above charge algebra is that it is non-linear.  In particular, the bracket of two boost generators is cubic in the charges of the subleading supertranslations.

That symmetry charges might form a non-linear algebra is not a surprise.  It is in fact more the rule than the exception, as clearly discussed in \cite{deBoer:1995cqx}.  In terms of the canonical transformations generated by the charges, non-linearities mean that the commutator of the corresponding transformations is a transformation that takes the same form, but with coefficients that are functions of the fields through the charges. Indeed, if $Q_a$ is a complete set of charges with $\{Q_a, Q_b\} = f_{ab}(Q)$ and $X_a$ the corresponding vector fields ($X_a F = \{Q_a, F\}$), then one finds $[X_a, X_b] = X_{\{Q_b, Q_a\}} = X_{f_{ba}(Q)} = \frac{\partial f_{ba}}{\partial Q_c} X_c$.  This is not a linearly independent vector field (over the functions), just as the corresponding charge-generator $f_{ab}$  is not independent from the $Q_a$'s. 

Thus, in our case, the non-linearities found in the BMS$_5$ brackets simply imply that the parameters $T^{(1)}$, $W^{(1)}$ and $I^{(1)}_A$ associated with the commutator of two asymptotic symmetries depend on the fields through the charges.  As we just observed, such transformations do not define independent asymptotic symmetries.   Examples of non-linear asymptotic symmetry algebras have been found earlier in \cite{Henneaux:1999ib,Henneaux:2010xg,Campoleoni:2010zq,Afshar:2013vka,Gonzalez:2013oaa,Fuentealba:2015jma,Henneaux:2015ywa,Fuentealba:2015wza,Henneaux:2015tar,Fuentealba:2020zkf,Fuentealba:2021xhn}.

There is a great flexibility in the presentation of non-linear algebras since one can make non-linear redefinition of the charges, under which  $f_{ab}(Q)$ changes.  That non-linear redefinitions may have a dramatic effect already in the Lie algebra case is illustrated in the recent paper \cite{Rodriguez:2021tcz}.

The space generated by the symmetry charges is a Poisson manifold, about which much is known in the finite-dimensional case \cite{Lichnerowicz,Weinstein,Arnold}.   In particular, one can bring the Poisson bracket to a canonical form that generalizes the Darboux canonical form by (non-linear) redefinitions of the charges (``Weinstein splitting theorem'' or ``Darboux-Weinstein theorem'').  Explicitly, one has

\vspace{.2 cm}

\noindent
{\bf Theorem \cite{Lichnerowicz,Weinstein}:} Let $P$ be an arbitrary point of a Poisson manifold.  One can find local coordinates ($q_i, p_i, y_\alpha$) in a neighborhood $U$ of $P$ with the properties:
\be
\{q_i, q_j \}= \{p_i, p_j \}= \{q_i, y_\alpha \} = \{p_i,  y_\alpha \} = 0, \quad, \{q_i, p_j \}= \delta_{ij} \quad \{y_\alpha, y_\beta \}= F_{\alpha \beta} (y)
\ee
 with 
\be
F_{\alpha \beta} (y) = 0 \qquad \textrm{at } \; P,
\ee
Here, $i, j = 1, \cdots, s$, $\alpha, \beta = 1, \cdots, r$, $2s + r = n$ where $n$ is the dimension of the Poisson manifold $M$.

\vspace{.2cm}
We can assume $P$ to have coordinates $q_i = p_i = y_\alpha = 0$, so that the last condition is $F_{\alpha \beta} (y=0) = 0$.  If one expands $F_{\alpha \beta}$ in Taylor series, one has
\be
F_{\alpha \beta} = {C^\gamma}_{\alpha \beta} y_\gamma + O(y^2)
\ee 
where the ${C^\gamma}_{\alpha \beta}$ are easily verified to fulfill the Jacobi identity and are thus necessarily the structure constants of a Lie algebra. An interesting question is whether one can get rid of the non linear terms by redefinitions.  This question is reviewed in \cite{FernandesMonnier} and can be expressed as a problem of Lie algebra cohomology (at least for the formal linearization problem) \cite{Weinstein}.  In particular, if the algebra is semi-simple, one can absorb the non-linear terms through redefinitions.

In our case, however, the space of the symmetry charges is infinite-dimensional, and furthermore, it is natural to restrict the redefinitions so as to maintain the structure of the new generators to be
\be
\int d^4 x (\xi \mathcal H + \xi^i \mathcal H_i) +  Q_{\xi}[g_{ij},\pi^{ij}]
\ee
where $(\xi, \xi^i)$ has the asymptotic behaviour (\ref{eq:Trans1})-(\ref{eq:Trans3}). Accordingly, the only redefinitions that we allow are field-dependent redefinitions of  the coefficients $b$, $T$, $W$,  $T^{(1)}$, $W^{(1)}$ and $I_A^{(1)}$ characterizing this asymptotic behaviour, in a way  that keeps integrability of the charges. This makes the problem different from the finite-dimensional one. 

It turns out to be possible, however, to make non-linear redefinitions of the above type such that (i) supertranslations commute up to central charges; and (ii) the Lorentz subalgebra is linearly realized. 

\subsection{Abelian description of the supertranslations (with central extension)}

In order to make the supertranslations abelian, we redefine the leading supertranslations $T$ by adding to them a subleading supertranslation with parameter 
\begin{equation}
I_{(T)}^{(1)A}=\frac{T}{\sqrt{\xbar g}}\kappa^{rA} = T K^A\,.\label{eq: Gauge1}
\end{equation}
Note that this parameter maintains integrability.  It modifies the leading time supertranslation
charge by a term proportional to the square $K^A K_A$ of the subleading space supertranslation charges. Thus,
we now have
\begin{equation}
Q_{T}=\oint d^{3}x\sqrt{\xbar g}\,T \mathcal{T}\,,
\end{equation}
where the new $\mathcal{T}$ (still denoted in the same way) is given by
\begin{equation}
\mathcal{T}=2k-2e^{(2)}+\frac{1}{4}\theta^{2}+\frac{1}{4}\theta_{B}^{A}\theta_{A}^{B} +\xbar g^{\,-1}\kappa^{rA}\kappa_{A}^{r} \,.
\end{equation}
and one easily checks that $\{Q_{T_1},Q_{T_2}\} = 0$.  

The extra improper gauge transformation added to $T$ does
not affect the remaining brackets between the other time and spatial supertranslation generators, which we take unchanged,
\be
Q_{W}=\oint d^3x\,W\, \mathcal W\, ,  \qquad
 \mathcal W =  2\Big( \xbar \pi^{rr}-\xbar D_A \xbar \pi^{rA}-\frac{1}{2}\xbar D^A(\theta_{AB}\kappa^{rB}) \Big)
\ee
\be
Q_{T^{(1)}}=-\oint d^3x \sqrt{\xbar g }\, T^{(1)}\Big((\xbar \triangle+3)U\Big)\,, \qquad
 Q_{I^{(1)}}=  -2\oint d^3x \sqrt{\xbar g} I^{(1)}V \, .
 \ee
The final supertranslation algebra is thus abelian with non-trivial central terms given by
\begin{align}
\{Q_{W},Q_{T^{(1)}}\} & = 2\oint d^{3}x\sqrt{\xbar g}\,W\, (\xbar \triangle+3) T^{(1)}\,, \\
 \{Q_{T},Q_{I^{(1)}}\}& =-2\oint d^{3}x\sqrt{\xbar g}\,I^{(1)}T\,,
\end{align}
while all the other Poisson brackets between supertranslation generators vanish.

On can view the elimination of the charge-dependent terms in the algebra of the supertranslations as an application of Darboux theorem since the central term defines a non-degenerate antisymmetric matrix.

\subsection{Linear realization of the Lorentz subalgebra}

In order to eliminate the cubic terms in the boost algebra, we add to the boosts a subleading supertranslation with parameters
\begin{align}
T_{(b)}^{(1)} & =2b\xbar g^{\,-1}\kappa^{rA}\kappa_{A}^{r}-\frac{b}{8}\theta^{2}\,,\label{eq:Gauge2.1}\\
I_{(b)}^{(1)A} & =-\frac{2b}{\sqrt{\xbar g}}\kappa^{rA}\theta\,.\label{eq:Gauge2.2}
\end{align}
Again, these parameters yield an integrable generator.  The new boost charges receive the following cubic contribution in the 
subleading supertranslation charges
$$
\oint d^{3}xb\left(-\frac{2}{\sqrt{\xbar g}}\kappa^{rA}\kappa_{A}^{r}\theta+\frac{\sqrt{\xbar g}}{24}\theta^{3}\right)
$$
and now read
\begin{align}
& B^i = \oint d^3x \,\sqrt{\xbar g} \, n^i \Big[
        2 k^{(2)} + \theta e^{(2)} - \frac{1}{24} \theta^3 - \frac{1}{3} \theta^A_C \theta^C_B\theta^B_A \nonumber \\
        &\qquad \qquad \qquad \quad+ \frac{1}{2} \theta^A_B \xbar h^B_A+\frac{4}{\xbar g}\Big(\kappa^r_A  \xbar \pi^{rA}+(\theta_{AB}- \frac{\theta}{2}  \xbar g_{AB})\kappa^{rA} 
        \kappa^{rB}\Big)\Big] 
\end{align}
while the rotation generators remain unchanged,
\be
 M^{ij} =  \oint d^3x  \,  x^{[i} e^{j]A} \, \xbar \pi^{(2)r}_A \, .
 \ee
The redefined boosts obey
\be 
\{Q_{b_1}, Q_{b_2} \}= \delta_{b_2} Q_{b_1} = Q_{\hat{Y}}
\ee
where $Q_{\hat{Y}}=\oint d^{3}x\,2\hat{Y}^{A}\pi_{A}^{(2)r}$ with
$\hat{Y}^{A}=b_{1}\xbar D^{A}b_{2}-b_{2}\xbar D^{A}b_{1}$, or equivalently,
\be
\{B^i, B^j\} =2 M^{ij}
\ee
with no non-linear contributions. 
Therefore, the Lorentz algebra becomes linearly realized.  That one can eliminate the non-linear terms in the Lorentz algebra is not too surprising either since it is a simple, finite-dimensional algebra.

One can summarize the redefinitions of the generators of the symmetry algebra by saying that the ``more'' terms in (\ref{eq:Trans1})-(\ref{eq:Trans3}) should contain the improper gauge transformations \eqref{eq: Gauge1},
 \eqref{eq:Gauge2.1} and \eqref{eq:Gauge2.2} in addition to the already displayed correcting terms. The final expression for the supertranslation parameters corresponding to the new form of the algebra
is thus
\begin{align}
\overline{I}^{A} & =\xbar D^{A}W+I_{(b)}^{A}\,,\\
\xbar T & =T+T_{(b)}\,,\\
\xbar W & =W \,,\\
\xbar T^{(1)} & =T^{(1)}+T_{(b,W)}^{(1)}\,,\\
\xbar I^{(1)A} & =I^{(1)A}+I_{(b,T,W)}^{(1)A}\,,
\end{align}
where
\begin{align}
I_{(b)}^{A} & =\frac{2b}{\sqrt{\xbar g}}\kappa^{rA}\,,\\
T_{(b)} & =-\frac{b\theta}{2}\,,\\
T_{(b,W)}^{(1)} & =-\xbar\lambda^{B}\partial_{B}b+(\xbar\triangle+3)^{-1}\left[\Big(\xbar D^{A}\xbar D^{B}+\xbar g^{AB}\Big)\Big(\frac{b}{2}\xbar h_{AB}+b\xbar D_{A}\xbar\lambda_{B}\Big)\right]\nonumber \\
 & \quad+2b\xbar g^{\,-1}\kappa^{rA}\kappa_{A}^{r}-\frac{b}{8}\theta^{2}+\frac{1}{\sqrt{\xbar g}}\xbar D_{A}W\kappa^{rA}\,,\\
I_{(b,T,W)}^{(1)A} & =\frac{2b}{\sqrt{\xbar g}}\big(\xbar\pi^{rA}+\theta_{B}^{A}\kappa^{rB}-\theta\kappa^{rA}\big)+\frac{T}{\sqrt{\xbar g}}\kappa^{rA}-\frac{1}{2}\theta_{B}^{A}\xbar D^{B}W\,.
\end{align}

The modification of the boost and leading supertranslation generators change the non-linear terms appearing in their Poisson bracket.  One now finds
\begin{eqnarray}
\Lambda_{\lbrace b,T\rbrace} & = & -\Lambda_{\lbrace T,b\rbrace}=-\oint d^{3}xb\kappa^{rA}\left(\theta_{A}^{B}\partial_{B}T-\theta\partial_{A}T\right)\,,\\
\Lambda_{\lbrace b,W\rbrace} & = & -\Lambda_{\lbrace W,b\rbrace} \nonumber \\
 & = & \oint d^{3}x\frac{1}{\sqrt{\xbar g}}\Big[bW(2\kappa^{rA}\kappa_{A}^{r}-2\xbar D_{A}\kappa^{rA}\xbar D_{B}\kappa^{rB}+\kappa^{AB}\kappa_{AB})+\partial_{A}b\xbar D^{A}W\kappa^{rA}\kappa_{A}^{r} \nonumber \\
 &  & +4b\xbar\triangle W\kappa^{rA}\kappa_{A}^{r}-4b\xbar D^{A}W\big(\kappa_{AB}\kappa^{rB}-\kappa_{B}^{r}\xbar D_{A}\kappa^{rB}+\kappa^{rB}\xbar D_{B}\kappa_{A}^{r}-\frac{1}{2}\kappa_{A}^{r}\xbar D_{B}\kappa^{rB}\big)\Big] \nonumber\\
 &  & +\oint d^{3}x\sqrt{\xbar g}\,\Big[\frac{1}{4}bW\big(5\theta^{2}-7\theta_{AB}\theta^{AB}+\xbar D_{A}\theta\xbar D^{A}\theta-\xbar D_{A}\theta_{BC}\xbar D^{A}\theta^{BC}\big)\nonumber \\
 &  & +\frac{1}{2}b\xbar D^{A}W\big(\theta_{AB}\xbar D^{B}\theta-2\theta^{BC}\xbar D_{A}\theta_{BC}\big)-\frac{1}{4}\partial_{A}b\xbar D^{A}W\theta_{BC}\theta^{BC} \nonumber \\
 &  & +\frac{1}{4}b\xbar\triangle W\big(2\theta^{2}-3\theta_{BC}\theta^{BC}\big)+\frac{1}{2}b\xbar D^{A}\xbar D^{B}W\big(\theta_{A}^{C}\theta_{CB}-\theta_{AB}\theta\big)\Big]\,.
\end{eqnarray}

The Poisson brackets of of the redefined boost and supertranslation charges are thus
\begin{align}
\{B^{i},\mathcal{T}\} & =-n^{i}\mathcal{W}+\Lambda_{\mathcal{T}}^{i}\,,\\
\{B^{i},\mathcal{W}\} & =-4n^{i}\mathcal{T}-\partial_{A}n^{i}\xbar D^{A}\mathcal{T}-n^{i}\xbar\triangle\mathcal{T}+\Lambda_{\mathcal{W}}^{i}\,.
\end{align}
where the nonlinear terms are given by
\begin{align}
\Lambda_{\mathcal{T}}^{i} & =\xbar D_{A}[K_{B}n^{i}(\xbar D^{A}\xbar D^{B}U+\xbar g^{AB}U)]-\xbar D_{A}[K^{A}n^{i}(\xbar\triangle U+3U)]\,,
\nonumber \\
\Lambda_{\mathcal{W}}^{i} & =2n^{i}K^{A}K_{A}-n^{i}(\xbar D_{C}K^{C})^{2}+n^{i}\xbar D^{A}K^{B}\xbar D_{A}K_{B}-\xbar D^{A}\left(\partial_{A}n^{i}K^{A}K_{A}\right)\nonumber \\
 & \quad+4\xbar\triangle(n^{i}K^{A}K_{A})+4\xbar D_{A}[n^{i}(\xbar D_{C}K^{C}K^{A}-\xbar D^{A}K^{B}K_{B}-\frac{1}{2}K_{A}\xbar D_{B}K^{B})]\nonumber \\
 & \quad+\frac{1}{4}n^{i}\Big[5(\xbar\triangle U+3U)]^{2}-7(\xbar D_{A}\xbar D_{B}U+\xbar g_{AB}U)(\xbar D^{A}\xbar D^{B}U+\xbar g^{AB}U)\nonumber \\
 & \quad+\xbar D_{A}(\xbar\triangle U+3U)\xbar D^{A}(\xbar\triangle U+3U)-\xbar D_{A}(\xbar D_{B}\xbar D_{C}U+\xbar g_{BC}U)\xbar D^{A}(\xbar D^{B}\xbar D^{C}U+\xbar g^{BC}U)\Big]\nonumber \\
 & \quad-\frac{1}{2}\xbar D^{A}\Big[n^{i}(\xbar D_{A}\xbar D_{B}U+\xbar g_{AB}U)\xbar D^{B}(\xbar\triangle U+3U)\Big]+\frac{1}{2}\xbar\triangle\Big[n^{i}(\xbar\triangle U+3U)^{2}\Big]\nonumber \\
 & \quad+\frac{1}{4}\xbar D^{A}\Big[\xbar D_{A}n^{i}(\xbar D_{B}\xbar D_{C}U+\xbar g_{BC}U)(\xbar D^{B}\xbar D^{C}U+\xbar g^{BC}U)\Big]\nonumber \\
 & \quad+\xbar D^{A}\Big[n^{i}(\xbar D^{B}\xbar D^{C}U+\xbar g^{BC}U)\xbar D_{A}(\xbar D_{B}\xbar D_{C}U+\xbar g_{BC}U)\Big]\nonumber \\
 & \quad+\frac{1}{2}\xbar D^{A}\xbar D^{B}\Big[n^{i}(\xbar D^{C}\xbar D_{A}U+\delta_{A}^{C}U)(\xbar D^{B}\xbar D^{C}U+\xbar g^{BC}U)\Big]\nonumber \\
 & \quad-\frac{3}{4}\xbar\triangle\Big[n^{i}(\xbar D_{B}\xbar D_{C}U+\xbar g_{BC}U)(\xbar D^{B}\xbar D^{C}U+\xbar g^{BC}U)\Big]\nonumber \\
 & \quad-\frac{1}{2}\xbar D^{A}\xbar D^{B}\Big[n^{i}(\xbar D_{A}\xbar D_{B}U+\xbar g_{AB}U)(\xbar\triangle U+3U)\Big]\,.
\end{align}

One might wonder whether one can eliminate the remaining non-linear terms $\Lambda_{\mathcal{T}}^{i}$ and $\Lambda_{\mathcal{W}}^{i}$ in the algebra.  We show in Appendix \ref{sec:Obstruc} that one can eliminate $\Lambda_{\mathcal{T}}^{i}$ but that once this is done along the lines we indicate, there is an obstruction to the elimination of $\Lambda_{\mathcal{W}}^{i}$.  Furthermore, the elimination of $\Lambda_{\mathcal{T}}^{i}$ itself leads to undesireable features in what concerns the transformation properties of the energy, which would be affected by supertranslation ambiguities (see Section \ref{sec:energy}).

\subsection{Lorentz transformation of the energy and the linear momentum}

It is to be pointed out that when $T$ reduces to its zero mode (constant time translations), the non-linear term $\Lambda_{\lbrace b,T\rbrace}$ vanishes.  The same is true when $W$ reduces to its vector harmonic ($k=1$), although it is less immediate to see.

The proof goes as follows.  It is convenient to split   $\Lambda_{\{b,W\}}$ as
follows,
\begin{equation}
\Lambda_{\{b,W\}}=\Lambda_{\{b,W\}}^{U}+\Lambda_{\{b,W\}}^{V}\,,
\end{equation}
where
\begin{eqnarray}
\Lambda_{\lbrace b,W\rbrace}^{U} & = & \oint d^{3}x\sqrt{\xbar g}\,\Big[\frac{1}{4}bW\big(5\theta^{2}-7\theta_{AB}\theta^{AB}+\xbar D_{A}\theta\xbar D^{A}\theta-\xbar D_{A}\theta_{BC}\xbar D^{A}\theta^{BC}\big) \nonumber \\
 &  & +\frac{1}{2}b\xbar D^{A}W\big(\theta_{AB}\xbar D^{B}\theta-2\theta^{BC}\xbar D_{A}\theta_{BC}\big)-\frac{1}{4}\partial_{A}b\xbar D^{A}W\theta_{BC}\theta^{BC} \nonumber \\ 
 &  & +\frac{1}{4}b\xbar\triangle W\big(2\theta^{2}-3\theta_{BC}\theta^{BC}\big)+\frac{1}{2}b\xbar D^{A}\xbar D^{B}W\big(\theta_{A}^{C}\theta_{CB}-\theta_{AB}\theta\big)\Big]\,,\label{eq:LambdaU3}
\end{eqnarray}
\begin{eqnarray}
\Lambda_{\lbrace b,W\rbrace}^{V} & = & \oint d^{3}x\frac{1}{\sqrt{\xbar g}}\Big[bW(2\kappa^{rA}\kappa_{A}^{r}-2\xbar D_{A}\kappa^{rA}\xbar D_{B}\kappa^{rB}+\kappa^{AB}\kappa_{AB})+\partial_{A}b\xbar D^{A}W\kappa^{rA}\kappa_{A}^{r} \nonumber\\
 &  & +4b\xbar\triangle W\kappa^{rA}\kappa_{A}^{r}-4b\xbar D^{A}W\big(\kappa_{AB}\kappa^{rB}-\kappa_{B}^{r}\xbar D_{A}\kappa^{rB}+\kappa^{rB}\xbar D_{B}\kappa_{A}^{r}-\frac{1}{2}\kappa_{A}^{r}\xbar D_{B}\kappa^{rB}\big)\Big]\,. \nonumber \\
\end{eqnarray}

First, let us focus on the term $\Lambda_{\lbrace b,W\rbrace}^{U}$.
We can directly apply the equation for the spatial translations parameter,
$\xbar D_{A}\xbar D_{B}W=-\xbar g_{AB}W$, on the second order derivative terms
in \eqref{eq:LambdaU3}. Then, we obtain that
\begin{align}
\Lambda_{\lbrace b,W\rbrace}^{U} & =\oint d^{3}x\sqrt{\xbar g}\Big[\frac{1}{4}bW\theta^{2}+\frac{1}{4}bW\left(\xbar D_{A}\theta\xbar D^{A}\theta-\xbar D_{A}\theta_{C}^{B}\xbar D^{A}\theta_{B}^{C}\right) \nonumber\\
 & \quad+\frac{1}{2}b\xbar D^{A}W\big(\theta_{AB}\xbar D^{B}\theta-2\theta^{BC}\xbar D_{A}\theta_{BC}\big)-\frac{1}{4}\partial_{A}b\xbar D^{A}W\theta_{BC}\theta^{BC}\Big]\,.\label{eq:LU2}
\end{align}

We will make use of the following identity
\begin{align}
\oint d^{3}x\sqrt{\xbar g}f\left(\xbar D_{A}\theta\xbar D^{A}\theta-\xbar D_{A}\theta_{C}^{B}\xbar D^{A}\theta_{B}^{C}\right) & =-\oint d^{3}x\sqrt{\xbar g}\Big[\partial_{B}f\left(\theta_{A}^{B}\xbar D^{C}\theta_{C}^{A}-\theta_{C}^{A}\xbar D^{C}\theta_{A}^{B}\right)  \nonumber \\ & \qquad \qquad +f\left(\theta^{2}-3\theta_{B}^{A}\theta_{A}^{B}\right)\Big]\,,
\end{align}
with $f$ an arbitrary function. This can be obtained  after integration by parts and taking into account the commutator of two covariant derivatives $[\xbar D_{C},\xbar D^{A}]\theta^{CB}=3\theta^{AB}-\theta\xbar g^{AB}$. If we now choose $f=bW/4$, and consider the equations for these parameters $(b,W)$, we have that
\begin{align}
\oint d^{3}x\sqrt{\xbar g}\frac{bW}{4}\left(\xbar D_{A}\theta\xbar D^{A}\theta-\xbar D_{A}\theta_{C}^{B}\xbar D^{A}\theta_{B}^{C}\right) & =\oint d^{3}x\sqrt{\xbar g}\Big[bW\left(\frac{1}{4}\theta^{2}+\frac{5}{4}\theta_{A}^{B}\theta_{B}^{A}\right) \nonumber\\
 & \quad-\frac{1}{2}\partial_{A}b\xbar D^{A}W\theta^{2}+\xbar D^{A}b\xbar D_{B}W\left(\theta_{A}^{B}\theta-\frac{1}{2}\theta_{C}^{B}\theta_{A}^{C}\right)\Big]\,.
\end{align}
Replacing this identity in the first line of \eqref{eq:LU2}, the non-linear term becomes
\begin{align}
\Lambda_{\lbrace b,W\rbrace}^{U} & =\oint d^{3}x\sqrt{\xbar g}\Big[\frac{1}{2}bW\theta^{2}+\frac{5}{4}bW\theta_{A}^{B}\theta_{B}^{A}-\frac{1}{2}\xbar D_{A}b\xbar D^{A}W\theta^{2} \nonumber\\
 & \quad+\xbar D_{A}b\xbar D^{B}W\theta_{B}^{A}\theta-\frac{1}{2}\xbar D^{A}b\xbar D_{B}W\theta_{C}^{B}\theta_{A}^{C}-\frac{1}{4}\xbar D_A b\xbar D^AW \theta^B_C\theta^C_B\Big] \nonumber\\
 & \quad+\frac{1}{2}\oint d^{3}x\sqrt{\xbar g}b\xbar D^{A}W\big(\theta_{AB}\xbar D^{B}\theta-2\theta^{BC}\xbar D_{A}\theta_{BC}\big)\,. \label{eq:LU3}
\end{align}
Integrating by parts the terms proportional to $\xbar D_{A}b$ in the expression above,
\eqref{eq:LU3} reduces to
\begin{align}
\Lambda_{\lbrace b,W\rbrace}^{U} & =\oint d^{3}x\sqrt{\xbar g}\Big(b\xbar D_{B}W\theta_{C}^{A}\xbar D_{A}\theta^{BC}-\frac{1}{2}b\xbar D_{A}W\theta^{AB}\xbar D_{B}\theta\Big)\nonumber\\
 & \quad+\frac{1}{2}\oint d^{3}x\sqrt{\xbar g}b\xbar D^{A}W\big(\theta_{AB}\xbar D^{B}\theta-2\theta^{BC}\xbar D_{A}\theta_{BC}\big) \nonumber\\
 &=0\,,
\end{align}
which identically vanish after the use of the identity $\xbar D_{A}\theta_{BC}=\xbar D_{B}\theta_{AC}$.

Let us now proceed to show how the non-linear term $\Lambda_{\lbrace b,W\rbrace}^{V}$ vanishes
for spatial translations. For this, we make use of the definitions of the functions $\kappa^{ij}$ in terms of $V$, which are given in \eqref{kapparA} and \eqref{kappaAB}. After some algebra and integration by parts, we get that $\Lambda_{\lbrace b,W\rbrace}^{V}$ becomes
\begin{align}
\Lambda_{\lbrace b,W\rbrace}^{V} & =-\oint d^{3}x\sqrt{\xbar g}\,V\Big[2\xbar D^{A}\left(W\partial_{A}\xbar\triangle V\right)+14\xbar D_{B}\left(W\partial_{A}b\xbar D^{A}\xbar D^{B}V\right) \nonumber\\
 & \quad-2\xbar D_{A}\left(bW\xbar D^{A}V\right)+8\xbar D_{A}\left(bW\xbar\triangle\,\xbar D^{A}V\right)-\xbar\triangle\left(bW\xbar\triangle V\right)-7\xbar D_{A}\xbar D_{B}\left(bW\xbar D^{A}\xbar D^{B}V\right)\Big]\,.
\end{align}
This can again be simplified by using the equation $\xbar D_{A}\xbar D_{B}W+\xbar g_{AB}W=0$. Thus,
$\Lambda_{\lbrace b,W \rbrace}^V$ reduces to
\begin{equation}
\Lambda_{\lbrace b,W\rbrace}^{V}=-\oint d^{3}x\sqrt{\xbar g}V\left[2\left(b\partial_{A}W-\partial_{A}bW\right)\xbar D^{A}V-8\left(b\partial_{A}W-\partial_{A}bW\right)\xbar\triangle\,\xbar D^{A}V\right]\,, \label{LambdaV2}
\end{equation}
where we have made use of the identity on the 3-sphere $\xbar \triangle \,\xbar D^AV=\xbar D^A \xbar \triangle V+2\xbar D^A V$.

It is straightforward to show that both terms in \eqref{LambdaV2} vanish (independently)
after integration by parts and using the equation for
the parameters $b$ and $W$. This can be seen for the first term of \eqref{LambdaV2}, where
\begin{eqnarray}
2\oint d^{3}x\sqrt{\xbar g}V\left(b\partial_{A}W-\partial_{A}bW\right)\xbar D^{A}V&=&\oint d^{3}x\sqrt{\xbar g}\left(b\partial_{A}W-\partial_{A}bW\right)\xbar D^{A}(V^2)\nonumber\\
&=&-\oint d^{3}x\sqrt{\xbar g}\left(b\xbar \triangle W-\xbar \triangle b W\right)V^2 \nonumber\\
&=&0\,.
\end{eqnarray}
For the second term in \eqref{LambdaV2}, we proceed in a similar way
\begin{eqnarray}
\oint d^{3}x\sqrt{\xbar g}\,V\left(b\partial_{A}W-\partial_{A}bW\right)\xbar\triangle\,\xbar D^{A}V&=&-\oint d^{3}x\sqrt{\xbar g}\,\big(b \xbar D_B\xbar D_{A}W-\xbar D_B\xbar D_{A}bW\big) V\xbar D^B \xbar D^{A}V \nonumber\\
&&-\frac{1}{2}\oint d^{3}x\sqrt{\xbar g}\,\left(b\partial_{A}W-\partial_{A}bW\right)  \xbar D^{A}(\xbar D_B V \xbar D^B V) \nonumber\\
&=&\frac{1}{2}\oint d^{3}x\sqrt{\xbar g}\,\left(b\xbar \triangle W-\xbar \triangle b W \right)  \xbar D_B V \xbar D^B V \nonumber\\
&=&0\,.
\end{eqnarray}
Hence, $\Lambda_{\lbrace b,W\rbrace}^{V}=0$ for spatial translations.
Accordingly, we have explicitly shown that the non-linear term $\Lambda_{\{b,W\}}$
is also zero for ordinary spatial translations.

It follows from this result that the energy $E$ and the linear momentum $P^i$ transform in the finite-dimensional vector representation of the Lorentz algebra since the non-linear terms appearing in the transformation of $\mathcal T$ and $\mathcal W$ vanish for them.  $E$ and $P_i$  are respectively defined as the values of the generators of constant translations in time ($T=T_0$) and space ($W= W_{k=1}$). 

\section{The importance of the non-linear terms in the energy}
\label{sec:energy}

The transformation properties of the energy $E$ and the linear momentum $P^i$ under leading and subleading supertranslations can also be directly read off from the algebra, by taking their bracket with the generators of supertranslations.    From the fact that the algebra is abelian modulo central terms that vanish for $T_0$ and $W_1$, we simply get that the energy and the linear momentum are invariant,
\be
\delta_{T,W,T^{(1)},I^{(1)}} E  = 0, \qquad \delta_{T,W,T^{(1)},I^{(1)}} P^i = 0 \, .
\ee
Energy and momentum are thus free from supertranslation ambiguities, as in four dimensions. 

This absence of ambiguities seems to be in conflict with the non-invariance of the energy under spatial supertranslations pointed out in \cite{Chrusciel0}.  There is in fact no contradiction, because the expression for the energy considered in that reference was the direct generalization to five dimensions of the expression for the  energy derived  in four dimensions in \cite{Arnowitt:1960zzc}, which is linear.  While this expression is correct in four dimensions, it is not so in five dimensions as soon as one takes (as here) boundary conditions compatible with the possibility of performing leading supertranslations.   The energy gets additional non-linear contributions,
\be
E = 2\oint_{S^3_\infty} d^3x \sqrt{\xbar g}\left[(1/2) \xbar h^A_A+D_A \xbar \lambda^A+3\xbar \lambda -(1/8) \theta_{AB} \theta^{AB}+ K^A K_A\right] \, .
\ee
Here the first three linear terms are the direct generalization to five dimensions of the four-dimensional expression of \cite{Arnowitt:1960zzc} (``ADM contribution''), while the other two terms are the non-linear (quadratic) corrections.  These non-linear terms acquire non-zero values under supertranslations and are crucial for ensuring the invariance of the energy. 

For instance if one performs the coordinate transformation $r = \rho + c$ on the flat metric $dr^2 + r^2 d\Omega^2$
where $c$ is a constant (see lecture notes  \cite{Chrusciel0}, pages 18-19 with $\alpha = 1$), which corresponds to a non-trivial leading spatial supertranslation with $W= c$, 
one finds after transformation that the fields are given by
\be \xbar h_{AB}=c^2 \xbar g_{AB}, \quad \theta_{AB}=2c \xbar g_{AB}, \quad  \xbar \lambda=0, \quad \xbar \lambda^A=0, \quad K_A = 0,
\ee
and therefore
\be E = 2\oint_{S^3_\infty} d^3x \sqrt{\xbar \gamma} \left(c^2 - c^2\right) = 0.
\ee
The first term is the ADM contribution while the compensating second term arises from the non-linear contribution.

The absence of supertranslation ambiguities for the energy and the linear momentum holds only because the supertranslation algebra is abelian modulo central terms that vanish for the energy and the momentum.  If the Poisson brackets between the leading supertranslations involved the subleading supertranslations, as did the original choice of $\mathcal T$ and $\mathcal W$, then the energy defined as the value of the zero mode $\mathcal T_0$ of the corresponding $\mathcal T$ would not be invariant under supertranslations.  This would mean that the energy could not be identified with the zero mode of that choice of $\mathcal T$, but rather of the improved $\mathcal T$ that fulfills a (centrally extended) abelian algebra.  It is for that reason that we redefined $\mathcal T$ above in order to achieve this property.

\section{Matching with null infinity}
\label{sec:Matching}

The Hamiltonian derivation of the ADM-BMS charges given here follows the general approach developed in \cite{Henneaux:2018cst,Henneaux:2018hdj,Henneaux:2019yax} for four dimensions. It is self-contained and makes no reference to null infinity.  It is thus independent from the dynamical question of the existence of a smooth null infinity, which is already a non trivial issue in four dimensions \cite{Friedrich2004,ValienteKroon:2002fa}.  In fact, the  parity conditions of \cite{Henneaux:2018hdj,Henneaux:2019yax} lead to the absence of the leading logarithmic singularities that develop as one goes to null infinity while those of \cite{Henneaux:2018cst} do not eliminate them, but from the point of view of the Hamiltonian formalism and initial data on Cauchy hypersurfaces, this smoothness condition is in fact not a necessary requirement.  

The four-dimensional boundary conditions adopted in  \cite{Henneaux:2018hdj,Henneaux:2019yax} leading to the absence of leading logarithmic singularities at null infinity where actually suggested by electromagnetism, where one encounters similar features in a simpler linear context  \cite{Henneaux:2018gfi}.  There also, appropriate parity conditions on the leading order of the fields at spatial infinity eliminate the leading logarithmic singularities that generically develop at null infinity.  In both cases, one can see how logarithmic singularities emerge as one goes to null infinity by (i) integrating the asymptotic equations of motion from spatial infinity to null infinity in hyperbolic coordinates \cite{Ashtekar:1978zz,BeigSchmidt,Beig:1983sw};  and (ii) following then the methods of Friedrich \cite{Fried1,Friedrich:1999wk,Friedrich:1999ax}, which are adapted to the ``critical sets where the two infinities meet''. The evolution of the leading terms as one goes from spatial infinity to null infinity is governed by the second order Legendre equation \cite{Troessaert:2017jcm} and the parity condition eliminates the term proportional to the Legendre function of the second kind, which becomes singular at the critical sets \cite{Henneaux:2018gfi,Henneaux:2018hdj} (see also \cite{Mohamed:2021rfg}). 

The approach just outlined has the advantage of matching explicitly the fields at spatial infinity and null infinities (in four dimensions).  In particular, one can derive the Strominger's antipodal matching conditions \cite{Strominger:2013jfa,He:2014cra},\cite{Strominger:2017zoo} from the parity conditions on the initial data of \cite{Henneaux:2018gfi} (for electromagnetism) and \cite{Henneaux:2018hdj,Henneaux:2019yax} (for gravity).  The equality of the past (future) limits of the soft U(1) charges and BMS charges with the corresponding Hamiltonian charges  follows from the matching of the fields but is more subtle to establish directly. This is because the natural bases in which the symmetry algebras are given are not the same at null infinity and at spatial  infinity, as shown in the earlier analysis of \cite{Troessaert:2017jcm} and further discussed in \cite{Henneaux:2018gfi,Henneaux:2018hdj}. [Matching conditions have been derived later from a different perspective in \cite{Prabhu:2019fsp,Prabhu:2021cgk,Capone:2022gme}, with \cite{Capone:2022gme} emphasizing also the connection between parity conditions and elimination of leading logarithmic singularities.]

Now in five spacetime dimensions, there is no parity condition on the leading orders in the asymptotic expansion of the fields as one goes to spatial infinity.  As shown in \cite{Henneaux:2019yqq} for electromagnetism, this leads to more general matching conditions between future and null infinities, involving both odd and even antipodal matchings that turn out to be associated with different inverse powers of $r$ as one goes to infinity  along null geodesics. The odd matching applies to the leading radiative branch which goes with the fractional power $r^{-\frac52}$ (in terms of the fields).  The even matching -- the only one in four dimensions -- applies to the Coulomb branch which goes with the power $r^{-3}$. We expect similar features to emerge in gravity.  The analysis is harder since spacetime is dynamical and a satisfactory definition of null infinity is more intricate.   Work on this question is in progress \cite{prep1}.

\section{Conclusions}
\label{sec:Conclusions}

In this paper, we have given boundary conditions for five-dimensional Einstein gravity at spatial infinity, which define asymptotic flatness.  These conditions lead to an infinite-dimensional asymptotic symmetry algebra, which we denoted BMS$_5$.  The property of this algebra, which turns out to be non-linear in terms of the chosen generators, have been studied.

 The boundary conditions are characterized by a relaxation of the decay rate as one goes to spatial infinity  with respect to the ``Coulomb rate''.   The new allowed terms take the form of a $\mathcal O(1)$ diffeomorphism. 
 These boundary conditions are rather general, since they contain the Myers-Perry metrics \cite{Myers:1986un}.  It is known thanks to the work of \cite{Corvino:2003sp} that under reasonable assumptions, rather general initial data can be put asymptotically to the Myers-Perry form by an appropriate change of coordinates. The coordinate transformation necessary to achieve this task might be improper, just as the boost necessary to bring a black hole at rest would be improper.  Our boundary conditions precisely allow for such improper gauge transformations, which it would be incorrect to gauge fix.  If we did not include the slowlier decay rate in our boundary conditions, we would see no sign of the supertranslations \cite{Tanabe:2009xb,Cameron:2021fhd}.
 
 Our work can be extended in principle to higher dimensions.   However, as one increases the dimension, the analysis becomes more and more technically intricate. This is because the gap between the pure (improper) diffeomorphism piece in the expansion of the fields (generated by $\mathcal O(1)$ vector fields) and the ``Coulomb'' piece widens by one power of $1/r$ as one increases the dimension by one.  Non-linearities (of increasing order) then proliferate.  Although these non-linearities are not seen in the linear Pauli-Fierz theory, it is nevertherless instructive to carry the analysis in this simplified context.  One finds that even though there are potentially more subleading supertranslations associated with the additional intermediate powers of $1/r$ in the expansion, preliminary analysis indicates that the size of the BMS group does not increase because the new terms in the diffeomorphism generators define proper gauge transformations.  Similar features hold for electromagnetism. 
 
Our analysis raises a number of questions, of which we will only list two. 

First, what is the meaning of the enlargement of the supertranslation subalgebra, parametrized now by four independent functions of the angles?  What are the corresponding Ward identities and soft theorems?  There are indications that a similar enlargement of the supertranslations -- and hence the similar problem of understanding the physical implications of these new symmetries -- might actually occur already in four dimensions, with the role of the leading supertranslations being played by logarithmic supertranslations, yielding an analogous structure for the supertranslation subalgebra \cite{Chandrasekaran:2021vyu}.  More work is required to understand this issue.  

Second, while there is an enlargement on the supertranslation side,  our analysis does not include the superrotations \cite{Barnich:2010eb,Campiglia:2015yka} as asymptotic symmetries.  How to include these and their higher dimensional generalization \cite{Capone:2021ouo} in a Hamiltonian treatment at spatial infinity remains an open question.

\section*{Acknowledgments}
It is a pleasure to thank Piotr Chru\'sciel, Hern\'an Gonz\'alez and Ricardo Troncoso for informative comments and useful discussions.  O. F. is grateful to the Coll\`ege de France for kind hospitality while this article was completed.  This work was partially supported by the ERC Advanced Grant ``High-Spin-Grav'', by FNRS-Belgium (conventions FRFC PDRT.1025.14 and IISN 4.4503.15), as well as by funds from the Solvay Family. JM is supported by the Austrian Science Fund (FWF) START project Y 1447-N.


\appendix

\section{Action of the diffeomorphisms on the fields (leading orders, unimproved)} \label{App0}
We write in this appendix the variations of the leading orders of the fields for a surface deformation (\ref{eq:Trans1})-(\ref{eq:Trans3}) without the correction terms (i.e., without the `more''). We consider successively the space-like diffeomorphisms ($\xi=0$) and the normal ones ($\xi^i = 0$). 

\subsection*{Action on the fields: I. Space-like diffeomorphisms (unimproved)}
Space-like diffeomorphisms $(\xi=0)$ generate the following transformation laws for the coefficients appearing in the spatial metric fall-off:

\begin{itemize}
\item Leading order (improper gauge part):
\begin{align}
\delta_{\xi^{i}}\theta_{AB} & =\mathcal{L}_{Y}\theta_{AB}+\left(\overline{D}_{A}I_{B}+\overline{D}_{B}I_{A}+2\overline{g}_{AB}W\right)\,,
\end{align}
where $I_{A}=\xbar D_{A}W$.
\item First subleading order (``core''):
\begin{align}
\delta_{\xi^{i}}\overline{\lambda} & =Y^{A}\partial_{A}\overline{\lambda}-W^{(1)}\,,\label{eq:dhrr-1-2}\\
\delta_{\xi^{i}}\overline{\lambda}_{A} & =\mathcal{L}_{Y}\overline{\lambda}_{A}-\theta_{AB}I^{B}+\overline{D}_{A}W^{(1)}-2I_{A}^{(1)}\,,\\
\delta_{\xi^{i}}\overline{h}_{AB} & =\mathcal{L}_{Y}\overline{h}_{AB}+\mathcal{L}_{I}\theta_{AB}+W\theta_{AB}+2\left(\overline{D}_{(A}I_{B)}^{(1)}+\overline{g}_{AB}W^{(1)}\right)\,.
\end{align}
\end{itemize}
where $\mathcal{L}_{I}$ stand for the Lie derivative along the vector
$I^{A}=\xbar D^{A}W$.

From $\delta_{\xi}\theta_{AB}$ we can read the transformation law for the field $U$, which is
\begin{equation}
\delta_{\xi^i}U=Y^A\partial_A U+2W\,.
\end{equation}
As mentioned above,   an arbitrary term of the form $\Sigma_{\ell, m}\alpha_{\ell, m}Y^{1\ell m}$ (spanning the kernel of the operator $\overline{D}_{A}\overline{D}_{B}+\overline{g}_{AB}$) can be added to the transformation law of $U$. This arbitrariness will play no role in the formulas giving the charges, as it should.

The transformation laws of the first terms in the expansion of the momentum are given
by
\begin{itemize}
\item Leading order (improper gauge part)
\begin{align}
\delta_{\xi^{i}}\kappa^{rr} & =\partial_{A}\left(Y^{A}\kappa^{rr}\right)\,,\\
\delta_{\xi^{i}}\kappa^{rA} & =\partial_{B}\left(Y^{B}\kappa^{rA}\right)-\partial_{B}Y^{A}\kappa^{rB}\,,\\
\delta_{\xi^{i}}\kappa^{AB} & =\partial_{C}\left(Y^{C}\kappa^{AB}\right)-\partial_{C}Y^{A}\kappa^{CB}-\partial_{C}Y^{A}\kappa^{CA}\,.
\end{align}

\item First subleading order (``core''):
\begin{align}
\delta_{\xi^{i}}\xbar\pi^{rr} & =\mathcal{L}_{Y}\xbar\pi^{rr}+\mathcal{L}_{I}\kappa^{rr}-2\kappa^{rA}\xbar D_{A}W+\kappa^{rr}W\,,\\
\delta_{\xi^{i}}\xbar\pi^{rA} & =\mathcal{L}_{Y}\xbar\pi^{rA}+\mathcal{L}_{I}\kappa^{rA}+I^{A}\kappa^{rr}-\kappa_{B}^{A}\xbar D^{B}W\,,\\
\delta_{\xi^{i}}\xbar\pi^{AB} & =\mathcal{L}_{Y}\xbar\pi^{AB}+\mathcal{L}_{I}\kappa^{AB}+I^{A}\kappa^{rB}+I^{B}\kappa^{rA}-W\kappa^{AB}\,.
\end{align}
\end{itemize}

From these transformation laws, we obtain that the field $V$ must transform as
\begin{equation}
\delta_{\xi^i} V=Y^A\partial_A V\,,
\end{equation}
an expression to which one can add an arbitrary constant, which parametrizes as we pointed out  the kernel of the operator relating $\kappa^{ij}$ to $V$.

\subsection*{Action on the fields: II. Time-like diffeomorphisms (unimproved)}
We will now write down the transformation law of the fields generated by time-like diffeomorphisms $(\xi^i=0)$. 

The coefficients of the spatial metric fall-off transform as
\begin{itemize}
\item Leading order (improper gauge part)
\begin{align}
\delta_{\xi}\theta_{AB} & =\frac{2b}{\sqrt{\overline{g}}}\left(\kappa_{AB}-\frac{1}{2}\overline{g}_{AB}\kappa_{C}^{C}\right)\,.
\end{align}
but $\delta_{\xi}\theta_{rA} \not= 0$ (see discussion at the end of this subsection).
\item First subleading order (``core''):
\begin{align}
\delta_{\xi}\overline{\lambda} & =\frac{b}{3\sqrt{\overline{g}}}\left(2\overline{\pi}^{rr}-\overline{\pi}_{A}^{A}-\theta_{AB}\kappa^{AB}\right)\,,\label{eq:dhrr-1-1-1}\\
\delta_{\xi}\overline{\lambda}_{A} & =\frac{2b}{\sqrt{\overline{g}}}\left(\overline{\pi}_{rA}+\theta_{AB}\kappa^{rB}-\frac{1}{2}\theta_{B}^{B}\kappa_{A}^{r}\right)+\frac{2T}{\sqrt{\overline{g}}}\kappa_{A}^{r}\,,\\
\delta_{\xi}\overline{h}_{AB} & =\frac{2b}{\sqrt{\overline{g}}}\left[\overline{\pi}_{AB}+2\theta_{C(A}\kappa_{B)}^{C}-\frac{1}{2}\theta_{AB}\kappa_{C}^{C}-\frac{1}{3}\overline{g}_{AB}\left(\overline{\pi}^{rr}+\overline{\pi}_{C}^{C}+\theta_{CD}\kappa^{CD}\right)\right] \nonumber\\
 & \quad-\frac{b}{\sqrt{\xbar g}}\theta_{D}^{D}\left(\kappa_{AB}-\frac{1}{2}\xbar g_{AB}\kappa_{C}^{C}\right)+\frac{2T}{\sqrt{\overline{g}}}\left(\kappa_{AB}-\frac{1}{2}\overline{g}_{AB}\kappa_{C}^{C}\right)\,.
\end{align}
\end{itemize}

The transformation law of the function $U$, which can be consistently extracted from the one of $\theta_{AB}$, is given by
\begin{equation}
    \delta_{\xi} U = 2bV  
\end{equation}
an expression to which one can again add an arbitrary element of  the kernel of the operator that relates $\theta_{AB}$ with $U$ in \eqref{thetaAB}.

Similarly, the transformation laws of the first orders of the momentum  in the asymptotic expansion turn out to be given by
\begin{itemize}
\item Leading order (improper gauge part)
\begin{align}
\delta_{\xi}\kappa^{rr}= & -\frac{\sqrt{\overline{g}}}{2}\left(b\theta_{A}^{A}-\partial_{A}b\overline{D}^{A}\theta_{B}^{B}\right)-\sqrt{\overline{g}}\,\overline{\triangle}T\,,\\
\delta_{\xi}\kappa^{rA}= & \frac{\sqrt{\overline{g}}}{2}\partial_{B}b\theta^{AB}-\sqrt{\overline{g}}\,\overline{D}^{A}T\,,\\
\delta_{\xi}\kappa^{AB}= & \frac{\sqrt{\overline{g}}}{2}\left[b\theta^{AB}-\partial_{C}b\overline{D}^{B}\theta^{AC}-\xbar g^{AB}\left(b\theta_{C}^{C}-\partial_{C}b\overline{D}^{C}\theta_{D}^{D}\right)\right]-\sqrt{\overline{g}}\left(\overline{g}^{AB}\overline{\triangle}T-\overline{D}^{A}\overline{D}^{B}T\right)\,.
\end{align}
\item First subleading order (``core''):
\begin{align}
\delta_{\xi}\xbar\pi^{rr} & =-\sqrt{\xbar g}\big(\xbar\triangle-3\big)T^{(1)}\nonumber \\
 & \quad+\frac{\sqrt{\xbar g}}{2}\big(-\theta\xbar\triangle\,T+\xbar D_{A}\theta\xbar D^{A}T+2\theta_{B}^{A}\xbar D_{A}\xbar D^{B}T\big)+\frac{\sqrt{\xbar g}}{4}\xbar D^{A}b\Big(12\xbar\lambda_{A}-2\xbar D_{A}\xbar h_{B}^{B}\nonumber \\
 & \quad-2\theta^{BC}\xbar D_{A}\theta_{BC}+\theta\xbar D_{A}\theta+4\xbar D_{B}\xbar h_{A}^{B}-2\theta_{A}^{B}\xbar D_{B}\theta\Big)+\frac{b}{24\sqrt{\xbar g}}\Big[-24(\kappa^{rr})^{2}+12\kappa^B_A\kappa^A_B\nonumber \\
 & \quad-24\kappa_{A}^{r}\kappa^{rA}+8\kappa^{rr}\kappa_{A}^{A}-4(\kappa_{A}^{A})^{2}+\xbar g\Big(288\xbar\lambda+24\xbar h_{A}^{A}+15\theta_{B}^{A}\theta_{A}^{B}-9\theta^{2}+72\xbar D_{A}\xbar\lambda^{A}\nonumber \\
 & \quad+12\xbar D_{A}\xbar D_{B}\xbar h^{AB}-3\xbar D^{A}\theta\xbar D_{B}\theta_{A}^{B}+3\xbar D_{A}\theta_{BC}\xbar D^{A}\theta^{BC}\Big)\Big]\,, \\
\delta_{\xi}\xbar\pi^{rA} & =\frac{\sqrt{\xbar g}}{2}\Big(3\theta_{B}^{A}\xbar D^{B}T-\theta\xbar D^{A}T\Big)-2\sqrt{\xbar g}\xbar D^{A}T^{(1)}\nonumber \\
 & \quad+\frac{\sqrt{\xbar g}}{4}\xbar D^{B}b\Big(4\xbar h_{B}^{A}+\theta\theta_{B}^{A}-4\theta_{B}^{C}\theta_{C}^{A}+2\xbar D_{B}\xbar\lambda^{A}-2\xbar D^{A}\xbar\lambda_{B}\Big)+\frac{b}{12\sqrt{\xbar g}}\Big[-16\kappa^{rr}\kappa^{rA}\nonumber \\
 & \quad+8\kappa_{B}^{B}\kappa^{rA}-24\kappa^{rB}\kappa_{B}^{A}+\xbar g\Big(12\xbar\lambda^{A}+12\xbar D_{B}\xbar h^{AB}-3\theta_{B}^{A}\xbar D^{B}\theta+6\xbar\triangle\xbar\,\lambda^{A}+6\xbar D_{B}\xbar D^{A}\xbar\lambda^{B}\nonumber \\
 & \quad-36\xbar D^{A}\xbar\lambda-12\xbar D^{A}\xbar h_{B}^{B}+3\theta_{C}^{B}\xbar D^{A}\theta_{B}^{C}-12\xbar D^{A}\xbar D_{B}\xbar\lambda^{B}\Big)\Big]\,,
\end{align}

\begin{align}
\delta_{\xi}\xbar\pi^{AB} & =\sqrt{\xbar g}\Big(\xbar D^{A}\xbar D^{B} T^{(1)}-\xbar g^{AB}\xbar\triangle\, T^{(1)}\Big)+\frac{\sqrt{\xbar g}}{2}\Big[2\theta^{AB}\xbar\triangle\, T-2\theta^{AC}\xbar D_{C}\xbar D^{B} T\nonumber \\
 & \quad-2\theta^{BC}\xbar D_{C}\xbar D^{A}T+\theta\xbar D^{A}\xbar D^{B} T-\xbar D^{A}\theta_{C}^{B}\xbar D^{C} T+\xbar g^{AB}\Big(2\theta_{AB}\xbar D^{A}\xbar D^{B} T\nonumber \\
 & \quad-\theta\xbar\triangle\,T-\xbar D_{A}\theta\xbar D^{A} T\Big)\Big]+\frac{b}{12\sqrt{\xbar g}}\Big[-24\kappa^{rA}\kappa^{rB}+8\kappa^{rr}\kappa^{AB}+8\kappa\kappa^{AB}\nonumber \\
 & \quad-24\kappa_{C}^{A}\kappa^{CB}+\xbar g\Big(24\xbar h^{AB}+12\theta\theta^{AB}-18\theta_{C}^{A}\theta^{CB}+6\xbar\triangle\,\xbar h^{AB}-6\xbar D_{C}\xbar D^{A}\xbar h^{BC}\nonumber \\
 & \quad-6\xbar D_{C}\xbar D^{B}\xbar h^{AC}+3\xbar D_{C}\theta\xbar D^{A}\theta^{BC}-6\xbar D^{A}\xbar\lambda^{B}-\xbar D^{B}\xbar\lambda^{A}+12\xbar D^{A}\xbar D^{B}\xbar\lambda\nonumber \\
 & \quad+6\xbar D^{A}\xbar D^{B}\xbar h-3\xbar D^{A}\theta^{CD}\xbar D^{B}\theta_{CD}\Big)+\frac{\xbar g^{AB}}{2}\Big(8(\kappa^{rr})^{2}+24\kappa^{rC}\kappa_{C}^{r}-8\kappa^{rr}\kappa\nonumber \\
 & \quad+12\kappa_{CD}\kappa^{CD}-4\kappa^{2}+\xbar g(-24\xbar h+21\theta_{CD}\theta^{CD}-9\theta^{2}+24\xbar D_{A}\xbar\lambda^{A}-24\xbar\triangle\,\xbar\lambda\nonumber \\
 & \quad+12\xbar D_{C}\xbar D_{D}\xbar h^{CD}-12\xbar\triangle\,\xbar h-3\xbar D^{C}\theta\xbar D_{C}\theta+3\xbar D_{C}\theta_{DF}\xbar D^{C}\theta^{DF})\Big)\Big]\nonumber \\
 & \quad+\frac{\sqrt{\xbar g}}{4}\partial_{C}b\Big[2\xbar D^{C}\xbar h^{AB}-2\theta^{AB}\xbar D^{C}\theta-2\xbar D^{A}\xbar h^{BC}-2\xbar D^{B}\xbar h^{AC}+2\theta_{D}^{A}\xbar D^{D}\theta^{BC}\nonumber \\
 & \quad+2\theta_{D}^{B}\xbar D^{D}\theta^{AC}-\theta\xbar D^{A}\theta^{BC}+2\theta^{CD}\xbar D^{A}\theta_{D}^{B}+\xbar g^{AB}\Big(4\xbar\lambda^{C}-4\xbar D^{C}\xbar\lambda-2\xbar D^{C}\xbar h\nonumber \\
 & \quad-\theta_{F}^{D}\xbar D^{C}\theta_{D}^{F}+\theta\xbar D^{C}\theta+4\xbar D_{D}\xbar h^{CD}-2\theta_{D}^{C}\xbar D^{D}\theta\Big)\Big]\,.
\end{align}
\end{itemize}

The transformations $\delta_{\xi}\kappa^{ij}$  induce the following transformation law for $V$
\begin{equation}
    \delta_{\xi} V = -T + \frac 1 2 \partial_A b  \xbar D^A U + \frac 1 2 b U\,,
\end{equation}
which, similarly to $\delta_\xi U$, can also differ by terms in the kernel of the operator that links $\kappa^{ij}$ and $V$, namely, an arbitrary constant. 

Now, as mentioned above, if one computes the variation of $g_{rA}$, one finds that it contains a $\mathcal O(1)$ term, in contradiction with the assumed boundary conditions.  In order to preserve the fall-off of the mixed radial-angular
component of the metric one must therefore perform a correcting spatial diffeomorphism
proporitonal to the Lorentz boost parameter
\begin{equation}
\xi^{r}=0\qquad\text{and\ensuremath{\qquad}}\xi^{A}=\frac{1}{r}I_{(b)}^{A}\,,
\end{equation}
where
\begin{equation}
    I^A _{(b)}= \frac {2b}{\sqrt {\xbar g}} \kappa^{rA} = 2 b \xbar D^A V\,. \label{Eq:CorrIA}
\end{equation}
This contribution is part of the ``more''-term in (\ref{eq:Trans3}).  It only affects the leading order $\theta_{AB}$ and the subleading
orders $\xbar h_{AB}$ of the metric,  as well as the subleading orders of the conjugate momentum.  Explicitly,
the variation of $\theta_{AB}$ acquires the extra contribution
\be
\frac{2}{\sqrt{\overline{g}}}\overline{D}_{A}\left(b\kappa_{B}^{r}\right)+\frac{2}{\sqrt{\overline{g}}}\overline{D}_{B}\left(b\kappa_{A}^{r}\right) \, ,
\ee
that of $\xbar h_{AB}$ is obtained by replacing $\mathcal{L}_I \theta_{AB}$ by $\mathcal{L}_{\xbar I}\theta_{AB}$ where $\mathcal{L}_{\xbar I}$ stands for the Lie derivative along
the vector $\xbar I^{A}=\xbar D^{A}W+I_{(b)}^{A}$, with similar replacements in the variations of $\xbar \pi^{ij}$ ($\mathcal{L}_I \kappa^{ij} \rightarrow \mathcal{L}_{\xbar I}\kappa^{ij}$).
The final form of the transformations will be shown explicitly in
Subsection \ref{subsec:improved} together with all the other contributions coming from the additional correcting gauge transformations.

\section{Action of the diffeomorphisms on the fields (leading orders, improved)}
\label{subsec:improved}

We now write down explicitly the transformations of the leading orders of the metric and its conjugate momentum, collecting all correction terms, i.e., including the ``more'' terms in (\ref{eq:Trans1})-(\ref{eq:Trans3}).

The correcting gauge transformations can most simply be displayed by shifting
the gauge parameters in the following way
\begin{align}
\xbar I^{A} & =\xbar D^{A}W+I_{(b)}^{A}\,,\\
\xbar T & =T+T_{(b)}\,,\\
\xbar T^{(1)} & =T^{(1)}+T_{(b,W)}^{(1)}\,,\\
\xbar I^{(1)A} & =I^{(1)A}+I_{(b,W)}^{(1)A}\,,
\end{align}
with $I_{(b)}^{A}$, $T_{(b)}$, $T_{(b,W)}^{(1)}$ and $I_{(b,W)}^{(1)A}$ 
given above.

This directly leads to the following somewhat cumbersome formulas, which play a central role in the derivation of the Poisson bracket of the generators.

\subsubsection*{Metric}
\begin{itemize}
\item Leading order
\begin{align}
\delta_{\xi,\xi^{i}}\theta_{AB} & =\mathcal{L}_{Y}\theta_{AB}+2\left(\xbar D_{A}\xbar D_{B}W+\overline{g}_{AB}W\right)\nonumber\\
 & \quad+\frac{2b}{\sqrt{\overline{g}}}\left(\kappa_{AB}-\frac{1}{2}\overline{g}_{AB}\kappa_{C}^{C}\right)+\frac{2}{\sqrt{\overline{g}}}\overline{D}_{A}\left(b\kappa_{B}^{r}\right)+\frac{2}{\sqrt{\overline{g}}}\overline{D}_{B}\left(b\kappa_{A}^{r}\right)\,.
\end{align}
From $\delta_{\xi}\theta_{AB}$ we can read the transformation law for the field $U$, which is
\begin{equation}
\delta_{\xi,\xi^{i}}U=Y^A\partial_A U+2W + 2bV\,.
\end{equation}
As mentioned above,   an arbitrary term of the form $\Sigma_{\ell, m}\alpha_{\ell, m}Y^{1\ell m}$ (spanning the kernel of the operator $\overline{D}_{A}\overline{D}_{B}+\overline{g}_{AB}$) can be added to the transformation law of $U$. This arbitrariness drops from the formulas giving the charges, as it should.
\item First subleading order (``core'')
\begin{align}
\delta_{\xi,\xi^{i}}\overline{\lambda} & =Y^{A}\partial_{A}\overline{\lambda}-W^{(1)}+\frac{b}{3\sqrt{\overline{g}}}\left(2\overline{\pi}^{rr}-\overline{\pi}_{A}^{A}-\theta_{AB}\kappa^{AB}\right)\,,\label{eq:dhrr-1-2-1}\\
\delta_{\xi,\xi^{i}}\overline{\lambda}_{A} & =\mathcal{L}_{Y}\overline{\lambda}_{A}+\overline{D}_{A}W^{(1)}-2I_{A}^{(1)}-\frac{2b}{\sqrt{\overline{g}}}\left(\xbar\pi_{A}^{r}+\theta_{B}^{A}\kappa^{rB}+\theta_{B}^{B}\kappa_{A}^{r}\right)+\frac{2T}{\sqrt{\overline{g}}}\kappa_{A}^{r}\,,\\
\delta_{\xi,\xi^{i}}\overline{h}_{AB} & =\mathcal{L}_{Y}\overline{h}_{AB}+\mathcal{L}_{\xbar I}\theta_{AB}+W\theta_{AB}+2\left(\overline{D}_{(A}\xbar I_{B)}^{(1)}+\overline{g}_{AB}W^{(1)}\right)\nonumber \\
 & \quad+\frac{2b}{\sqrt{\overline{g}}}\left[\overline{\pi}_{AB}+2\theta_{C(A}\kappa_{B)}^{C}-\frac{1}{2}\theta_{AB}\kappa_{C}^{C}-\frac{1}{3}\overline{g}_{AB}\left(\overline{\pi}^{rr}+\overline{\pi}_{C}^{C}+\theta_{CD}\kappa^{CD}\right)\right]\nonumber \\
 & \quad-\frac{2b}{\sqrt{\xbar g}}\theta_{D}^{D}\left(\kappa_{AB}-\frac{1}{2}\xbar g_{AB}\kappa_{C}^{C}\right)+\frac{2T}{\sqrt{\overline{g}}}\left(\kappa_{AB}-\frac{1}{2}\overline{g}_{AB}\kappa_{C}^{C}\right)\,.
\end{align}
\end{itemize}
where $\mathcal{L}_{\xbar I}$ stands for the Lie derivative along
the vector $\xbar I^{A}=\xbar D^{A}W+I_{(b)}^{A}$.

\subsubsection*{Conjugate momentum}
\begin{itemize}
\item Leading order
\begin{align}
\delta_{\xi,\xi^{i}}\kappa^{rr} & =\partial_{A}\left(Y^{A}\kappa^{rr}\right)-\frac{\sqrt{\overline{g}}}{2}\left(4b\theta_{A}^{A}-3\partial_{A}b\overline{D}^{A}\theta_{B}^{B}-b\xbar\triangle\theta_{A}^{A}\right)-\sqrt{\overline{g}}\,\overline{\triangle}T\,,\\
\delta_{\xi,\xi^{i}}\kappa^{rA} & =\partial_{B}\left(Y^{B}\kappa^{rA}\right)-\partial_{B}Y^{A}\kappa^{rB}+\frac{\sqrt{\overline{g}}}{2}\left[\partial_{B}b\theta^{AB}+\xbar D^{A}\left(b\theta_{B}^{B}\right)\right]-\sqrt{\overline{g}}\,\overline{D}^{A}T\,,\\
\delta_{\xi,\xi^{i}}\kappa^{AB} & =\partial_{C}\left(Y^{C}\kappa^{AB}\right)-\partial_{C}Y^{A}\kappa^{CB}-\partial_{C}Y^{A}\kappa^{CA}\nonumber \\
 & \quad+\frac{\sqrt{\overline{g}}}{2}\left[b\theta^{AB}-\partial_{C}b\overline{D}^{B}\theta^{AC}-\xbar g^{AB}\left(2b\theta_{C}^{C}-3\partial_{C}b\overline{D}^{C}\theta_{D}^{D}-b\xbar\triangle\theta_{C}^{C}\right)\right]\nonumber \\
 & \quad-\frac{\sqrt{\overline{g}}}{2}\left(2\xbar D^{(A}b\xbar D^{B)}\theta_{C}^{C}+b\xbar D^{A}\xbar D^{B}\theta_{C}^{C}\right)-\sqrt{\overline{g}}\left(\overline{g}^{AB}\overline{\triangle}T-\overline{D}^{A}\overline{D}^{B}T\right)\,.
\end{align}
From these transformation laws, we obtain that the field $V$ must transform as
\begin{equation}
\delta_{\xi,\xi^{i}} V=Y^A\partial_A V-T+2bU + \frac 1 2 \partial_A b  \xbar D^A U + \frac 1 2 b\xbar \triangle U\,,
\end{equation}
an expression to which one can add an arbitrary constant, which parametrizes as we pointed out  the kernel of the operator relating $\kappa^{ij}$ to $V$, and which again drops from the expression of the charges.

\item First subleading order (``core'')
\begin{align}
\delta_{\xi,\xi^{i}}\xbar\pi^{rr} & =\mathcal{L}_{Y}\xbar\pi^{rr}+\mathcal{L}_{\xbar I}\kappa^{rr}-2\kappa_{A}^{r}\xbar D^{A}W+\kappa^{rr}W-\sqrt{\xbar g}\big(\xbar\triangle-3\big)\xbar T^{(1)}\nonumber \\
 & \quad+\frac{\sqrt{\xbar g}}{2}\big(-\theta\xbar\triangle\,\xbar T+\xbar D_{A}\theta\xbar D^{A}\xbar T+2\theta_{B}^{A}\xbar D_{A}\xbar D^{B}\xbar T\big)+\frac{\sqrt{\xbar g}}{4}\xbar D^{A}b\Big(12\xbar\lambda_{A}-2\xbar D_{A}\xbar h_{B}^{B}\nonumber \\
 & \quad-2\theta^{BC}\xbar D_{A}\theta_{BC}+\theta\xbar D_{A}\theta+4\xbar D_{B}\xbar h_{A}^{B}-2\theta_{A}^{B}\xbar D_{B}\theta\Big)+\frac{b}{24\sqrt{\xbar g}}\Big[-24(\kappa^{rr})^{2}+12\kappa^B_A\kappa^A_B\nonumber \\
 & \quad-24\kappa_{A}^{r}\kappa^{rA}+8\kappa^{rr}\kappa_{A}^{A}-4(\kappa_{A}^{A})^{2}+\xbar g\Big(288\xbar\lambda+24\xbar h_{A}^{A}+15\theta_{B}^{A}\theta_{A}^{B}-9\theta^{2}+72\xbar D_{A}\xbar\lambda^{A}\nonumber \\
 & \quad+12\xbar D_{A}\xbar D_{B}\xbar h^{AB}-3\xbar D^{A}\theta\xbar D_{B}\theta_{A}^{B}+3\xbar D_{A}\theta_{BC}\xbar D^{A}\theta^{BC}\Big)\Big]\,,\\
\nonumber \\
\delta_{\xi,\xi^{i}}\xbar\pi^{rA} & =\mathcal{L}_{Y}\xbar\pi^{rA}+\mathcal{L}_{\xbar I}\kappa^{rA}+\xbar I^{A}\kappa^{rr}-\kappa_{B}^{A}\xbar D^{B}W+\frac{\sqrt{\xbar g}}{2}\Big(3\theta_{B}^{A}\xbar D^{B}\xbar T-\theta\xbar D^{A}\xbar T\Big)-2\sqrt{\xbar g}\xbar D^{A}\xbar T^{(1)}\nonumber \\
 & \quad+\frac{\sqrt{\xbar g}}{4}\xbar D^{B}b\Big(4\xbar h_{B}^{A}+\theta\theta_{B}^{A}-4\theta_{B}^{C}\theta_{C}^{A}+2\xbar D_{B}\xbar\lambda^{A}-2\xbar D^{A}\xbar\lambda_{B}\Big)+\frac{b}{12\sqrt{\xbar g}}\Big[-16\kappa^{rr}\kappa^{rA}\nonumber \\
 & \quad+8\kappa_{B}^{B}\kappa^{rA}-24\kappa^{rB}\kappa_{B}^{A}+\xbar g\Big(12\xbar\lambda^{A}+12\xbar D_{B}\xbar h^{AB}-3\theta_{B}^{A}\xbar D^{B}\theta+6\xbar\triangle\,\xbar\lambda^{A}+6\xbar D_{B}\xbar D^{A}\xbar\lambda^{B}\nonumber \\
 & \quad-36\xbar D^{A}\xbar\lambda-12\xbar D^{A}\xbar h_{B}^{B}+3\theta_{C}^{B}\xbar D^{A}\theta_{B}^{C}-12\xbar D^{A}\xbar D_{B}\xbar\lambda^{B}\Big)\Big]\,,
\end{align}
\end{itemize}
\begin{align}
\delta_{\xi,\xi^{i}}\xbar\pi^{AB} & =\mathcal{L}_{Y}\xbar\pi^{AB}+\mathcal{L}_{\xbar I}\kappa^{AB}+\xbar I^{A}\kappa^{rB}+\xbar I^{B}\kappa^{rA}-W\kappa^{AB}\nonumber \\
 & \quad+\sqrt{\xbar g}\Big(\xbar D^{A}\xbar D^{B}\xbar T^{(1)}-\xbar g^{AB}\xbar\triangle\,\xbar T^{(1)}\Big)+\frac{\sqrt{\xbar g}}{2}\Big[2\theta^{AB}\xbar\triangle\,\xbar T-2\theta^{AC}\xbar D_{C}\xbar D^{B}\xbar T\nonumber \\
 & \quad-2\theta^{BC}\xbar D_{C}\xbar D^{A}\xbar T+\theta\xbar D^{A}\xbar D^{B}\xbar T-\xbar D^{A}\theta_{C}^{B}\xbar D^{C}\xbar T+\xbar g^{AB}\Big(2\theta_{AB}\xbar D^{A}\xbar D^{B}\xbar T\nonumber \\
 & \quad-\theta\xbar\triangle\,\xbar T-\xbar D_{A}\theta\xbar D^{A}\xbar T\Big)\Big]+\frac{b}{12\sqrt{\xbar g}}\Big[-24\kappa^{rA}\kappa^{rB}+8\kappa^{rr}\kappa^{AB}+8\kappa\kappa^{AB}\nonumber \\
 & \quad-24\kappa_{C}^{A}\kappa^{CB}+\xbar g\Big(24\xbar h^{AB}+12\theta\theta^{AB}-18\theta_{C}^{A}\theta^{CB}+6\xbar\triangle\,\xbar h^{AB}-6\xbar D_{C}\xbar D^{A}\xbar h^{BC}\nonumber \\
 & \quad-6\xbar D_{C}\xbar D^{B}\xbar h^{AC}+3\xbar D_{C}\theta\xbar D^{A}\theta^{BC}-6\xbar D^{A}\xbar\lambda^{B}-\xbar D^{B}\xbar\lambda^{A}+12\xbar D^{A}\xbar D^{B}\xbar\lambda\nonumber \\
 & \quad+6\xbar D^{A}\xbar D^{B}\xbar h-3\xbar D^{A}\theta^{CD}\xbar D^{B}\theta_{CD}\Big)+\frac{\xbar g^{AB}}{2}\Big(8(\kappa^{rr})^{2}+24\kappa^{rC}\kappa_{C}^{r}-8\kappa^{rr}\kappa\nonumber \\
 & \quad+12\kappa_{CD}\kappa^{CD}-4\kappa^{2}+\xbar g(-24\xbar h+21\theta_{CD}\theta^{CD}-9\theta^{2}+24\xbar D_{A}\xbar\lambda^{A}-24\xbar\triangle\,\xbar\lambda\nonumber \\
 & \quad+12\xbar D_{C}\xbar D_{D}\xbar h^{CD}-12\xbar\triangle\,\xbar h-3\xbar D^{C}\theta\xbar D_{C}\theta+3\xbar D_{C}\theta_{DF}\xbar D^{C}\theta^{DF})\Big)\Big]\nonumber \\
 & \quad+\frac{\sqrt{\xbar g}}{4}\partial_{C}b\Big[2\xbar D^{C}\xbar h^{AB}-2\theta^{AB}\xbar D^{C}\theta-2\xbar D^{A}\xbar h^{BC}-2\xbar D^{B}\xbar h^{AC}+2\theta_{D}^{A}\xbar D^{D}\theta^{BC}\nonumber \\
 & \quad+2\theta_{D}^{B}\xbar D^{D}\theta^{AC}-\theta\xbar D^{A}\theta^{BC}+2\theta^{CD}\xbar D^{A}\theta_{D}^{B}+\xbar g^{AB}\Big(4\xbar\lambda^{C}-4\xbar D^{C}\xbar\lambda-2\xbar D^{C}\xbar h\nonumber \\
 & \quad-\theta_{F}^{D}\xbar D^{C}\theta_{D}^{F}+\theta\xbar D^{C}\theta+4\xbar D_{D}\xbar h^{CD}-2\theta_{D}^{C}\xbar D^{D}\theta\Big)\Big]\,.
\end{align}

The transformation laws of the next subleading terms in the asymptotic expansion of the fields can be found in appendix \ref{App1}.

\section{Decomposition of the spatial metric and spatial curvature} \label{App2}

We denote in this appendix polar coordinates as $x^i=(r,x^A)$,
where $x^A$ are coordinates on the $3$-sphere. We introduce the ``lapse'' $\lambda$ and the ``shift'' $\lambda^A$ adapted to the slicing of space by the spheres of constant radius $r$, 
\begin{equation}
\gamma_{AB} \equiv g_{AB}, \quad \lambda_A \equiv g_{rA}, \quad \lambda \equiv \frac{1}{\sqrt{g^{rr}}}.
\end{equation}
In terms of these, the metric and its inverse take the form:
\begin{equation}
g_{ij}= \left(\begin{array}{cc}
\lambda^2 + \lambda_C\lambda^C & \lambda_B \\
\lambda_A & \gamma_{AB}
  \end{array}
\right),\quad 
g^{ij}= \left(\begin{array}{cc}
\frac{1}{\lambda^2} & -\frac{\lambda^B}{\lambda^2} \\
-\frac{\lambda^A}{\lambda^2} & \gamma^{AB}+ \frac{\lambda^A\lambda^B}{\lambda^2}
  \end{array}
\right),
\end{equation}
where we used $\gamma_{AB}$ and its inverse $\gamma^{AB}$ to raise and
lower the angular indices $A, B, ...$
Introducing the extrinsic curvature of the 3-spheres $K_{AB}$, we can write
all the Christoffel symbols:
\begin{eqnarray}
K_{AB} & =  & \frac{1}{2 \lambda} \left( - \d_r g_{AB} + D_A \lambda_B
  + D_B \lambda_A\right)\\
\Gamma^r_{AB} & = & \frac{1}{\lambda} K_{AB} \\
\Gamma^A_{BC} & = & {}^\gamma\Gamma^A_{BC} - \frac{\lambda^A}{\lambda} K_{BC} \\
\Gamma^r_{rA} & = & \frac{1}{\lambda} \left( \d_A \lambda + K_{AB} \lambda^B\right) \\
\Gamma^r_{rr} & = &\frac{1}{\lambda} \d_r \lambda
+\frac{\lambda^A}{\lambda} \left( \d_A \lambda + K_{AB}
  \lambda^B\right) \\ 
\Gamma^A_{rB} & = & -\frac{\lambda^A}{\lambda} \left( \d_B \lambda + K_{BC}
  \lambda^C\right) + D_B \lambda^A - \lambda K^A_B \\
\Gamma^A_{rr} & = & -\lambda \left( \gamma^{AB} + \frac{\lambda^A
    \lambda^B}{\lambda^2}\right) \left( \d_B \lambda +
  K_{BC}\lambda^C\right) - \lambda^C \left( D^A \lambda_C - \lambda
  K^A_C\right) \nonumber \\ && \qquad- \frac{\lambda^A}{\lambda} \d_r \lambda + \gamma^{AB}
\d_r \lambda_B
\end{eqnarray}
where $D_A$ is the covariant derivative associated to $\gamma_{AB}$.The Ricci tensor is given by:
\begin{eqnarray}
	{}^{(4)}R_{AB} & = & \frac{1}{\lambda} \d_r K_{AB} + 2K_{AC}K^C_B -
	K K_{AB} - \frac{1}{\lambda} D_AD_B \lambda \nonumber \\ 
	&& \quad + {}^{\gamma} R_{AB} - \frac{1}{\lambda} 
	\cL_\lambda K_{AB},\\
	{}^{(4)}R_{rA} & = & \lambda \left( \d_A K - D_B
	K^B_A\right) + {}^{(4)}R_{AB} \lambda^B,\\
	{}^{(4)}R_{rr} & = & \lambda(\d_rK - \lambda^A\d_A K) - \lambda^2
	K^A_B K^B_A - \lambda D_AD^A\lambda \nonumber \\ && \quad-
	{}^{(4)}R_{AB}\lambda^A\lambda^B + 2 \, {}^{(4)}R_{rB}\lambda^B,
\end{eqnarray}
which implies that the Ricci scalar takes the form
\begin{equation}
	{}^{(4)}R=\frac{2}{\lambda}(\d_rK -\lambda^A\d_AK) + {}^{\gamma}R-
	K^A_BK^B_A-K^2-\frac{2}{\lambda}D_AD^A\lambda.
\end{equation}

\section{Vanishing of the linear divergence in the canonical generator} \label{App3}
In this appendix we will explicitly show that the linear divergence of the canonical generator
\begin{align}
I_{\text{linear}}&= \oint d^3x\Big[\sqrt{\xbar g}b\left(2\delta \xbar k+\frac{1}{2}\theta\delta \theta+\frac{1}{2}\theta_{AB}\delta \theta^{AB}\right)+\frac{4b}{\sqrt{\xbar g}}\kappa^r_A\delta \kappa^{rA}\label{Lin1}\\
&\qquad \qquad\qquad+2Y^A\delta \left(\xbar \pi^r_A+\theta_{AB}\kappa^{rB}\right)+2\xbar D^A W\delta \kappa^r_A+2W\delta \kappa^{rr}\Big]\label{Lin2}\,,
\end{align}
vanishes by virtue of the faster fall-off of the constraint equations. In fact, by using the fact that $\xbar D_A\kappa^{rA}-\kappa^{rr}=0$, we can see that the two last terms terms in \eqref{Lin2} cancel each other. For the remaining terms of \eqref{Lin2}, we make use of the faster fall-off of the radial component of momentum constraint $\mathcal{H}_A\sim \mathcal O\left(r^{-1}\right)$, which implies that
\begin{equation}
 \xbar D_B \xbar \pi^B_A +\xbar g_{AB} \xbar \pi^{rB}
            - \theta_{AB} \kappa^{rB}+ \frac 1 2 (2\xbar D_B \theta_{AC}
            - \xbar D_A \theta_{BC}) \kappa^{BC}=0\,.
\end{equation}
Then, we have that
\begin{equation}
\oint d^3x Y^A\delta \left(\xbar \pi^r_A+\theta_{AB}\kappa^{rB}\right)=\delta\oint d^3x \sqrt{\xbar g} \left(Y^A\theta_{AB}\xbar D^B V-\frac{1}{2}\xbar D^BY^A\xbar D_A \xbar D_C U\xbar D_B\xbar D^C V\right)\,.
\end{equation}
It can be shown this last integral vanishes (after some integration by parts) by virtue of the following properties of the Killing vector $Y^A$:
\begin{equation}
\xbar D^A Y^B=\xbar D^{[A} Y^{B]}\,,\quad \xbar D_A\xbar D_BY_C=-\xbar g_{AB}Y_C+\xbar g_{AC}Y_B\,.
\end{equation}

The remaining terms in \eqref{Lin1} along the boost parameter $b$ become
\begin{equation}
    \delta\oint d^3x \,\sqrt {\xbar g} \,b \Big(2\xbar h + 6 \xbar \lambda
        + \frac {1} 4 \theta^2 - \frac 3 4 \theta^A_B \theta^B_A + 2 \xbar D_AV \xbar D^A V + 2 \xbar D_A \xbar \lambda^A \Big). 
\end{equation}
Using the properties of $b$, we have
\begin{equation}
    \delta\oint d^3x \,\sqrt {\xbar g} \,b \Big(2\xbar h + 6 \xbar \lambda\Big) = 
    \delta\oint d^3x \,\sqrt {\xbar g} \,b \Big(\xbar D^A\xbar D^B \xbar h_{AB} - \xbar D^A\xbar D_A \xbar h 
        - 2 \xbar D_A \xbar D^A\xbar \lambda\Big) \,.
\end{equation}
Using the asymptotic constraint coming from the fact that $\mathcal H \sim \mathcal O \left(r^{-2}\right)$, we have that
\begin{multline}
    \xbar D^A(\xbar D^B \xbar h_{AB} - \xbar D_A \xbar h)- 2 \xbar D^A\xbar D_A \xbar \lambda + 2\xbar D_A \xbar \lambda^A \\
        =  2 \xbar D_AV \xbar D^BV 
         + \xbar D^A \xbar D^B V \xbar D_A \xbar D_B V - \xbar D_A \xbar D^A V \xbar D_B \xbar D^B V \\
        +  \frac 1 4 \theta^2 - \frac 3 4 \theta^{AB} \theta_{AB}
        - \frac 1 4 \xbar D^B \theta^{AC} \xbar D_A \theta_{BC} + \frac 1 4 \xbar D_A \theta^{AC} \xbar D^B \theta_{BC},
\end{multline}
the boundary term then reduces to two independent terms which are both zero:
\begin{flalign}
    \mathcal B_1 & = \delta\oint d^3x \,\sqrt {\xbar g} \,b \Big(\frac {1} 2 \theta^2 - \frac 3 2 \theta^A_B \theta^B_A - \frac 1 4 \xbar D^B \theta^{AC} \xbar D_A \theta_{BC} + \frac 1 4 \xbar D_A \theta^{AC} \xbar D^B \theta_{BC}\Big)
        \nonumber \\ & 
        = \frac 1 4 \delta\oint d^3x \,\sqrt {\xbar g} \,\xbar D^B\Big(b (-  \theta^{AC} \xbar D_A \theta_{BC} + \theta^C_B \xbar D^A \theta_{AC})\Big)
        \nonumber \\ & \qquad \qquad 
        + \frac 1 4 \delta\oint d^3x \,\sqrt {\xbar g} \left\{b \Big( \theta^2 - 3 \theta^A_B \theta^B_A\Big)
         +  \xbar D^B b \Big(\xbar D_B (\theta^{AC} \theta_{AC})  -\xbar D^A ( \theta^C_B \theta_{AC})\Big) \right\}
        \nonumber \\ & 
        = \frac 1 4 \delta\oint d^3x \,\sqrt {\xbar g} \,\xbar D^B\Big(b (-  \theta^{AC} \xbar D_A \theta_{BC} + \theta^C_B \xbar D^A \theta_{AC}) + \xbar D_B b \,\theta^{AC} \theta_{AC}  -\xbar D_A b \,\theta^A_C \theta^C_B\Big)
        \nonumber \\ & \qquad \qquad 
        + \frac 1 4 \delta\oint d^3x \,\sqrt {\xbar g} \, b \Big(\theta^2 - \theta^A_B \theta^B_A\Big) =0,
        \\
    \mathcal B_2 & = \delta\oint d^3x \,\sqrt {\xbar g} \,b \Big(4 \xbar D_AV \xbar D^A V+ \xbar D^A \xbar D^B V \xbar D_A \xbar D_B V - \xbar D_A \xbar D^A V \xbar D_B \xbar D^B V \Big)
    \nonumber \\ & 
        = \delta\oint d^3x \,\sqrt {\xbar g} \,\xbar D^A \Big((b \xbar D^B V - V\xbar D^B b ) (\xbar D_A \xbar D_B V - \xbar \gamma_{AB} \xbar D_C \xbar D^C V)\Big) = 0,
\end{flalign}
where we have used
\begin{multline}
    b(\theta^2 - \theta^A_B\theta^B_A) 
    = -b (\xbar D^A \xbar D^B U \xbar D_A\xbar D_B U - \xbar D_A \xbar D^A U \xbar D_B \xbar D^B U)
        + 4 bU \xbar D_A \xbar D^A U + 6b U^2\\
    = \xbar D^A \Big((b\xbar D^A U - U\xbar D^Ab) \xbar D_B \xbar D^B U - (b\xbar D^B U - U\xbar D^Bb) \xbar D_A\xbar D_B U + b \xbar D_A U^2 - 2 \xbar D_A b U^2\Big).
\end{multline}

\section{Transformation laws of the subleading terms} \label{App1}

In this appendix, we list the transformation laws of the subleading terms of the fields. These are necessary to obtain the asymptotic symmetry algebra. The subleading components of the metric fall-off transform as follows
\begin{align}
\delta_{\xi,\xi^i}h^{(2)}_{rr}&=Y^A \partial_A h^{(2)}_{rr}-2\xbar D^A  W \xbar \lambda_A+2\xbar D^A  W \partial_A\xbar \lambda-4 W \xbar \lambda +\frac{2T}{3\sqrt{\xbar g}}\Big(2\xbar \pi^{rr}-\xbar \pi^A_A-\theta_{AB}\kappa^{AB}\Big) \nonumber\\
&\quad -\frac{2b}{3\sqrt{\xbar g}}\theta \Big(2\xbar \pi^{rr}-\xbar \pi^A_A-\theta_{AB}\kappa^{AB}\Big)+\frac{2b}{3\sqrt{\xbar g}}\Big(2 \pi^{(2)rr}-\pi^{(2)A}_A+4\xbar \lambda\, \kappa^{rr}+6\partial_A \xbar \lambda \kappa^{rA}\nonumber\\
&\quad -2\xbar \lambda_A \kappa^{rA}-\xbar h_{AB}\kappa^{AB}-\theta_{AB}\xbar \pi^{AB}\Big)\,,\\
\delta_{\xi,\xi^i}h^{(2)}_{rA}&=\mathcal L_{Y}h^{(2)}_{rA}-2I^{(1)B}\theta_{AB}-W\xbar \lambda_A+2\xbar \lambda \partial_A W+\xbar \lambda_B \xbar D_A \xbar D^B W-\xbar h_{A}^B\xbar D_B W+\theta_A^B\theta_{B}^C\xbar D_C W \nonumber\\
&\quad+\frac{2}{\xbar g}\kappa^r_A\kappa^r_B \xbar D^B W +\xbar D_B \xbar \lambda_A \xbar D^B W +\frac{2}{\sqrt{\xbar g}}T^{(1)}\kappa^r_{A}+\frac{T}{\sqrt{\xbar g}}\Big[-\theta \kappa^r_A+2\big(\xbar \pi^r_A+\theta_{AB}\kappa^{rB}\big)\Big]\nonumber\\
&\quad+\frac{2}{\sqrt{\xbar g}}\Big(\partial_A b\,\xbar \lambda_B\kappa^{rB}-\partial_B b\, \xbar \lambda^B \kappa^r_A\Big)+\frac{2b}{3\sqrt{\xbar g}}\Big[3\xbar \lambda-\frac{3}{2}\xbar h^A _A+\frac{9}{8}\theta^2+\frac{3}{4}\theta_{AB}\theta^{AB}-6\theta_A^B\theta^C_B\kappa^r_C\nonumber \\
&\quad +2\kappa^{rr}\xbar \lambda_A-\kappa^B_B\xbar \lambda_A+3\kappa_{A}^B\xbar  \lambda_B+3\pi^{(2)r}_A-3\theta \xbar \pi^r_A-3\theta_{AB}\big(\xbar \pi^{rB}+\theta \kappa^{rB}\big)+3\xbar \lambda_B \xbar D_A\kappa^{rB} \nonumber \\
&\quad+3\kappa^{rB}\xbar D_B \xbar \lambda_A\Big]+\frac{2}{\sqrt{\xbar g}}\kappa^r_A \left(\xbar \triangle+3\right)^{-1}\left[\Big(\xbar D^C\xbar D^D+\xbar g^{CD}\Big)\Big(\frac{b}{2}\xbar h_{CD}+b \xbar D_C\xbar \lambda_D\Big)\right]\,,
\end{align}
\begin{align}
\delta_{\xi,\xi^i} h^{(2)}_{AB}&=\mathcal L_Y h^{(2)}_{AB}+\mathcal L_{I^{(1)}}\theta_{AB}+W^{(1)}\theta_{AB}-\frac{2}{\xbar g}\xbar g_{AB}\kappa^{rr}\kappa^r_C\xbar D^C W+\frac{2}{\xbar g}\kappa_{AB}\kappa^r_C \xbar D^C W+\xbar D_C \xbar h_{AB}\xbar D^C W\nonumber\\
&\quad+2\xbar \lambda_{(A}\xbar D_{B)}W-\theta^C_{(A}\xbar D^DW \xbar D_{B)}\theta_{CD}+2\xbar h_{C(A}\xbar D_{B)}\xbar D^C W-\theta_{CD}\theta^C_{(A}\xbar D_{B)}\xbar D^DW\nonumber\\
&\quad  -\frac{1}{2}\theta^C_D\xbar D_C \theta_{AB}\xbar D^D W+\frac{2T^{(1)}}{\sqrt{\xbar g}}\Big(\kappa_{AB}-2\kappa^{rr}\xbar g_{AB}\Big)+\frac{T}{3\sqrt{\xbar g}}\Big[6\xbar \pi_{AB}-6\kappa^{rr}\theta_{AB} \nonumber\\ 
&\quad+12\kappa_{C(A}\theta^C_{B)} -3\theta \kappa_{AB}+3\xbar g_{AB}\kappa^{rr}\theta-\xbar g_{AB}\big(2\xbar \pi^{rr}+2\xbar \pi^C_C+2\theta^C_D\kappa^D_C\big)\Big]\nonumber\\
&\quad  +\frac{2}{\sqrt{\xbar g}}\xbar g_{AB}\kappa^{rr}\xbar \lambda_C \xbar D^C b-\frac{2}{\sqrt{\xbar g}}\kappa_{AB}\xbar \lambda_C \xbar D^C b+\frac{4}{\sqrt{\xbar g}}\xbar h_{C(A}\kappa^{rC}\xbar D_{B)}b+\frac{4}{\sqrt{\xbar g}}\theta_{CD}\theta^C_{(A}\kappa^{rD}\xbar D_{B)}b\nonumber\\
&\quad+\frac{4}{\sqrt{\xbar g}}\theta_{C(A}\xbar \pi^{rC}\xbar D_{B)}b-\frac{b}{12\sqrt{\xbar g}}\Big[-24\kappa^{rr}\xbar h_{AB}-8\xbar \pi^{rr}\theta_{AB}+24\kappa^{rr}\theta\theta_{AB}-8\kappa^D_C\theta^C_D\theta_{AB} \nonumber \\
&\quad+24\theta^C_A\theta^D_B\kappa_{CD}+\xbar g_{AB}\big(8\kappa^{rr}\xbar \lambda-8\pi^{(2)rr}+12\kappa^{rr}\xbar h^C_C-9\kappa^{rr}\theta^2 +8\xbar \pi^{rr}\theta-6\kappa^{rr}\theta^C_D\theta^D_C-8\xbar h^C_D\kappa^D_C\nonumber \\
&\quad+8\theta \theta^C_D\kappa^D_C-16\kappa^r_B\xbar \lambda^B-8\pi^{(2)rr}-8\theta^C_D\xbar \pi^D_C+8\theta \xbar \pi^C_C\big)+48\xbar h^C_{(A}\kappa_{B)C}-48\theta \theta^C_{(A}\kappa_{B)C} \nonumber \\
&\quad-24\xbar \lambda \kappa_{AB}-12\xbar h^C_C \kappa_{AB}+6\theta^C_D\theta^D_C\kappa_{AB}+\theta^2\kappa_{AB}+48\kappa^r_{(A}\xbar \lambda_{B)}+24\pi^{(2)}_{AB}-8\theta_{AB}\xbar \pi^C_C\nonumber \\
&\quad +48\theta ^C_{(A}\xbar \pi_{B)C} -24\theta \xbar \pi_{AB}+24\kappa^r_C\xbar D^C\xbar h_{AB}+24\xbar \pi^r_C\xbar D^C\theta_{AB}+48\theta^C_{(A}\kappa^{rD}\xbar D_{B)}\theta_{C}^D\nonumber \\
&\quad +48\xbar h_{C(A} \xbar D_{B)}\kappa^{rC}+48\theta_{CD}\theta^C_{(A}\xbar D_{B)}\kappa^{rD}+48\theta^C_{(A}\xbar D_{B)}\xbar \pi^r_C\Big] \nonumber \\ 
&\quad+\frac{2}{\sqrt{\xbar g}}\big(\kappa_{AB}-\kappa^{rr}\xbar g_{AB} \big) \left(\xbar \triangle+3\right)^{-1}\left[\Big(\xbar D^C\xbar D^D+\xbar g^{CD}\Big)\Big(\frac{b}{2}\xbar h_{CD}+b \xbar D_C\xbar \lambda_D\Big)\right] \,,
\end{align}
while the transformation laws of the coefficientes of the momentum fall-off are given by
\begin{align}
\delta_{\xi,\xi^i}\pi^{(2)rA}&=\mathcal L_Y  \pi^{(2)rA}+\mathcal L_{\xbar I} \xbar \pi^{rA}+\xbar I^A \xbar \pi^{rr}+\mathcal L_{\xbar I^{(1)}} \kappa^{rA}+2\xbar I^{(1)A} \kappa^{rr}-\xbar \pi^A_B\xbar D^BW-\xbar \pi^{rA}W \nonumber \\
&\quad-\kappa^A_B\xbar D^BW^{(1)}+\frac{\sqrt{\xbar g}}{2}\Big(5\theta^A_B\xbar D^B \xbar T^{(1)}-2\theta \xbar D^A \xbar T^{(1)} \Big)+\frac{\sqrt{\xbar g}}{8}\Big( 8\xbar \lambda^A \xbar \triangle \, \xbar T+16 \xbar h^A_B \xbar D^B \xbar T \nonumber \\
&\quad+6\theta \theta^A_B \xbar D^B \xbar T-16\theta^A_B\theta^B_C\xbar D^C \xbar T +4\xbar D_B \xbar \lambda^A \xbar D^B \xbar T-8\xbar \lambda_B\xbar D^A  \xbar D^B \xbar T+8 \xbar \lambda  \xbar D^A \xbar T\nonumber \\
&\quad-4\xbar h ^B_B \xbar D^A \xbar T-\theta^2  \xbar D^A \xbar T-4\xbar D^A \xbar \lambda_B \xbar D^B \xbar T \Big)+\frac{2\xbar T}{3\sqrt{\xbar g}}\Big[\kappa^{rA}\big(-2\kappa^{rr}+\kappa^B_B\big)-3\kappa^A_B\kappa^{rB}\nonumber \\
&\quad+\frac{3\xbar g}{8}\Big(-4\xbar D_B \xbar h^{AB}+\theta^A_B\xbar D^B\theta-2\xbar \triangle\,\xbar \lambda^A-2\xbar D_B\xbar D^A\xbar \lambda^B+8\xbar D^A \xbar \lambda+4 \xbar D^A \xbar h^B_B \nonumber \\
&\quad-\theta^B_C\xbar D^A \theta^C_B+4\xbar D^A \xbar D_B\xbar \lambda^B\Big)\Big]+\frac{\sqrt{\xbar g}}{16}\xbar D^B b\Big(24 h^{(2)A}_B-24\xbar h^A_C\theta^C_B +8\theta \xbar h^A_B-8\xbar \lambda \theta^A_B\nonumber \\
&\quad+\xbar h^C_C \theta^A_B-2\theta^C_D\theta^D_C\theta^A_B +\theta^2 \theta^A_B-24\theta^A_C \xbar h^C_B+24\theta^A_C\theta^C_D\theta^D_B -8\theta \theta^A_C\theta^C_B+8\xbar D_Bh^{(2)A}_r\nonumber \\
&\quad-8\xbar \lambda^A \xbar D_B \theta-8 \theta^A_C \xbar D_B \xbar \lambda^C+4\theta \xbar D_B \xbar \lambda^A+8\theta^A_C\xbar D^C\xbar \lambda_B-8\theta^C_B\xbar D_C \xbar \lambda^A-8\xbar D^A \xbar h^{(2)}_{rB} \nonumber \\
&\quad+8\xbar \lambda_C \xbar D^A \theta^C_B-4\theta \xbar D^A\xbar \lambda_B +8\theta^C_B\xbar D^A\xbar \lambda_C\Big)+\frac{b}{24\sqrt{\xbar g}}\Big[24\theta \kappa^{rB}\kappa^A_B-48\theta^B_C\kappa^{rC}\kappa^A_B\nonumber \\
&\quad-48\kappa^A_B\xbar \pi^{rB}-32\kappa^{rr}\xbar \pi^{rA}+16\kappa^B_B\xbar \pi^{rA}-8\kappa^{rA}\Big(4\xbar \pi^{rr}-\theta_{BC}\kappa^{BC}-2\xbar \pi^B_B \nonumber \\
&\quad +\theta(-2\kappa^{rr}+\kappa^B_B)\Big)-48\kappa^{rB}\xbar \pi ^A_B+\xbar g\Big(24 h^{(2)A}_r-48\theta^A_B \xbar \lambda^B+24\theta \xbar \lambda^A+30\theta^A_B\xbar D^B \xbar h^C_C \nonumber \\
&\quad+36\xbar D_B \xbar h^{(2)A}_B-12\xbar h^A_B \xbar D^B\theta-3\theta \theta^A_B\xbar D^B \theta+12 \xbar \triangle h^{(2)A}_r+6\theta \xbar \triangle \,\xbar \lambda^A-12\theta^A_B\xbar D_C\xbar D^B \xbar \lambda^C \nonumber \\
&\quad+12\xbar D_B \xbar D^A h^{(2)B}_r+6\theta \xbar D_B\xbar D^A \xbar \lambda^B+60\theta^A_B\xbar D^B \xbar \lambda-6\xbar D_B \theta \xbar D^B \xbar \lambda^A-36\theta^A_B\xbar D^C\xbar h^B_C\nonumber \\
&\quad-24\theta^B_C\xbar D^C\xbar h^A_B+12\theta \xbar D^B\xbar h^A_B+12\theta^A_B\theta^B_C\xbar D^C \theta-12\theta^A_B \xbar \triangle \xbar \,\xbar \lambda^B-12 \xbar \lambda^B\xbar D_C\xbar D^A\theta^C_B \nonumber \\
&\quad+24\theta^A_B \xbar D^B\xbar D_C \xbar \lambda^C-12\theta^B_C\xbar D^C \xbar D_B \xbar \lambda^A-12\theta^B_C\xbar D^C \xbar D^A\xbar \lambda_B-36\xbar D^Ah^{(2)}_{rr}-24\theta \xbar D^A\xbar \lambda \nonumber \\
&\quad+30\theta^B_C \xbar D^A \xbar h^C_B-12 \theta \xbar D^A \xbar h^B_B-36\xbar D^Ah^{(2)B}_B+12\xbar h^B_C \xbar D^A \theta^C_B-12\theta^B_C\theta^C_D\xbar D^A\theta^D_B \nonumber \\
&\quad+3\theta \theta^B_C \xbar D^A\theta^C_B+6\xbar D_B \theta \xbar D^A \xbar \lambda^B-24\xbar D^A \xbar D_B h^{(2)B}_r+12\xbar \lambda^B\xbar D^A\xbar D_B\theta-12\theta \xbar D^A\xbar D_B \xbar \lambda^B \nonumber \\
&\quad+24\theta^B_C \xbar D^A \xbar D_B \xbar \lambda^C\Big)\Big] \,, \nonumber \\
\end{align}
with
\begin{align}
\xbar I^A&=\xbar D^A W+\frac{2b}{\sqrt{\xbar g}}\kappa^{rA}\,,\\
\xbar T&= T+T_{(b)}\,,\\
\xbar T^{(1)}&= T^{(1)}+T^{(1)}_{(b,W)}\,,\\
\xbar I ^{(1)A}&= I ^{(1)A}+ I ^{(1)A}_{(b,W)}\,.
\end{align}
where $T_{(b)}$, $T^{(1)}_{(b,W)}$ and $I^{(1)A}_{(b,W)}$ are given in \eqref{Tb}, \eqref{T1bW} and \eqref{I1bW}, respectively.

\section{An obstruction to the elimination of the non-linear terms in the algebra}
\label{sec:Obstruc}

We have suceeded in the main text in redefining away the non-linearities appearing in the BMS$_5$ algebra, except for the brackets 
between boosts and leading supertranslations. Specifically, these brackets
read
\begin{align}
\{Q_{b},Q_{T}\} & =Q_{\hat{W}}+\Lambda_{\lbrace b,T \rbrace}\,,\\
\{Q_{b},Q_{W}\} & =Q_{\hat{T}}+\Lambda_{\lbrace b,W \rbrace}\,,
\end{align}
where
\begin{equation}
\hat{W}=-bT\,,\quad\hat{T}=-4bW-\partial_{A}b\xbar D^{A}W-b\xbar\triangle W\,,
\end{equation}
and where the non-linear terms are given by
\begin{eqnarray}
\Lambda_{\lbrace b,T\rbrace} & = & -\oint d^{3}xb\kappa^{rA}\left(\theta_{A}^{B}\partial_{B}T-\theta\partial_{A}T\right)\,,\\
\Lambda_{\lbrace b,W\rbrace} & = & \oint d^{3}x\frac{1}{\sqrt{\xbar g}}\Big[bW(2\kappa^{rA}\kappa_{A}^{r}-2\xbar D_{A}\kappa^{rA}\xbar D_{B}\kappa^{rB}+\kappa^{AB}\kappa_{AB})+\partial_{A}b\xbar D^{A}W\kappa^{rA}\kappa_{A}^{r}\nonumber \\
 &  & +4b\xbar\triangle W\kappa^{rA}\kappa_{A}^{r}-4b\xbar D^{A}W\big(\kappa_{AB}\kappa^{rB}-\kappa_{B}^{r}\xbar D_{A}\kappa^{rB}+\kappa^{rB}\xbar D_{B}\kappa_{A}^{r}-\frac{1}{2}\kappa_{A}^{r}\xbar D_{B}\kappa^{rB}\big)\Big]\nonumber \\
 &  & +\oint d^{3}x\sqrt{\xbar g}\,\Big[\frac{1}{4}bW\big(5\theta^{2}-7\theta_{AB}\theta^{AB}+\xbar D_{A}\theta\xbar D^{A}\theta-\xbar D_{A}\theta_{BC}\xbar D^{A}\theta^{BC}\big)\nonumber \\
 &  & +\frac{1}{2}b\xbar D^{A}W\big(\theta_{AB}\xbar D^{B}\theta-2\theta^{BC}\xbar D_{A}\theta_{BC}\big)-\frac{1}{4}\partial_{A}b\xbar D^{A}W\theta_{BC}\theta^{BC}\nonumber \\
 &  & +\frac{1}{4}b\xbar\triangle W\big(2\theta^{2}-3\theta_{BC}\theta^{BC}\big)+\frac{1}{2}b\xbar D^{A}\xbar D^{B}W\big(\theta_{A}^{C}\theta_{CB}-\theta_{AB}\theta\big)\Big]\,.
\end{eqnarray}

The problem of linearizing a Poisson structure can be reformulated as a problem of Lie algebra cohomology \cite{Weinstein} (see review in the paper \cite{FernandesMonnier}).   In our case, the algebra is the semi-direct product of the homogeneous Lorentz algebra with its infinite-dimensional representation spanned by the supertranslations, which is non semi-simple. The cohomology is therefore expected to be non-trivial so one cannot invoke general theorems to infer that the non-linearities could be removed.  

Since the non-linearities appear only in the brackets  between boosts and leading supertranslations, one can reformulate the problem of eliminating them as a problem of Lie agebra cohomology for the finite-dimensional Lorentz subalgebra in the representation given by the supertranslations.  Because this representation is infinite-dimensional (and reducible but indecomposable), however, this does not provide much insight.  Furthermore, locality requirements on the form of the generators must be preserved.

Given the intricacy of this question, we shall examine it in a restricted context, namely, we shall only consider redefinitions of the supertranslation charges that take the form
\begin{align}
\tilde{Q}_{T} & =Q_{T}+\oint d^{3}x\sqrt{\xbar g}\,TF\,,\\
\tilde{Q}_{W} & =Q_{W}+\oint d^{3}x\sqrt{\xbar g}\,WG\,.
\end{align}
where the functions $F$ and $G$ are assumed to be the most general quadratic, rotation-invariant homogeneous polynomials in $\theta^{AB}$ and $\xbar D_{A}\kappa^{rB}$, i.e., 
\begin{align}
F & =A_{1}\theta_{B}^{A}\theta_{A}^{B}+B_{1}\theta^{2}+\frac{C_{1}}{\sqrt{\xbar g}}\xbar D_{A}\kappa_{B}^{r}\theta^{AB}+\frac{D_{1}}{\sqrt{\xbar g}}\xbar D_{A}\kappa^{rA}\theta+\frac{F_{1}}{\xbar g}\xbar D_{A}\kappa^{rA}\xbar D_{B}\kappa^{rB}+\frac{G_{1}}{\xbar g}\xbar D_{A}\kappa_{B}^{r}\xbar D^{A}\kappa^{rB}\,,\\
G & =A_{2}\theta_{B}^{A}\theta_{A}^{B}+B_{2}\theta^{2}+\frac{C_{2}}{\sqrt{\xbar g}}\xbar D_{A}\kappa_{B}^{r}\theta^{AB}+\frac{D_{2}}{\sqrt{\xbar g}}\xbar D_{A}\kappa^{rA}\theta+\frac{F_{2}}{\xbar g}\xbar D_{A}\kappa^{rA}\xbar D_{B}\kappa^{rB}+\frac{G_{2}}{\xbar g}\xbar D_{A}\kappa_{B}^{r}\xbar D^{A}\kappa^{rB}\,.
\end{align}
Here $A_i$, $B_i$, $C_i$, $D_i$, $F_i$ and $G_i$ are at this stage arbitrary constants.  We consider quadratic polynomials because the non-linearities that we want to eliminate are themselves quadratic, and the number of derivatives is also taken to match.

One can view the new terms as generating improper gauge transformations with parameters
\begin{align}
T_{T}^{(1)} & =-A_{1}\left(\xbar\triangle+3\right)^{-1}\left[\left(\xbar D^{A}\xbar D^{B}+\xbar g^{AB}\right)\left(2T\theta_{AB}\right)\right]-2B_{1}T\theta \nonumber \\
 & \quad-\frac{C_{1}}{\sqrt{\xbar g}}\left(\xbar\triangle+3\right)^{-1}\left[\left(\xbar D^{A}\xbar D^{B}+\xbar g^{AB}\right)\left(T\xbar D_{A}\kappa_{B}^{r}\right)\right]-\frac{D_{1}}{\sqrt{\xbar g}}T\xbar D_{A}\kappa^{rA}\,,\label{eq:NewT1} \\
I_{T}^{(1)} & =\xbar D_{A}\left[-\frac{1}{2}C_{1}\xbar D_{B}\left(T\theta^{AB}\right)-\frac{1}{2}D_{1}\xbar D^{A}\left(T\theta\right)-\frac{F_{1}}{\sqrt{\xbar g}}\xbar D^{A}\left(T\xbar D_{B}\kappa^{rB}\right)-\frac{G_{1}}{\sqrt{\xbar g}}\xbar D_{B}\left(T\xbar D^{B}\kappa^{rA}\right)\right]\,,
\end{align}
for $F$ and 
\begin{align}
T_{W}^{(1)} & =-A_{2}\left(\xbar\triangle+3\right)^{-1}\left[\left(\xbar D^{A}\xbar D^{B}+\xbar g^{AB}\right)\left(2W\theta_{AB}\right)\right]-2B_{2}W\theta \nonumber \\
 & \quad-\frac{C_{2}}{\sqrt{\xbar g}}\left(\xbar\triangle+3\right)^{-1}\left[\left(\xbar D^{A}\xbar D^{B}+\xbar g^{AB}\right)\left(W\xbar D_{A}\kappa_{B}^{r}\right)\right]-\frac{D_{2}}{\sqrt{\xbar g}}W\xbar D_{A}\kappa^{rA}\,,\\
I_{W}^{(1)} & =\xbar D_{A}\left[-\frac{1}{2}C_{2}\xbar D_{B}\left(W\theta^{AB}\right)-\frac{1}{2}D_{2}\xbar D^{A}\left(W\theta\right)-\frac{F_{2}}{\sqrt{\xbar g}}\xbar D^{A}\left(W\xbar D_{B}\kappa^{rB}\right)-\frac{G_{2}}{\sqrt{\xbar g}}\xbar D_{B}\left(W\xbar D^{B}\kappa^{rA}\right)\right]\,.\label{eq:NewI1}
\end{align}
for $G$. 
Of course, the surface integrals are accompanied by the appropriate bulk terms proportional to the constraints.

\subsection*{The non-linearities cannot be eliminated}

We now investigate whether it is possible to determine the arbitrary coefficients in $F$ and $G$ in such a way that
\begin{align}
\delta_{b}\tilde{Q}_{T} & =\tilde{Q}_{\hat{W}}\,,\label{eq:Eq1}\\
\delta_{b}\tilde{Q}_{W} & =\tilde{Q}_{\hat{T}}\,,\label{eq:Eq2}
\end{align}
where $\hat{W}=bT$ and $\hat{T}=4bW+\partial_{A}b\xbar D^{A}W+b\xbar\triangle W$. 

We start taking the transformation of $\tilde{Q}_{T}$ under Lorentz
boosts
\begin{align}
\delta_{b}\tilde{Q}_{T} & =\delta_{b}Q_{T}+\oint d^{3}x\sqrt{\xbar g}\,T\delta_{b}F\\
 & =Q_{\hat{W}}+\Lambda_{\{T,b\}}+\oint d^{3}x\sqrt{\xbar g}\,T\delta_{b}F\\
 & =\tilde{Q}_{\hat{W}}-\oint d^{3}x\sqrt{\xbar g}\,\hat{W}G+\Lambda_{\{T,b\}}+\oint d^{3}x\sqrt{\xbar g}\,T\delta_{b}F\,,
\end{align}
where the relation $\delta_{b}Q_{T}=Q_{\hat{W}}+\Lambda_{\{T,b\}}$ was used.
Note that it is possible to write the non-linear term as
\begin{equation}
\Lambda_{\{T,b\}}=-\oint d^{3}x\sqrt{\xbar g}\,T\xbar D_{A}f^{A}\,,
\end{equation}
with $f^{A}=b\left(\kappa^{rB}\theta_{B}^{A}-\kappa^{rA}\theta\right)$.
Then, if we demand \eqref{eq:Eq1}, we obtain the first condition
on the functions $F$ and G:
\begin{equation}
\oint d^{3}x\sqrt{\xbar g}\,T\left(\delta_{b}F-\xbar D_{A}f^{A}-bG\right)=0\,.
\end{equation}

We now proceed similarly for $\tilde{Q}_{W}$:
\begin{align}
\delta_{b}\tilde{Q}_{W} & =\delta_{b}Q_{W}+\oint d^{3}x\sqrt{\xbar g}\,W\delta_{b}G\\
 & =Q_{\hat{T}}+\Lambda_{\{W,b\}}+\oint d^{3}x\sqrt{\xbar g}\,W\delta_{b}G\\
 & =\tilde{Q}_{\hat{T}}-\oint d^{3}x\sqrt{\xbar g}\,\hat{T}F+\Lambda_{\{W,b\}}+\oint d^{3}x\sqrt{\xbar g}\,W\delta_{b}G\,,
\end{align}
where the relation $\delta_{b}Q_{W}=Q_{\hat{T}}+\Lambda_{\{W,b\}}$ was used.
Note that the non-linear term in this case can be written as
\begin{equation}
\Lambda_{\{W,b\}}=\oint d^{3}x\sqrt{\xbar g}\,W\left(\xbar D_{A}\xbar D_{B}+\xbar g_{AB}\right)f^{AB}\,,
\end{equation}
with 
\begin{equation}
f^{AB}=-b\Big[\frac{3}{4}\theta^A_C\theta^{BC}-\theta^{AB}\theta +6V\xbar D^A\xbar D^BV+\xbar g^{AB}\Big(\frac{3}{4}\theta^2-\frac{1}{2}\theta^D_C\theta^C_D+V^2+V\xbar \triangle V -\frac{7}{2}\xbar D_C V\xbar D^CV\Big)\Big]\,.
\end{equation}
Thus, if we now demand \eqref{eq:Eq2}, we obtain the second condition
\begin{equation}
\oint d^{3}x\sqrt{\xbar g}\,W\left[\delta_{b}G+\left(\xbar D_{A}\xbar D_{B}+\xbar g_{AB}\right)f^{AB}-4bF-\partial_{A}b\xbar D^{A}F-b\xbar\triangle F\right]=0\,.
\end{equation}

In summary, we have that the two following conditions must be simultaneously
satisfied
\begin{align}
\delta_{b}F-\xbar D_{A}f^{A}-bG & =0\,,\\
\delta_{b}G+\left(\xbar D_{A}\xbar D_{B}+\xbar g_{AB}\right)f^{AB}-4bF-\partial_{A}b\xbar D^{A}F-b\xbar\triangle F & =0\,.
\end{align}
Now, if we solve the first condition, we find upon computing $\delta_{b}F$ and $\delta_{b}G$ from $\delta_{b}\theta_{AB}$ and $\delta_{b}\kappa^{rA}$ given in Appendix \ref{subsec:improved}, that the coefficients $A_i$, $B_i$, $C_i$, $D_i$, $F_i$ and $G_i$ are completely determined.  The only non-vanishing
ones are
\begin{equation}
A_{1}=\frac{1}{8}\,,\quad B_{1}=-\frac{1}{8}\,,\quad C_{2}=-\frac{1}{2}\,,\quad D_{2}=\frac{1}{2}\,. \label{eq:ValueCoeff}
\end{equation}
This yields
\begin{align}
\tilde{Q}_{T} & =Q_{T}+\frac{1}{8}\oint d^{3}x\sqrt{\xbar g}\,T\left(\theta^{AB}\theta_{AB}-\theta^{2}\right)\,,\\
\tilde{Q}_{W} & =Q_{W}-\frac{1}{2}\oint d^{3}x\sqrt{\xbar g}\,W\xbar D_{A}\kappa_{B}^{r}\left(\theta^{AB}-\theta\xbar g^{AB}\right)\,.
\end{align}

The bracket with the boosts now read
\begin{align}
\{\tilde{Q}_{b},\tilde{Q}_{T}\} & =\tilde{Q}_{\hat{W}}\,,\\
\{\tilde{Q}_{b},\tilde{Q}_{W}\} & =\tilde{Q}_{\hat{T}}+\tilde{\Lambda}_{\lbrace b,W \rbrace}\,,
\end{align}
with
\begin{align}
\tilde{\Lambda}_{\lbrace b,W \rbrace} & =-2\oint d^{3}x\frac{bW}{\sqrt{\xbar g}}\kappa^{rA}\kappa_{A}^{r}+\tilde{\Lambda}_{\lbrace b,W \rbrace}^{\theta}\,.
\end{align}
Here,
\begin{eqnarray}
\tilde{\Lambda}_{\lbrace b,W\rbrace}^{\theta} & = & \oint d^{3}x\sqrt{\xbar g}\,\Big[\frac{1}{4}bW\big(5\theta^{2}-7\theta_{AB}\theta^{AB}+\xbar D_{A}\theta\xbar D^{A}\theta-\xbar D_{A}\theta_{BC}\xbar D^{A}\theta^{BC}\big)\nonumber \\
 &  & +\frac{1}{2}b\xbar D^{A}W\big(\theta_{AB}\xbar D^{B}\theta-2\theta^{BC}\xbar D_{A}\theta_{BC}\big)-\frac{1}{4}\partial_{A}b\xbar D^{A}W\theta_{BC}\theta^{BC}\nonumber \\
 &  & +\frac{1}{4}b\xbar\triangle W\big(2\theta^{2}-3\theta_{BC}\theta^{BC}\big)+\frac{1}{2}b\xbar D^{A}\xbar D^{B}W\big(\theta_{A}^{C}\theta_{CB}-\theta_{AB}\theta\big)\nonumber \\
 &  & -\frac{1}{4}\left(\partial_{C}b\theta_{B}^{C}+\partial_{B}(b\theta)\right)\xbar D_{A}W\left(\theta^{AB}-\xbar g^{AB}\theta\right)\nonumber \\
 &  & +\frac{1}{8}\left(4bW+\partial_{C}b\xbar D^{C}W+b\xbar\triangle W\right)\left(\theta^{AB}\theta_{AB}-\theta^{2}\right)\Big]\,.
\end{eqnarray}

The non-linearity in $\{\tilde{Q}_{b},\tilde{Q}_{W}\}$ has not been eliminated. It is therefore not possible to eliminate simultaneously all the non-linearities within the class of redefinitions of the generators that we have considered.

\subsection*{Insight from the Jacobi identity}

The above redefinition (\ref{eq:ValueCoeff}) makes at the same time the supertranslations non-abelian.
Indeed, the new supertranslation algebra is given by
\begin{align}
\{\tilde{Q}_{T_{1}},\tilde{Q}_{T_{2}}\} & =0\,,\\
\{\tilde{Q}_{T},\tilde{Q}_{W}\} & =Q_{\hat{T}^{(1)}}\,,\\
\{\tilde{Q}_{W_{1}},\tilde{Q}_{W_{2}}\} & =Q_{\hat{I}^{(1)}}\,,
\end{align}
where
\begin{align}
\hat{T}^{(1)} & =-\frac{1}{2}W\xbar\triangle T+\frac{1}{2}T\xbar\triangle W+\frac{3}{2}TW\nonumber \\
 & \quad+(\xbar\triangle+3)^{-1}\left[\left(\xbar D^{A}\xbar D^{B}+\xbar g^{AB}\right)\left(\frac{1}{2}W\xbar D_{A}\xbar D_{B}T-\frac{1}{2}T(\xbar D_{A}\xbar D_{B}W+\xbar g_{AB}W)\right)\right]\,,\label{eq:T1hat}\\
\hat{I}^{(1)} & =W_{1}\xbar\triangle W_{2}-W_{2}\xbar\triangle W_{1}\,.\label{eq:I1hat}
\end{align}

The presence of the non-abelian terms in the new supertranslation algebra forces in fact the presence of non-linearities in the brackets with the boosts.  Indeed, the Jacobi identity would otherwise be violated.

One can for instance check explicitly that the Jacobi identity involving
boosts, time supertranslations and spatial supertranslations,
\begin{equation}
\{Q_{b},\{\tilde{Q}_{T},\tilde{Q}_{W}\}\}+\{\tilde{Q}_{W},\{Q_{b},\tilde{Q}_{T}\}\}+\{\tilde{Q}_{T},\{\tilde{Q}_{W},Q_{b}\}\}=0\,,\label{eq:Jacobi}
\end{equation}
is satisfied only if the non-linear term $\tilde{\Lambda}_{(W,b)}$ is
present. In fact, we have that the first term of the Jacobi identity
reduces to
\begin{equation}
\{Q_{b},\{\tilde{Q}_{T},\tilde{Q}_{W}\}\}=\{Q_{b},Q_{\hat{T}^{(1)}}\}=Q_{\hat{I}^{(1)}}=2\oint d^{3}x\sqrt{\xbar g}\,b(\xbar\triangle+3)\hat{T}^{(1)}V\,,
\end{equation}
where we used that $\hat{I}^{(1)}=-b(\xbar\triangle+3)\hat{T}^{(1)}$,
with $\hat{T}^{(1)}$ given in \eqref{eq:T1hat}.

The second term in the the Jacobi identity gives
\begin{align}
\{\tilde{Q}_{W},\{Q_{b},\tilde{Q}_{T}\}\}&=\{\tilde{Q}_{W},\tilde{Q}_{\hat{W}}\}\\
&=Q_{\hat{I}^{(1)}}\\
&= 2\oint d^{3}x\sqrt{\xbar g}\left(\xbar\triangle W\hat{W}-\xbar\triangle\hat{W}W\right)V\,,\\
&= 2\oint d^{3}x\sqrt{\xbar g}\left[b\left(W\xbar\triangle T-T\xbar\triangle W-3TW\right)+2\partial_{A}b\xbar D^{A}TW\right]V\label{eq:2ndTermJacobi}
\end{align}
where we used that $\hat{W}=-bT$ and the expression for $\hat{I}^{(1)}$
given in \eqref{eq:I1hat}. 

Finally, the third term in Jacobi identity is given by
\begin{align}
\{\tilde{Q}_{T},\{\tilde{Q}_{W},Q_{b}\}\}&=-\{\tilde{Q}_{T},\tilde{Q}_{\hat{T}}+\tilde{\Lambda}_{\lbrace b,W \rbrace}\}\\
 & =-\{\tilde{Q}_{T},\tilde{\Lambda}_{\lbrace b,W \rbrace}\}\\
 & =4\oint d^{3}x\sqrt{\xbar g}bW\xbar D_{A}T\xbar D^{A}V\\
 & =-4\oint d^{3}x\sqrt{\xbar g}\left(\partial_{A}bW\xbar D^{A}T+b\partial_{A}W\xbar D^{A}T+bW\xbar\triangle T\right)V\,.
\end{align}
One can immediately see that the non-linear term $\tilde{\Lambda}_{\lbrace b,W \rbrace}$
is needed to eliminate the term along $\partial_{A}b$ in \eqref{eq:2ndTermJacobi}.
Summing over the three components of the Jacobi identity \eqref{eq:Jacobi},
we conclude that it is identically satisfied, as it should, but only because $\tilde{\Lambda}_{(b,W)}$ is present.

\subsection*{Abelian supertranslations}
As we argued in the text, the condition that the leading supertranslations are abelian plays an important role in the invariance properties of the energy and is thus a desirable feature.
One can ask which values of the coefficients $A_i$, $B_i$, $C_i$, $D_i$, $F_i$ and $G_i$ preserve this property, i.e., are such that 
\begin{equation}
\delta_{T_{2}}\tilde{Q}_{T_{1}}=\delta_{T}\tilde{Q}_{W}\left(=-\delta_{T}\tilde{Q}_{W}\right)=\delta_{W_{2}}\tilde{Q}_{W_{1}}=0\,.
\end{equation}

A direct computation yields
\begin{align}
\delta_{T_{2}}\tilde{Q}_{T_{1}} & =\delta_{I_{T_{2}}^{(1)}}Q_{T}+\oint d^{3}x\sqrt{\xbar g}\,T_{1}\delta_{T_{2}}F=\oint d^{3}x\sqrt{\xbar g}\,T_{1}\left(\delta_{T_{2}}F-2I_{T_{2}}^{(1)}\right)=0\,,\\
\delta_{W}\tilde{Q}_{T} & =\delta_{I_{W}^{(1)}}Q_{T}+\oint d^{3}x\sqrt{\xbar g}\,T\delta_{W}F=\oint d^{3}x\sqrt{\xbar g}\,T\left(\delta_{W}F-2I_{W}^{(1)}\right)=0\,,\\
\delta_{T}\tilde{Q}_{W} & =\delta_{T_{T}^{(1)}}Q_{W}+\oint d^{3}x\sqrt{\xbar g}\,W\delta_{T}G=\oint d^{3}x\sqrt{\xbar g}\,W\left(\delta_{T}G+2\left(\xbar\triangle+3\right)T_{T}^{(1)}\right)=0\,,\\
\delta_{W_{2}}\tilde{Q}_{W_{1}} & =\delta_{T_{W_{2}}^{(1)}}Q_{W_{1}}+\oint d^{3}x\sqrt{\xbar g}\,W_{1}\delta_{W_{2}}G=\oint d^{3}x\sqrt{\xbar g}\,W_{1}\left(\delta_{W_{2}}G+2\left(\xbar\triangle+3\right)T_{W_{2}}^{(1)}\right)=0\,.
\end{align}

Replacing the above parameters into these conditions and using (\ref{eq:NewT1})-(\ref{eq:NewI1}), we get that the only non-vanishing parameters
turn out to be $A_{2}$ and $B_{2}$, subject to the condition
\begin{equation}
A_{2}+B_{2}=0\,.
\end{equation}
Thus $F=0$ and $G=B \left(\theta^{2}-\theta_{AB}\theta^{AB}\right)$ with $B$ an arbitrary constant, so that the most general abelian supertranslation charges are given
by
\begin{align}
\tilde{Q}_{T} & =Q_{T}\,,\\
\tilde{Q}_{W} & =Q_{W}+B_{2}\oint d^{3}x\sqrt{\xbar g}\,W\left(\theta^{2}-\theta_{AB}\theta^{AB}\right)\,.
\end{align}

The abelian condition is therefore rather strong and in any case, incompatible with the elimination on the non-linearities, which needs different values of the coefficients $A_i$, $B_i$, $C_i$, $D_i$, $F_i$ and $G_i$.


\end{document}